\documentclass[11pt,oneside,letterpaper]{article}
\usepackage{amssymb}
\usepackage{amsmath}
\usepackage[dvips]{graphicx}
\usepackage{setspace}
\usepackage{fancyhdr}
\usepackage{ifpdf}
\usepackage{graphicx}

\newcommand{\bi}{\begin{itemize}}
\newcommand{\ei}{\end{itemize}}
\newcommand{\bea}{\begin{eqnarray}}
\newcommand{\eea}{\end{eqnarray}}
\newcommand{\be}{\begin{equation}}
\newcommand{\ee}{\end{equation}}

\numberwithin{equation}{section}
\allowdisplaybreaks  

\begin{document}

\vspace*{-2cm}
\begin{flushright}
\end{flushright}

\vspace*{2.5cm}
\begin{center}
{  \textbf{Thermodynamics of the Maxwell-Gauss-Bonnet anti-de Sitter Black Hole with Higher Derivative Gauge Corrections }\\}
\vspace*{1.7cm}
Dionysios Anninos and Georgios Pastras\\
\vspace*{1.0cm}
Jefferson Physical Laboratory, Harvard University, Cambridge, MA 02138,
USA
\vspace*{0.8cm}
\tt anninos@physics.harvard.edu, pastras@fas.harvard.edu
\end{center}
\vspace*{1.5cm}

\begin{abstract}

\noindent

\end{abstract}
The local and global thermal phase structure for asymptotically anti-de Sitter black holes charged under an abelian gauge group, with both Gauss-Bonnet and quartic field strength corrections, is mapped out for all parameter space. We work in the grand canonical ensemble where the external electric potential is held fixed. The analysis is performed in an arbitrary number of dimensions, for all three possible horizon topologies - spherical, flat or hyperbolic. For spherical horizons, new metastable configurations are exhibited both for the pure Gauss-Bonnet theory as well as the pure higher derivative gauge theory and combinations thereof. In the pure Gauss-Bonnet theory with negative coefficient and five or more spatial dimensions, two locally thermally stable black hole solutions are found for a given temperature. Either one or both of them may be thermally favored over the anti-de Sitter vacuum - corresponding to a single or a double decay channel for the metastable black hole. Similar metastable configurations are uncovered for the theory with pure quartic field strength corrections, as well combinations of the two types of corrections, in three or more spatial dimensions. Finally, a secondary Hawking-Page transition between the smaller thermally favored black hole and thermal anti-de Sitter space is observed when both corrections are turned on and their couplings are both positive.

\newpage
\setcounter{page}{1}
\pagenumbering{arabic}

\tableofcontents
\setcounter{tocdepth}{2}

\onehalfspacing


\section{Introduction}

A common feature of UV completions of our current theoretical framework, the Einstein action coupled to the matter content of the Standard Model, are higher derivative corrections to both the gravitational and matter low energy effective actions \cite{Boulware:1985wk,Zwiebach:1985uq,Gross:1986iv,Grisaru:1986vi,Freeman:1986zh}. These corrections are suppressed by the scale where the new effective description becomes relevant, which may typically be the SUSY breaking scale or the Planck scale to give but two examples. It is therefore important to study such higher derivative corrections in order to better understand the properties of the low energy effective actions of possible UV completions to the Standard Model coupled to Einstein gravity.

Of the large variety of possible gravitational corrections, there arises a particularly interesting case, where the additional term is given by the Gauss-Bonnet correction \cite{Boulware:1985wk,Zwiebach:1985uq, Wheeler:1985qd}. In four dimensions, this is a topologically invariant term that is analogous to the Euler number in two dimensions. Its variation is a total derivative and thus does not affect the classical bulk equations of motion. Upon studying gravitational perturbations of theories modified by higher derivative terms in any dimension, it has been discovered that only the Gauss-Bonnet term does not give rise to ghost-like propagating modes with modified kinetic terms \cite{Boulware:1985wk,Zwiebach:1985uq} at the level of Ricci squared corrections. The Gauss-Bonnet term, for example, is the $\alpha'$ order $R^2$ correction to the Einstein action in the ten dimensional heterotic string theory \cite{Gross:1986mw,Bergshoeff:1989qk,Chemissany:2007he}, and may arise in the lower dimensional actions of various compactified string theories\footnote{In pure IIB string theory on $AdS_5 \times S^5$, there are no $R^2$ corrections \cite{Gross:1986iv,Grisaru:1986vi,Freeman:1986zh}, but they may arise when we compactify on less symmetric manifolds.} \cite{Chamblin:1999tk,Cvetic:1999ne,Liu:2008kt}. Furthermore, it can also result from $\alpha'$ corrections to the bosonic sector of the $\mathcal{N}=2$ gauged supergravity that arises in the case of spinning D3-branes in type IIB superstring theory or spinning M2-branes in M-theory. In the dual conformal field theory, these corrections would be of order $1/\sqrt{\lambda}$ in the strong coupling expansion.

Another endless source of fascination has been the discovery that black holes have a very natural thermodynamic description \cite{Hawking:1974sw,Hartle:1976tp}. In particular, they emit a thermal spectrum and have an entropy related to their surface area. Of significant interest are black holes in asymptotically anti-de Sitter space, since these black holes have the remarkable feature of being thermodynamically favored for high enough temperatures, as was shown in the classic paper \cite{Hawking:1982dh}. With the advent of the anti-de Sitter/conformal field theory correspondence \cite{Maldacena:1997re,Witten:1998qj,Gubser:1998bc}, such anti-de Sitter black holes have become even more interesting, since their thermodynamic structure is related to that of the dual conformal theory \cite{Sundborg:1999ue,Sundborg:2000wp,Polyakov:2001af,Aharony:2003sx,Liu:2004vy,AlvarezGaume:2005fv,Witten:1998zw,Dey:2006ds,Dey:2007vt}. Specifically, the high temperature black hole phase is dual to a deconfining phase, whereas the low temperature thermal anti-de Sitter phase is dual to a confining phase in the boundary theory \cite{Witten:1998zw}.

In this paper we will study asymptotically anti-de Sitter black hole solutions, which are electrically charged under a gauged $U(1)$. Our gravitational action is corrected by a Gauss-Bonnet term, and we further include an $F^4$ correction to the electromagnetic field strength in the matter action. These corrections are a subset of all possible $\alpha'$ corrections to the bosonic sector of $\mathcal{N}=2$ supergravity in four or five spacetime dimensions \cite{Liu:2008kt}, and also a subset of $\alpha'$ corrections for higher dimensional anti-de Sitter space with a gauged $U(1)$. We obtain explicit black hole solutions and study their respective global and local thermodynamic stability in the grand canonical ensemble, where the external electric potential is held fixed. The thermodynamic phase structure we obtain should, in principle, be related to the thermodynamic phase structure of the dual boundary gauge theory, with a constant chemical potential. We will build our results from the well studied cases of Reissner-N\"{o}rdstrom anti-de Sitter black holes \cite{Chamblin:1999tk,Cvetic:1999ne,Chamblin:1999hg, Peca:1998cs} and Gauss-Bonnet black holes \cite{Dey:2006ds,Dey:2007vt,Nojiri:2001aj,Cvetic:2001bk,Cai:2001dz,Cho:2002hq}, and generalize to the Gauss-Bonnet-$F^4$ black hole.

There are various features that are novel to this discussion. As far as we know, the thermodynamics and thermal phase structure of the Gauss-Bonnet-$F^4$ black hole has never been studied before. It is thus of interest to study the thermodynamics of three types of limits in order to understand what the features of the thermodynamic structure are attributed to. The first system we will explore is the charged Gauss-Bonnet black hole with no $F^4$ corrections. Then we will study the $F^4$ black hole with no Gauss-Bonnet corrections and finally we will obtain the thermodynamic structure of the black hole with both corrections turned on. Furthermore, we will study black holes with hyperbolic, spherical and flat spatial horizons \cite{Brill:1997mf,Birmingham:1998nr,Horowitz:1998ha,Surya:2001vj,Cai:2004pz,Cai:2007wz} - known as topological black holes - and express our results for an arbitrary number of dimensions.

Due to the various systems we are studying, we will organize the paper in three parts. The first part of the paper will discuss the action of our theory and the explicit solutions to the ansatz. We will further identify the parameter space giving rise to actual black holes. In the second part, we will review the Gauss-Bonnet black hole with no $F^4$ corrections. The results will be presented in a parametrization that is most convenient when studying the $F^4$ black hole for the grand canonical ensemble, which is the ensemble most relevant to holography. In the third part of the paper we present the thermodynamic structure of the Gauss-Bonnet-$F^4$ black hole, again for the grand canonical ensemble. We conclude the paper with a brief discussion and summary of our results.

\newpage
\begin{center}
\section*{PART I - The Black Hole Solution}
\end{center}
\addcontentsline{toc}{part}{PART I - The Black Hole Solution}
\section{Framework}
We will now develop the basic setup of the gravitational and matter actions. This will be the goal of the section below, where we derive the equations of motion given a stationary spherically symmetric ansatz for the metric and the assumption of purely electrically charged matter.

\subsection{$R^2$ Corrections and the Gauss-Bonnet Term}
Our story begins with the action for Einstein-Maxwell gravity with negative cosmological constant and higher derivative corrections. In our conventions we have the following action:
\be
I_{tot} = I_{grav} + I_{matter},
\ee
where
\be
I_{grav} =  \int_{\mathcal{M}} d^{d+1} x \sqrt{-g} \left( \frac{1}{\kappa^2}R - \Lambda + a R^2 + b R^{\mu\nu}R_{\mu\nu} + cR^{\mu\nu\rho\sigma}R_{\mu\nu\rho\sigma} \right).
\ee
As we shall see, the Hawking-Gibbons boundary term will not play a role in our discussion and so we do not include it in the action.

From the point of view of a stringy UV completion these $R^2$ corrections would have coefficients of order $\alpha'$. A particular combination of the coefficients, given by  $a=c=-b/4$, gives rise to the Gauss-Bonnet term. In four dimensions, this term gives rise to a total derivative upon variation of the action with respect to the metric and thus is a topological invariant for four dimensions. It is interesting to note that of all ten-dimensional string theories, it is only the heterotic string theory that picks up $R^2$ corrections, and they always come in the form of the Gauss-Bonnet invariant. Furthermore, it is only the Gauss-Bonnet invariant that does not give rise to ghost-like propagation for gravitational perturbations \cite{Zwiebach:1985uq}. For such reasons we will work exclusively with the Gauss-Bonnet correction throughout the rest of the discussion.

Having specified our gravitational action, we can obtain the equations of motion for some arbitrary matter content. We will use the following convention for our stress-energy tensor:
\be
T_{\mu \nu }  \equiv - \frac{1}{{\sqrt { - g} }}\frac{{\delta I_{matter} }}{{\delta g^{\mu \nu } }}.
\label{eq:stress}
\ee
Upon variation of the action $I_{grav}$ with respect to $g_{\mu\nu}$ we obtain the equations of motion for our gravitational theory with Gauss-Bonnet corrections:
\begin{multline}
\label{eq:gb}
  T_{\mu \nu } = - \frac{1}{2}g_{\mu \nu } \left[ {\frac{1}{{\kappa ^2 }}R - \Lambda  + c \left( R^2  -4 R_{\mu \nu } R^{\mu \nu }  + R_{\mu \nu \rho \sigma } R^{\mu \nu \rho \sigma } \right) } \right] \\+ \frac{1}{{\kappa ^2 }}R_{\mu \nu }
  + c \left( {  2RR_{\mu \nu }  - 4R_{\mu \rho } R_\nu  ^\rho   - 4R_{\mu \rho \nu \sigma } R^{\rho \sigma }  + 2R_\mu  ^{\lambda \rho \sigma } R_{\nu \lambda \rho \sigma } } \right).
\end{multline}
For the benefit of the reader we have included a rather lengthy derivation of this and subsequent results, which demand extensive computations, in  appendix \ref{section:appA}. In what follows we will search exclusively for spherically symmetric solutions. In particular, we will be using the stationary spherically symmetric ansatz
\begin{equation}
ds^2  =  - e^{2\nu \left( r \right)} dt^2  + e^{2\lambda \left( r \right)} dr^2  + r^2 \sum\limits_{i,j = 1}^{d - 1} {\tilde g_{ij} dx^i dx^j },
\end{equation}
where we have that $\tilde{g}_{ij}$ is an Einstein metric whose Ricci tensor satisfies $R_{ij}=k \tilde{g}_{ij}$, and furthermore, we rescale the $x^i$ such that $\text{det}[\tilde{g}_{ij}]=1$. Many of the steps required to obtain the equations of motion are given in appendix \ref{sec:intermediate}. We reproduce here only the $tt$-equation and $rr$-equation which will be of more immediate use in the discussion that follows. The $tt$-equation of motion is given by:
\begin{multline}
T^{tt} = - \frac{\left( {d - 1} \right)}{\kappa^2} \left[ e^{ - 2\left( {\nu  + \lambda } \right)} \left( { - \frac{{\lambda '}}{r} + \frac{{\left( {d - 2} \right)}}{{2r^2 }}} \right) + e^{ - 2\nu } \frac{{k}}{{2r^2 }} \right] -\frac{1}{2} e^{-2 \nu} \Lambda \\
- \left( {d - 1} \right)\left( {d - 3} \right) c\left[ \frac{1}{2}e^{ - 2\nu } \frac{{\left( {d - 4} \right)k^2 }}{{\left( {d - 2} \right)r^4 }} \right.
  + e^{ - 2\nu  - 2\lambda } \left( { - \frac{{2k\lambda '}}{{r^3 }} + \frac{{\left( {d - 4} \right)k}}{{r^4 }}} \right) \\
 \left.  + \left( {d - 2} \right) e^{ - 2\nu  - 4\lambda } \left( {\frac{{2\lambda '}}{{r^3 }} - \frac{1}{2}\frac{{\left( {d - 4} \right)}}{{r^4 }}} \right) \right]
 \label{eq:tt}
\end{multline}
and the $rr$-equation of motion is given by:
\begin{multline}
T^{rr}=\frac{d - 1}{\kappa^2} \left[ e^{ - 4\lambda } \left( {\frac{{\nu '}}{r} + \frac{{\left( {d - 2} \right)}}{{2r^2 }}} \right) - e^{ - 2\lambda } \frac{{k}}{{2r^2 }}\right] +\frac{1}{2} e^{-2 \lambda} \Lambda \\
    - \left( {d - 1} \right)\left( {d - 3} \right) c \left[ \frac{1}{2}e^{ - 2\lambda } \frac{{\left( {d - 4} \right)k^2 }}{{\left( {d - 2} \right)r^4 }} \right.
  + e^{ - 4\lambda } \left( { - \frac{{2k\nu '}}{{r^3 }} - \frac{{\left( {d - 4} \right)k}}{{r^4 }}} \right) \\
  \left. + \left( {d - 2} \right) e^{ - 6\lambda } \left( {\frac{{2\nu '}}{{r^3 }} + \frac{1}{2}\frac{{\left( {d - 4} \right)}}{{r^4 }}} \right) \right].
\end{multline}
Our next task will be to obtain the equations of motion and stress-energy tensor following from the matter Lagrangian.

\subsection{Field Strength Corrections and the $F^4$ Term}
Our matter content is described by the usual Einstein-Maxwell matter Lagrangian with a quartic field strength correction, given as follows:
\begin{multline}
I_{matter}  =  - \frac{1}{{4g^2 }}\int {d^{d + 1} x\sqrt { - g} F_{\mu \nu } F^{\mu \nu }
 }  \\+ \int {d^{d + 1} x \sqrt { - g} \left[ {c_1 \left( {F_{\mu \nu } F^{\mu \nu } } \right)^2  + c_2 F_{\mu \nu } F^{\nu \lambda } F_{\lambda \rho } F^{\rho \mu } } \right] }.
 \label{eq:matter}
\end{multline}
Once again from the stringy point of view, the $F^4$ term would be of order $\alpha'$, which is the same order as the Gauss-Bonnet term. In fact, it is more general to include other $\alpha'$ terms of the forms $F^2 R$ and $(\nabla F)^2$ (see for example \cite{Liu:2008kt,Kats:2006xp}). Furthermore, there is a potential Chern-Simons term of the form $CS(A_\mu) = \varepsilon^{\mu\nu\rho\sigma
\tau} A_{\mu} F_{\nu\rho}F_{\sigma\tau}$ which could be included. For purely electric solutions, however, this term will not contribute to our equations of motion\footnote{More precisely, we will be working with only $F_{tr}(r)$ non-zero from which it is evident that there is no non-vanishing Chern-Simons term.}. In any case, it is not our purpose here to obtain the most general $\alpha'$ corrections to the Maxwell-Einstein theory, but to study the thermodynamics of a theory with both Gauss-Bonnet and gauge field strength corrections in asymptotic anti-de Sitter space\footnote{In fact as pointed out in \cite{Kats:2006xp}, we can take an example of this situation to be the ten-dimensional Lagrangian of the $E_8\times E_8$ or $SO(32)$ heterotic string theory. The $E_8\times E_8$ or $SO(32)$ gauge group has a $U(1)$ subgroup with field strength $F_{\mu\nu}$. Turning off all other gauge fields and higher forms and keeping the dilaton $\phi_0$ constant leaves us with the ten dimensional effective quartic order Lagrangian \cite{Gross:1986mw,Bergshoeff:1989qk,Chemissany:2007he} (albeit lacking a cosmological constant):
\begin{multline}
\nonumber\mathcal{L}_{het} = \frac{1}{2\kappa^2}R - \frac{1}{4}F_{\mu\nu}F^{\mu\nu} + \frac{\alpha'h}{16\kappa^2}(R_{\mu\nu\rho\sigma}R^{\mu\nu\rho\sigma}- 4R_{\mu\nu}R^{\mu\nu}+R^2)\\-\frac{3\alpha' h \kappa^2}{64}((F_{\mu\nu}F^{\mu\nu})^2-4F^{\mu\nu}F_{\nu\rho}F^{\rho\sigma}F_{\sigma\mu}),
\end{multline}
where $h \equiv e^{-\kappa \phi_0/\sqrt{2}}$.}. At the linearized level, however, it was shown in \cite{Liu:2008kt} that field redefinitions of the metric and $U(1)$ gauge field allow us to eliminate four of the eight possible $\mathcal{O}(\alpha')$ terms in the effective Lagrangian, at least for $d=4$ spatial dimensions.

Having established our matter Lagrangian we can compute the stress energy-tensor by applying (\ref{eq:stress}) on $I_{matter}$:
\begin{multline}
T^{\mu \nu }  =  - \frac{1}{{4g^2 }}\left( {\frac{1}{2}g^{\mu \nu } F^{\rho \sigma } F_{\rho \sigma }  - 2{F_\lambda}^ \mu  F^{\lambda \nu } } \right) \\+ c_1 \left( {\frac{1}{2}g^{\mu \nu } \left( {F^{\rho \sigma } F_{\rho \sigma } } \right)^2  - 4 {F_\lambda}^ \mu  F^{\lambda \nu }  F^{\rho \sigma } F_{\rho \sigma } } \right) \\+ c_2 \left( {\frac{1}{2}g^{\mu \nu } F^{\lambda \rho } F_{\rho \sigma } F^{\sigma \tau } F_{\tau \lambda }  - 4 F^{\mu \rho } F_{\rho \sigma } F^{\sigma \tau } {F_\tau} ^\nu  } \right).
\end{multline}
We proceed to obtain the equations of motion for the Maxwell field strength. These are a modified form of the Maxwell equations in a curved background:
\begin{equation}
\partial _\nu  \left[ {\sqrt { - g} \left( { - \frac{1}{{g^2 }}F^{\mu \nu }  + 8 c_1 \left( F^{\rho \sigma } F_{\rho \sigma }\right) F^{\mu \nu }  + 8 c_2 F^{\mu \rho } F_{\rho \sigma } F^{\sigma \nu } } \right)} \right] = 0.
\end{equation}
We note that the equations will depend only on the combination $(2c_1+c_2)$, which we denote by $\varepsilon$ from now on:
\be
\varepsilon \equiv 2 c_1 + c_2.
\ee
Our next task is to solve these equations for the spherically symmetric case, and we will further assume that only $F_{tr}$ is non-vanishing. For convenience we define the following:
\be
F_{tr}(r) = e^{\nu+\lambda}f(r),
\label{eq:fans}
\ee
such that the non-vanishing components of the stress-energy tensor become:
\begin{eqnarray}
 T^{tt}  &=& e^{ - 2\nu } \left( {\frac{1}{{4g^2 }}f\left( r \right)^2  + 3\varepsilon f\left( r \right)^4 } \right), \\
 T^{rr}  &=&  - e^{ - 2\lambda } \left( {\frac{1}{{4g^2 }}f\left( r \right)^2  + 3\varepsilon f\left( r \right)^4 } \right), \\
 T^{ij}  &=& \frac{{\tilde g^{ij} }}{{r^2 }}\left({\frac{1}{{4g^2 }}f\left( r \right)^2  + \varepsilon f\left( r \right)^4 } \right).
\end{eqnarray}
Inserting the ansatz (\ref{eq:fans}) into the modified Maxwell equations gives rise to the following equation for $f(r)$:
\begin{equation}
8\varepsilon f\left( r \right)^3  + \frac{1}{{g^2 }}f\left( r \right) - \frac{Q}{{r^{d - 1} }} = 0.
\end{equation}
This cubic equation can be solved exactly, and will have either three or one real solutions, from which we have to choose the physical one. By physical we simply mean that solution which leads to the usual Coulomb's law when $\varepsilon$ approaches zero and which does not diverge as $r$ approaches infinity. The general solution to the cubic is given as follows:
\begin{equation}
f_n(r) = \frac{1}{2}\left(-\frac{e^{i\frac{2 \pi n}{3}}}{6^{1/3} \Delta\left( r \right)} + \frac{\Delta\left( r \right)}{e^{ i\frac{2 \pi n}{3}} 6^{2/3} g^2 \varepsilon }\right),
\end{equation}
where $n\in\{0,1,2\}$ and we have defined:
\begin{equation}
\Delta\left( r \right) = \frac{\left(18 g^6 Q r^{1+2 d} \varepsilon ^2+\sqrt{6\left(g^6 r^{6 d} \varepsilon ^3+54 g^{12} Q^2 r^{2+4 d} \varepsilon ^4 \right)}\right)^{1/3}}{r^{d} }.
\end{equation}
When $\varepsilon>0$ the physical solution corresponds to the only real solution which is given by $n=0$. When $\varepsilon<0$ the physical solution is given by $n=1$. This solution is real only for radii larger than a critical value, $r_{min}$, obtained by requiring that the expression in the root within $\Delta(r)$ is positive. In particular, we find that
\begin{equation}
r^{2d - 2}_{min}  =  - 54g^6 Q^2 \varepsilon.
\label{eq:epsbound}
\end{equation}
For both of the physical solutions there are two complementary expansions to consider. For $\chi \equiv (8g^6Q^2  \varepsilon)/r^{2(d-1)}\ll 1$ we can expand as follows:
\begin{equation}
f\left( r \right) = \frac{{g^2 Q}}{{r^{d - 1} }}\left[ 1 - \chi +
\mathcal{O}\left( \chi^2 \right)  \right].
\end{equation}
\begin{figure}[h]
\begin{center}
$\begin{array}{c@{\hspace{0.1in}}c}
\includegraphics[angle=0,width=0.5\textwidth]{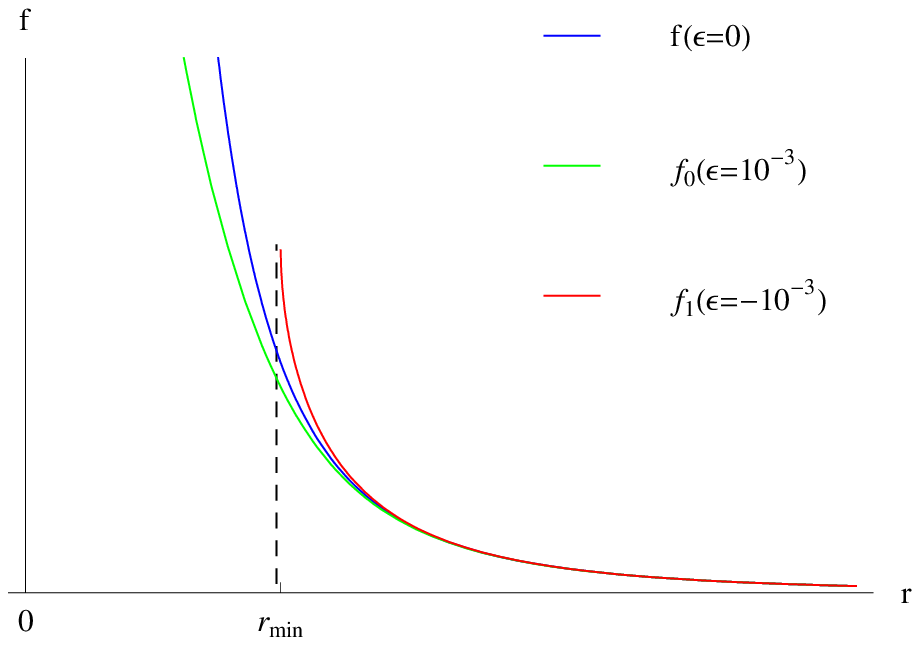} &
 \includegraphics[angle=0,width=0.5\textwidth]{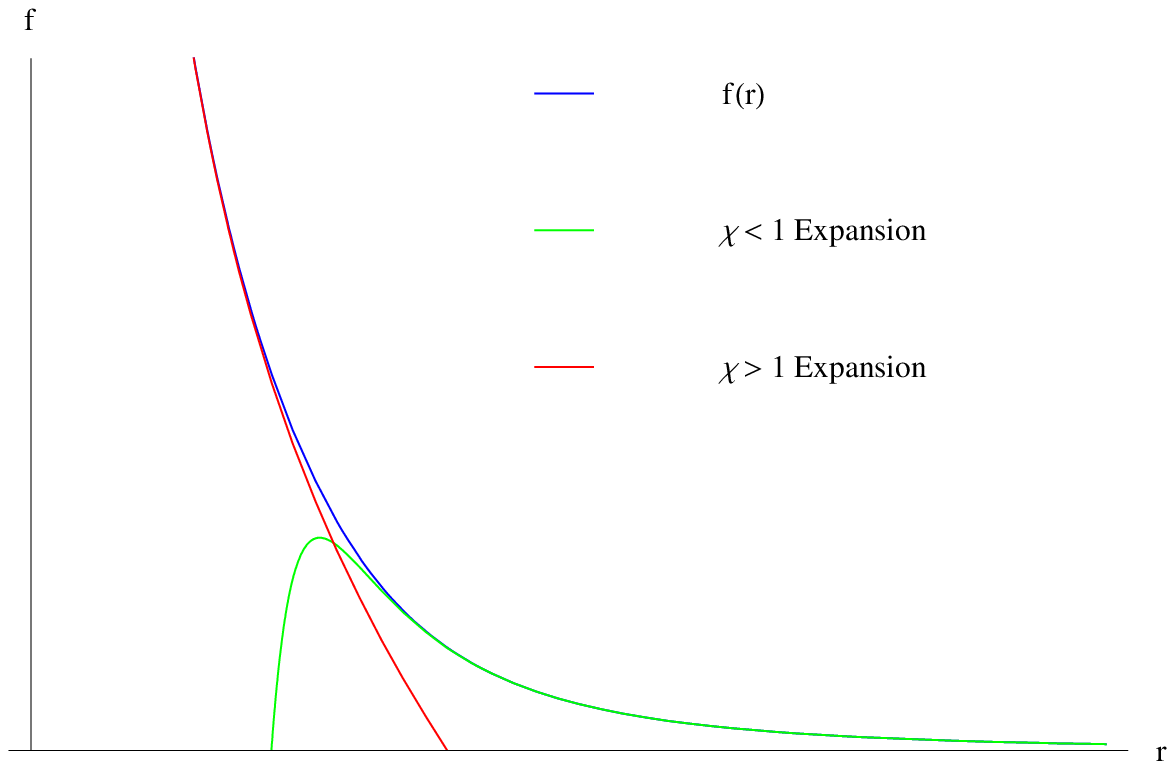} \\
   \text{(a)}&\text{(b)}
   \end{array}$
\end{center}
\caption{(a) The physical solution for $\varepsilon=0$, $\varepsilon>0$ and $\varepsilon<0$, and (b) the validity of the $\chi \gg 1$ and $\chi \ll 1$ expansions for $f(r)$.}
\label{fig:solns}
\end{figure}
We immediately recognize the first term as Coulomb's law which is subsequently corrected by $\varepsilon$ effects. When $\chi^{-1} = r^{2(d-1)}/(8g^6Q^2 \varepsilon) \ll 1$ then the above expansion breaks down and we should consider the following expansion:
\be
f\left( r \right) = \left( {\frac{Q}{{8\varepsilon r^{d - 1} }}}
\right)^{1/3} \left[ 1 - \frac{{\chi}^{-1/3}}{3}  +
\mathcal{O}\left( ({\chi}^{-1/3})^2 \right)  \right].
\ee
The physical solutions and the validity of the two expansions is displayed in figures \ref{fig:solns}(a) and (b). When $\varepsilon < 0$, the region $\chi^{-1} < 1$ is excluded due to the physical bound (\ref{eq:epsbound}), so we should always consider the $\chi$ expansion in this case. In what follows $\chi$ is always evaluated at $r=r_H$, the horizon radius. Finally, when $Q=0$ the physical solution corresponds to $f(r)=0$, since the other two blow up as $\varepsilon$ tends to zero.

Having established our basic conventions and formalism for the higher derivative corrections to the Einstein-Maxwell theory, we are ready to find the corresponding black hole solutions. This will be the main goal of the next section.

\section{The Gauss-Bonnet-$F^4$  Topological AdS$_{d+1}$ Black Hole}

We have obtained the dynamical equations for our action with a given matter content and we are now in the position to obtain an explicit solution for our ansatz. We will identify this solution as a static charged black hole in asymptotic anti-de Sitter space, and we evaluate its Hawking temperature and horizon radius.

\subsection{Exact Solution}
In order to obtain the black hole solutions to the action we have specified, we can take advantage of some of the equations of motion we derived in the previous section. In particular, we find the following useful relation:
\begin{multline}
0=e^{2\nu } T^{tt}  + e^{2\lambda } T^{rr} \\ = \left( {\nu ' + \lambda '} \right) \left[ \frac{\left( {d - 1} \right)}{{\kappa ^2 }} \frac{{e^{ - 2\lambda } }}{r} - 2 \left( {d - 1} \right)\left( {d - 3} \right) c\frac{{\left( {d - 2} \right)e^{ - 4\lambda } - k e^{ - 2\lambda }}}{{r^3 }} \right].
\end{multline}
The following result is immediately yielded:
\be
\nu' + \lambda' = 0.
\ee
This allows us to set $\nu=-\lambda$ up to a constant shift in $\nu$, which can always be reabsorbed into our definition of time. Having obtained this result, some algebra on the $tt$-equation of motion given in (\ref{eq:tt}) gives rise to the equation:
\begin{multline}
  - \frac{\left( {d - 1} \right)}{{\kappa ^2 }}\left[ {\frac{d}{{dr}}\left( { - e^{ - 2\lambda } r^{d - 2} } \right) + kr^{d - 3} } \right] + r^{d - 1} \Lambda\\
  + \left( {d - 1} \right)\left( {d - 3} \right) c\left[ { - \frac{{\left( {d - 4} \right)k^2 r^{d - 5} }}{{\left( {d - 2} \right)}} + 2\frac{d}{{dr}}\left( {e^{ - 2\lambda } kr^{d - 4} } \right) - \frac{d}{{dr}}\left( {e^{ - 4\lambda } \left( {d - 2} \right)r^{d - 4} } \right)} \right] \\
   + 2 r^{d - 1} \left[ {\frac{1}{{4g^2 }}f\left( r \right)^2  + 3\varepsilon f\left( r \right)^4 } \right]=0.
\end{multline}
It is a simple task to integrate over $r$ and find:
\begin{multline}
   - \frac{\left( {d - 1} \right)}{{\kappa ^2 }}\left[ { - e^{ - 2\lambda } r^{d - 2} + \frac{{1}}{{d - 2}}kr^{d - 2} } \right] + \frac{{r^d }}{d}\Lambda \\
  +\left( {d - 1} \right)\left( {d - 3} \right) c\left[ { - \frac{{k^2 r^{d - 4} }}{{\left( {d - 2} \right)}} + 2e^{ - 2\lambda } kr^{d - 4} - e^{ - 4\lambda } \left( {d - 2} \right)r^{d - 4} } \right] + I\left( r \right) = \mu,
  \label{eq:quad}
\end{multline}
where we have defined:
\begin{equation}
I\left( r \right) \equiv 2\int {dr} r^{d - 1} \left( {\frac{1}{{4g^2 }}f\left( r \right)^2  + 3\varepsilon f\left( r \right)^4 } \right)
\end{equation}
and $\mu$ is an integration constant, which we will soon relate to the mass of the black hole. We select the integration constant so that the limit of $I(r)$ at infinity vanishes, in order to make connection with the usual Reissner-N\"{o}rdstrom solution in the absence of $F^4$ corrections. The above integral $I(r)$ will play a crucial role for the remainder of the paper, for it encompasses the $\varepsilon$ dependence of the metric in a highly non-trivial way. Its expansion for $\chi\ll 1$ will be of use in later sections and is given by:
\be
I(r_H) = -\frac{g^2Q^2}{2(d-2)r^{d-2}_H}\left[1+\frac{(d-2)}{2(4-3d)}\chi + \frac{6(d-2)}{6-5d}\chi^2+\mathcal{O}(\chi^3)\right].
\label{eq:intexp}
\ee

Notice that (\ref{eq:quad}) is a quadratic equation in $e^{-2\lambda}$, which can be solved quite simply to give our metric component:
\begin{equation}
e^{ - 2\lambda }  = \frac{k}{{d - 2}} + \frac{{\left( {d - 1} \right)r^2 }}{{2\kappa ^2 D}}  \pm \sqrt {\frac{{K^2 r^4 }}{{4D^2 }} + \frac{{\left( {I\left( r \right) - \mu } \right) }}{D r^{d - 4}}},
\label{eq:sol}
\end{equation}
where
\begin{eqnarray}
D &\equiv& c\left( {d - 1} \right)\left( {d - 2} \right)\left( {d - 3} \right),\\
K^2 &\equiv&  {\frac{{\left( {d - 1} \right)^2 }}{{\kappa ^4 }} + \frac{4D\Lambda}{d} }.
\end{eqnarray}
For particular regions of parameter space, the solution we have obtained contains a physical singularity at the origin hidden by event horizons and is thus a black hole. We proceed to examine the relevant parameter space.

\subsubsection*{$\bold{c >0}$}

Let us discuss first the case $c>0$ for our solution (\ref{eq:sol}). We take $k>0$ for the discussion in this subsection, and leave the case of $k<0$ with $c>0$ for the next subsection. Of the two roots in (\ref{eq:sol}) only the negative one reduces to the usual anti-de Sitter Reissner-N\"{o}rdstrom solution as $c$ and $\varepsilon$ go to zero. This will be the one we will work with from now on for $c>0$. It has actually been shown for the $\varepsilon = 0$ case that the positive root solution (with $ck>0$) has gravitons that have ghost-like propagation \cite{Boulware:1985wk,Zwiebach:1985uq}, and we take this as further reason to stick to the negative root.

It is also clear from the form of the metric that arbitrarily large values of $c>0$ are not acceptable. In fact, it is clearly seen that as $r \to \infty$ the values of $c$ that give rise to a real metric must satisfy the following bound:
\begin{equation}
c < \frac{{d\left( {d - 1} \right)}}{{\left( {d - 2} \right)\left( {d - 3} \right)}}\left( { - \frac{1}{{\kappa ^4 \Lambda }}} \right).
\label{eq:cbound}
\end{equation}
If $c$ is above this bound our metric becomes complex for the particular ansatz we have chosen and is thus rendered unphysical. It was, however, pointed out in \cite{Cvetic:2001bk} that exactly where the asymptotically anti-de Sitter metric breaks down the de Sitter metric with positive $\Lambda$ yields an acceptable solution. We will not consider the implications of this observation and simply end our parameter space where the bound (\ref{eq:cbound}) is saturated.

\subsubsection*{$\bold{ck <0}$}

For the case of $c<0$ with $k>0$ our solution (\ref{eq:sol}) tends to the usual anti-de Sitter Reissner-N\"{o}rdstrom black hole as $c \to 0$ for the positive root. Furthermore, we must be particularly careful as to when the positive root solution of (\ref{eq:sol}) gives rise to black holes with horizons. The equation for the horizon radius $r_H$ is given by solving the equation $g_{tt}=0$. When we choose the positive root, $g_{tt}$ can only vanish if the sum of the first two terms in (\ref{eq:sol}) is negative for values of $r$ smaller than $r_H$, in order to cancel the positive contribution of the square root as $r$ tends to $r_H$. This condition implies that
\begin{equation}
r_H^2  \ge   -2\left( {d - 3} \right)\kappa ^2 c k.
\label{eq:kcbound}
\end{equation}
Furthermore, we should mention that the above condition holds whenever $c > 0$ and $k<0$ due to an analogous reasoning, taking into account that we must take the solution with the negative root.
\newline
\newline
We would like to emphasize the above point: once we set $ck<0$ and choose the appropriate sign for the root, then the black hole solutions of our theory have a \emph{minimum} horizon radius given by (\ref{eq:kcbound}).

\subsection{Vacuum Solutions}

If we consider the vacuum black hole solution, $\mu=Q=0$, it allows for horizons but no singularities whenever $k<0$. The horizon radius of these geometries is given by:
\be
r_H^2 = \frac{|k|}{(d-2)} \left|\frac{K}{2|D|} - \frac{(d-1)}{2\kappa^2 D}\right|^{-1}.
\label{eq:kcbound2}
\ee

When $k>0$ there is no horizon and the vacuum solutions correspond to regular anti-de Sitter space with spherical spatial slices. When $k<0$ then the $\mu=Q=0$ is not actually the vacuum solution, as there are negative energy black holes.

\subsection{Horizon Geometries}

Taking $k=\pm(d-2)$ or $k=0$ gives rise to three qualitatively different asymptotic spacetimes. In particular, the asymptotic structure we obtain is as follows:
\begin{eqnarray}
k &=& +(d-2) \to S^d\times \mathbb{R}^1\\
k &=& 0 \quad \quad \quad \text{  } \to \mathbb{R}^{d}\times \mathbb{R}^1 \\
k &=& -(d-2) \to \mathbb{H}^d \times \mathbb{R}^1,
\end{eqnarray}
where the $\mathbb{R}^1$ denotes the timelike direction and $\mathbb{H}^d$ denotes the spatial $d$-dimensional hyperboloid. We will be studying the thermodynamics of all three branches in the sections that follow.

\subsection{Horizon, Temperature and Extremality}

We will proceed to study the horizon structure and Hawking temperature $T_H$ of our black hole solutions. The horizon radius $r_H$ is given by the value of $r$ that leads to the vanishing of the $g_{tt}$ component of the metric. This condition implies that $r_H$ obeys\footnote{It may seem that (\ref{eq:hor}) has solutions for $kc<0$ that violate the bound (\ref{eq:kcbound}); however, these solutions correspond to the metric (\ref{eq:sol}) with the opposite root.}:
\begin{equation}
 - \frac{1}{{\kappa ^2 }}\frac{{d - 1}}{{d - 2}}k{r_H} ^{d - 2}  - c\frac{{\left( {d - 1} \right)\left( {d - 3} \right)}}{{d - 2}}k^2 {r_H} ^{d - 4}  + \frac{\Lambda }{d}{r_H} ^d  + I\left( {r_H } \right) - \mu = 0.
 \label{eq:hor}
\end{equation}
The above equation will have multiple solutions for $r_H$ corresponding to multiple horizons. The true event horizon is given by the largest positive root of $(\ref{eq:hor})$.

The Hawking temperature is most readily obtained by the well known result \cite{Hartle:1976tp} which reads $4\pi T_H = (e^{-2\lambda})'|_{r=r_H}$. Applying this formula to our metric gives the following equation for the temperature:
\begin{multline}
4\pi T_H \left[ {\frac{{\left( {d - 1} \right){r_H} ^2 }}{{\kappa ^2 }} + \frac{{2kD}}{{d - 2}}} \right]{r_H} ^{d - 3}  \\- \frac{2}{{\kappa ^2 }}\frac{{d - 1}}{{d - 2}}k{r_H} ^{d - 2}  + \frac{{4\Lambda }}{d}{r_H} ^d  + \left( {d - 4} \right)\mu  + {r_H} ^{d - 3} \left( {\frac{{I\left( r \right)}}{{r^{d - 4} }}} \right)' = 0.
\label{eq:mut}
\end{multline}

We can always choose to parameterize our thermodynamics with two of the four variables $r_H$, $T_H$, $Q$ and $\mu$, since we have a constraint equation relating the temperature to the derivative of $g_{tt}$, as well as the above constraint that $r_H$ must obey. The most convenient choice for our discussion will be $r_H$ and $Q$. The reason for this choice is evident from the complicated form of $I(r)$, which can only be written explicitly as a function of $Q$ and $r_H$. In this language, we can express the $T_H$ and  $\mu$ as functions of $r_H$ and $Q$ as follows:
\begin{eqnarray}
T_H &=& \frac{\frac{1}{{\kappa ^2 }}\left( {d - 1} \right)k{r_H} ^{2}+c\frac{{\left(
{d - 1} \right)\left( {d - 3} \right)\left( {d - 4} \right)}}{{d - 2}}k^2
-\Lambda {r_H} ^4 - {r_H}^{5-d} I'\left( {{r_H} } \right)}{4\pi \left[
{\frac{{\left( {d - 1} \right){r_H} ^2 }}{{\kappa ^2 }} + \frac{{2kD}}{{d -
2}}} \right]{r_H} },\label{eq:htrq}\\
\mu &=& - \frac{1}{{\kappa ^2 }}\frac{{d - 1}}{{d - 2}}k{r_H} ^{d - 2}  -
c\frac{{\left( {d - 1} \right)\left( {d - 3} \right)}}{{d - 2}}k^2 {r_H} ^{d -
4}  + \frac{\Lambda }{d}{r_H} ^d  + I\left( {r_H } \right).\label{eq:murq}
\end{eqnarray}
We can obtain the horizon radius for the extremal black hole by solving $T_H=0$. We will mostly be interested in the explicit form for the $Q=0$ extremal black hole with $g_{tt}$ component follows from (\ref{eq:sol}):
\be
- g_{tt}(r) \equiv e^{ - 2\lambda }  = \frac{k}{{d - 2}} + \frac{{\left( {d - 1} \right)r^2 }}{{2\kappa ^2 D}}  \pm \sqrt {\frac{{K^2 r^4 }}{{4D^2 }} - \frac{\mu_{e}}{D r^{d - 4}}}
\label{eq:extsol}
\ee
and horizon radius satisfying:
\be
g_{tt}(r_e) = 0
\label{eq:extradius}
\ee
The parameter $\mu_e$ is obtained by evaluating (\ref{eq:murq}) at $r_e$ with $Q=0$:
\be
\mu_e = - \frac{1}{{\kappa ^2 }}\frac{{d - 1}}{{d - 2}}k{r_e} ^{d - 2}  -
c\frac{{\left( {d - 1} \right)\left( {d - 3} \right)}}{{d - 2}}k^2 {r_e} ^{d -
4}  + \frac{\Lambda }{d}{r_e}^d.
\ee

\subsection{Asymptotic Solution}

For far away observers, we can take the $r \to \infty$ limit of our metric to obtain an expansion in powers of $r$:
\begin{equation}
e^{ - 2\lambda }  = \frac{r^2}{\ell^2} + \frac{k}{{d - 2}}  -\frac{m}{ r^{d - 2}}  + \frac{q^2}{ r^{2 d - 4}}  + \mathcal{O}\left( \frac{1}{{r^{4 d - 6} }} \right),
\end{equation}
where we have made the following relations between the asymptotic parameters and those in our action:
\begin{eqnarray}
\Lambda  &=& - \frac{{d\left( {d - 1} \right)}}{{\ell^2 \kappa ^2 }} + \frac{{d D}}{{\ell^4 }}, \\
\mu  &=&  -m K, \\
Q^2   &=& 2(d - 2)K q^2/g^2.
\end{eqnarray}
We immediately note that the above asymptotic form is exactly that of the charged black hole with a negative cosmological constant. It is important, however, to recognize that the parameters of our theory are $Q$, $\mu$ and $\Lambda$ and not $q$, $m$ and $\ell$, and it is with these parameters that we will continue to work for the rest of the paper.

Finally, we can write down the mass $\mathcal{M}$ and charge $\mathcal{Q}$ of the black hole as measured by a faraway observer:
\begin{eqnarray}
\mathcal{M} &=& -\Sigma_k\times \frac{(d-1)\mu }{K \kappa^2},  \\
\mathcal{Q}^2 &=& \Sigma^2_k\times \frac{4{ (d-1)(g Q)^2}}{K\kappa^2},
\label{eq:faraway}
\end{eqnarray}
where we have that $\Sigma_k$ is the $(d-1)$-dimensional volume of a hypersurface with curvature given by $(d-1)k$. Note that $c$ is not required to be small in the above expressions. In section \ref{sec:thquant} we will compute the thermodynamic mass and charge of the black hole in the grand canonical ensemble and find that they agree favorably with the above expressions when $|c| \ll 1$.

\newpage
\begin{center}
\section*{PART II - Gauss-Bonnet Thermodynamics}
\end{center}
\addcontentsline{toc}{part}{PART II - Gauss-Bonnet Thermodynamics}

\section{Global Stability}

Having completed the discussion of our black hole solutions, we can proceed to examine their thermodynamic phase structure. Hawking and Page demonstrated in their seminal paper \cite{Hawking:1982dh} that asymptotically anti-de Sitter Schwarzschild black holes are thermally favored when their temperature is sufficiently high, whereas the pure anti-de Sitter background was thermally favored for low temperatures. As the temperature is decreased there is a first order phase transition leading the black hole spacetime to the pure anti-de Sitter spacetime. Subsequently, the effect has been studied for various other asymptotically anti-de Sitter black holes, including those with Gauss-Bonnet terms \cite{Chamblin:1999tk,Cvetic:1999ne,Dey:2006ds,Dey:2007vt,Chamblin:1999hg,Peca:1998cs,Nojiri:2001aj,Cvetic:2001bk,Cai:2001dz,Cho:2002hq,Cai:2004pz}. Our aim in this work is to contribute to this discussion the case for charged black holes, in theories with both Gauss-Bonnet terms as well as $F^4$ terms. We begin by reviewing and expanding the $\varepsilon = 0$ case which was considered in \cite{Dey:2007vt,Cvetic:2001bk}. Our results are formulated in the language of the $r_H$, $Q$ parametrization, which is best suited for the subsequent $\varepsilon \neq 0$ analysis.

\subsection{Free Energy}
\label{sec:free}
The free energy for our black hole solution is obtained from the thermal partition function, $Z$, of our theory. The Euclidean metric is obtained by continuing to imaginary time and identifying it periodically. The period is given by the inverse Hawking temperature.

Using the saddle point approximation we can evaluate the partition function by evaluating the classical action of the black hole metric. In particular, the free energy is given as follows:
\be
F = -k_B T_H \log Z \approx -k_B T_H  I^{BH}_E.
\ee
We should pause for a moment and note that we are working in the \emph{grand canonical ensemble}, where we keep that electric potential fixed rather than the electric charge and thus the free energy we are computing is in fact the Gibbs free energy. We hope to study the thermodynamics of the canonical ensemble, where the electric charge of the black hole is kept fixed, in a future work.

The Euclidean action evaluated for the black hole solution is in fact divergent. In order to regulate this divergence, we cut off our space at finite yet large $r=r_{max}$ and subtract off the Euclidean action of the reference background before taking the cutoff back to infinity. We note that one can obtain the free energy of the black hole through the method of counterterms (see for example \cite{Henningson:1998gx,Balasubramanian:1999re,Emparan:1999pm,Kofinas:2006hr,Astefanesei:2008wz}), which is particularly convenient if there is no natural reference background to subtract.
\newline
\newline
The reference background for the $\bold{k>0}$ is given by the pure thermal anti-de Sitter space obtained from (\ref{eq:sol}) with $k>0$ and $\mu=Q=0$, along with a uniform external potential taking the value of (minus) the internal electric potential at the horizon of the black hole \cite{Braden:1990hw}. Such a uniform electric potential leads to a vanishing Maxwell tensor, and consequently the Euclidean action remains that of the original thermal anti-de Sitter space.
\newline
\newline
Our reference background for the case of $\bold{k=0}$ is given by the pure thermal anti-de Sitter space obtained from (\ref{eq:sol}) with $k=0$ and $\mu=Q=0$. Once again, there is a uniform external potential as in the $k>0$ case.\footnote{However, we should make an important remark about the $k=0$ regularization procedure. For the case of vanishing Gauss-Bonnet coefficient it was shown in \cite{Horowitz:1998ha} that there exists a vacuum solution with energy lower than pure anti-de Sitter space, known as the anti-de Sitter soliton, that is also asymptotically anti-de Sitter space. This solution can be regarded as the  reference background for the Hawking-Page transition as long as we compactify one of the directions of the Ricci flat black hole in order to embed it asymptotically in the anti-de Sitter soliton background. The thermodynamics with the anti-de Sitter soliton vacuum were studied with no Gauss-Bonnet coefficient in \cite{Surya:2001vj} and in \cite{Cai:2007wz} for finite Gauss-Bonnet coefficient but electrically neutral black holes. In this paper we will not consider the anti-de Sitter soliton vacuum as a reference, we will only explore the Hawking-Page structure for the unmodified asymptotically anti-de Sitter charged Gauss-Bonnet black holes with $F^4$ corrections. It would be interesting to explore the thermodynamics of the anti-de Sitter soliton reference background for the theories we are considering.}
\newline
\newline
Finally, for the case of black holes with curvature $\bold{k<0}$ the reference background is rather different. One possible regularization procedure is that of \cite{Brill:1997mf,Cai:2004pz,Braden:1990hw}, where we subtract the (\ref{eq:sol}) solution with $k<0$ and $\mu=Q=0$. This solution is still asymptotically anti-de Sitter space; however, it also contains a horizon and thus a fixed temperature. In order to avoid the problem of our reference background having a fixed temperature, we will choose our reference background to be the neutral extremal black hole (\ref{eq:extsol}) which can have an arbitrary periodicity in Euclidean time. Note, however, that black holes with non-zero external potential cannot decay to this reference background, since the reference background acquires a non-vanishing Wilson loop around the Euclidean time for an arbitrary external potential. Thus we cannot discuss Hawking-Page transitions in the $k<0$ case, and we will only study the local phase structure with our free energy regulated by (\ref{eq:extsol}). We refer to \cite{Braden:1990hw} for more details (see also \cite{Neupane:2003vz}).
\newline
\newline
There is one more subtlety as to how we regulate the Euclidean action. We must ensure that the thermal circle of both the black hole solution and the reference background are the same at the cutoff radius $r_{max}$ in order to have the same geometry at $r=r_{max}$. This is possible since, even though the Euclidean black hole can only have a particular temperature to avoid conical singularities, Euclidean anti-de Sitter space or the Euclidean extremal black hole can have any temperature and no conical singularities.

We are now ready to compute the free energy. It is very helpful to take the trace of (\ref{eq:gb}) which yields the expression:
\begin{multline}
\frac{\left( {d - 1} \right)}{{\kappa ^2 }}R - \left( {d + 1} \right)\Lambda  + c\left( {d - 3} \right)\left[ {R^2  + R_{\mu \nu } R^{\mu \nu }  + R_{\mu \nu \lambda \rho } R^{\mu \nu \lambda \rho } } \right] \\
-\frac{\left( {d - 3} \right)}{{4 g^2 }}F_{\mu \nu } F^{\mu \nu }  + c_1 \left( {d - 7} \right) \left( {F_{\mu \nu } F^{\mu \nu } } \right)^2  + c_2 \left( {d - 7} \right) F_{\mu \nu } F^{\nu \lambda } F_{\lambda \rho } F^{\rho \mu }=0.
\end{multline}
Using the above result allows us to decompose our Euclidean action into the following two pieces:
\begin{equation}
I_{tot} = I_1  + I_2,
\end{equation}
where we have defined:
\begin{eqnarray}
I_1& \equiv & -\frac{2}{(d-3)}\int d^{d + 1} x\sqrt { - g} \left( \frac{1}{{\kappa ^2 }}R - 2\Lambda \right) \approx -\beta_H F_1 \label{eq:I1},\\
I_2& \equiv &\frac{4}{{d - 3}} \int {d^{d + 1} x\sqrt { - g}  \left[ {c_1 \left( {F_{\mu \nu } F^{\mu \nu } } \right)^2  + c_2 F_{\mu \nu } F^{\nu \lambda } F_{\lambda \rho } F^{\rho \mu } } \right]} \approx -\beta_H F_2\label{eq:I2}
\end{eqnarray}
and $\beta_H \equiv T^{-1}_H$. Notice that we have eliminated the Gauss-Bonnet term from our action.

We regularize our Gibbs free energy by subtracting away that of the pure thermal anti-de Sitter space or any other appropriate reference background with the thermal circles identified at the cutoff boundary. This leaves us with the expression:
\begin{multline}
  - \frac{{F_1 }}{{\Sigma_k }} = -\frac{2}{(d-3)}\mathop {\lim }\limits_{r_{\max }  \to \infty } \left[ \int_{r_H }^{r_{\max } } {drr^{d - 1} \left( \frac{1}{{\kappa ^2 }}R - 2\Lambda  \right)}  \right.\\
  \left. - e^{ - \lambda \left( {r_{\max } } \right) + \lambda _0 \left( {r_{\max } } \right)} \int_{r_{vac}}^{r_{\max } } {drr^{d - 1} \left( \frac{1}{{\kappa ^2 }}R_0  - 2\Lambda  \right)}  \right].
  \label{eq:f1}
\end{multline}
The factors of $\beta_H$ have canceled since they also appear as part of the Euclidean time integral. $R_0$ is the Ricci scalar of the reference background solution. The vacuum horizon radius, $r_{vac}$, vanishes for $k\ge0$, whereas it equals $r_e$ in (\ref{eq:extradius}) for $k<0$. We can now use the following relation for the curvature scalar:
\be
R =  - \frac{1}{{r^{d - 1} }}\left( {r^{d - 1} e^{ - 2\lambda } } \right)'' + \frac{{\left( {d - 1} \right)k}}{{r^2 }}
\ee
to obtain
\begin{multline}
\frac{{F_1 - F_{vac}}}{{\Sigma_k }} = \\  \frac{2}{{d - 3}}\left[ {\frac{1}{{\kappa ^2 }}\left( { 4\pi {r_H} ^{d - 1} T_H(r_H,Q)  - \frac{{d - 1}}{{d - 2}}k{r_H} ^{d - 2} } \right) - \frac{\mu(r_H,Q)}{2} + \frac{2}{d}\Lambda {r_H} ^d }\right],
  \label{eq:free1}
\end{multline}
where $F_{vac}=0$ for $k\ge0$ and
\be
\frac{F_{vac}}{\Sigma_k} = \frac{2}{{d - 3}}\left[ -{\frac{1}{{\kappa ^2 }} {\frac{{d - 1}}{{d - 2}}k{r_{vac}} ^{d - 2} }  - \frac{\mu_{vac}}{2} + \frac{2}{d}\Lambda {r_{vac}}^d }\right]
\label{eq:fvac}
\ee
for $k<0$. The Hawking temperature $T_H$ as a function of $r_H$ and $Q$ is given in (\ref{eq:htrq}). We finally note that $F_2=0$ for the case of vanishing $\varepsilon$ in accordance with \cite{Cvetic:2001bk}, so that in fact (\ref{eq:free1}) is the total Gibbs free energy of our thermodynamic system.

In the grand canonical ensemble we are considering systems with a fixed external electric potential, $\phi$. Imposing that $\phi$ be fixed to cancel the electric potential at the horizon of the black hole, $\phi_{BH}$, leads to the following expression:
\be
\phi_{BH}(r_H,Q)= -\frac{g Q}{2(d-2)r_H^{d-2}} = -\phi(r_H,Q).
\label{eq:phi}
\ee
This expression will be of use when computing the thermodynamic energy, entropy and charge of the black hole. Furthermore, we will study the global and local stability of the system along slices of constant $\phi$.

\subsection{Thermodynamic Quantities}
\label{sec:thquant}
The Gibbs free energy contains all the information required to compute these quantities which are given by the usual thermodynamic relations. In particular, the thermodynamic energy (which we will also call mass) is given by:
\begin{eqnarray}
\nonumber E &=& \left(\frac{\partial I_1}{\partial \beta_H}\right)_\phi - \frac{\phi}{\beta_H}\left( \frac{\partial I_1}{\partial \phi} \right)_{\beta_H} \\
\nonumber  &=& \Sigma_k\times \left[ \frac{{\left( {d - 1} \right)k r_H ^{d - 2}
}}{{\left( {d - 2} \right)\kappa ^2 }} + \frac{{g^2 Q^2 r_H ^{2 - d} }}{{2\left(
{d - 2} \right)}} - \frac{{r_H ^d \Lambda }}{d} + \frac{{\left( {d - 1}
\right)\left( {d - 3} \right)ck^2 r_H ^{d - 4} }}{{\left( {d - 2} \right)}}\right]
\\&=& -\Sigma_k \times \mu(r_H,Q),
\end{eqnarray}
where we recall that $I_1 = -\beta_H F_1$ was given in (\ref{eq:free1}) and our electric potential $\phi$ is given in (\ref{eq:phi}). The thermodynamic charge and entropy are given by:
\begin{eqnarray}
\mathcal{Q} &=& -\frac{1}{\beta_H}\left( \frac{\partial I_1}{\partial \phi} \right)_{\beta_H} = \Sigma_k \times 2 g Q, \\
S &=& \beta_H \left(\frac{\partial I_1}{\partial \beta_H}\right)_\phi - I_1 =
\Sigma_k \times \left(\frac{{4\pi r_H^{d - 1} }}{{\kappa ^2 }} + 8\pi c k\left( {d - 1} \right)r_H^{d - 3} \right).
\end{eqnarray}
Our thermodynamic electric charge has a factor of $2 g$ that may differ from other expressions in the literature. This is due to our normalization of the kinetic term of the field strength in (\ref{eq:matter}). We also notice that the thermodynamic entropy can become negative when $c k <0$ (see for example \cite{Neupane:2002bf}). In this case we find that the minimum radius of a black hole with positive entropy is given by:
\be
r_H^2(S=0) = -2ck(d-1)\kappa^2.
\label{eq:entropybound}
\ee

It is worth noting that since our free energy is parameterized by $r_H$ and $Q$, the above expressions are formal and should be computed as derivatives in $r_H$ and $Q$. We give the explicit expressions for the thermodynamic derivatives in appendix \ref{sec:thder}. It is also a point of interest to notice that the thermodynamic energy and charge differ from the ones we obtain from the asymptotic metric in (\ref{eq:faraway}) when $c$ is not small. This should not come as a surprise since the faraway observables are not the conserved ADM charges of the metric. On the other, hand the thermodynamic quantities computed above do in fact agree with the conserved ADM charges (see for example \cite{Torii:2005nh}) as expected.

\subsection{Hawking-Page Transitions}

Since we are parameterizing our thermodynamic variables with $r_H$ and $Q$, we must express $\mu$ and $T_H$ in terms of the horizon radius and charge from (\ref{eq:mut}) by setting $\varepsilon=0$:
\begin{eqnarray}
\mu &=& - \frac{k}{{\kappa ^2 }}\frac{{(d - 1)}}{{(d - 2)}}r_H ^{d - 2}  -
c\frac{{\left( {d - 1} \right)\left( {d - 3} \right)}}{{(d - 2)}}k^2 r_H^{d -
4}  + \frac{\Lambda }{d}r_H^d - \frac{{Q^2 g^2 }}{{2\left( {d - 2}
\right)r_H^{d - 2} }},\quad\\
T_H&=&\frac{\frac{k}{{\kappa ^2 }}\left( {d - 1} \right)r_H^{2}+c\frac{{\left(
{d - 1} \right)\left( {d - 3} \right)\left( {d - 4} \right)}}{{d - 2}}k^2
-\Lambda r_H^4 - Q^2 g^2 \frac{1}{{2r_H^{2\left( d - 3 \right)} }}}{{4\pi
\left[ {\frac{{\left( {d - 1} \right)r_H^2 }}{{\kappa ^2 }} +
\frac{{2kD}}{{d - 2}}} \right]r_H}}. \label{eq:tgb}
\end{eqnarray}

As in the usual Hawking-Page transition we are essentially going to search for regions in our parameter space where the free energy is negative, and identify these regions with black holes that are thermally favored over the reference background. We further note, as discussed in \cite{Chamblin:1999tk,Cvetic:1999ne,Braden:1990hw}, that there may be regions where black holes are favored globally over the reference background, but are locally unstable in the sense that the specific heat of the thermodynamic system is negative for the globally stable configuration. This implies that even though we have located regions where the black holes are globally favored, they may still decay into some (unknown) configuration that presumably escapes our ansatz \cite{Chamblin:1999tk,Cvetic:1999ne}. We will not attempt to find what these solutions are but we will still attempt to identify such regions in the analysis that follows.

We should also mention once again that the $\varepsilon=0$ analysis has been performed in \cite{Dey:2007vt,Cvetic:2001bk}, although we present it in slightly more generality since we are concentrating on arbitrary dimensions and arbitrary horizon curvature $k$. Furthermore, we present the results in the $r_H$, $Q$ parametrization for $d=4$ in order to readily compare the vanishing $\varepsilon$ results with the non-vanishing $\varepsilon$ ones. We have done the $d\ge4$ analysis in the $r_H$, $T_H$ parametrization and the results are presented in appendix \ref{sec:rt}.

In order to study the global thermodynamic stability of our system we can obtain an explicit form for the critical temperature for any $c$ in any number of dimensions by solving (\ref{eq:free1}) at $F_1=0$ for $T_H$. When working in the $r_H$, $T_H$ parametrization for pure Gauss-Bonnet corrections, we show in appendix \ref{sec:rt} that this results in:
\be
T_c = - \frac{1}{{4\pi }}\frac{{\frac{\Lambda }{d}r_H ^4 + \frac{{\left( {d -
1} \right)\left( {d - 3} \right)k^2 c}}{{d - 2}}}}{{\frac{{r_H ^3 }}{{2\kappa ^2
}} - kc\left( {d - 1} \right)r_H }}.
\label{eq:tcgc}
\ee
The significance of the critical temperature is as follows. For temperatures above the critical temperature with electric potential fixed at infinity, we find that the black hole solution is thermally favored globally with respect to the reference background solution with constant electric potential. Conversely, for temperatures below the critical temperature we will have that the reference background solution is globally favored. This is precisely the Hawking-Phase transition.

We can also express the transition between globally stable black holes versus the reference background solution in terms of a critical charge which is relevant to the $r_H$, $Q$ parametrization we are working in. In particular, we find that the critical charge $Q_c$ is given by:
\begin{multline}
{Q_c}^2=
\frac{2(d-1)}{{g^2 }} \left[ \frac{\frac{1}{\kappa ^2}k{r_H}^2 + ck^2 \left(
\left(d - 7\right) + 2c\left({d - 1}\right)\left(d - 3\right)\frac{\kappa ^2 k}{r_{\rm H} ^2}\right)}{{{1 - 2c\left(d - 1
\right)\frac{\kappa ^2 k}{r_{\rm H} ^2}}}}\right.
\\
+ \left. \frac{{\left(d - 2\right)\Lambda}{r_H}^4 \left(1 + 6c\left(d - 1\right)\frac{\kappa ^2
k}{r_{\rm H} ^2}\right)}{{{d\left(d - 1\right)\left(1 - 2c\left( {d - 1} \right)\frac{{\kappa ^2
k}}{r_{\rm H} ^2}\right)}}}\right]{r_H}^{2\left( d - 3 \right)}.
\label{eq:qcgc}
\end{multline}
Whenever we are considering a configuration with $Q > Q_c$ we find that the black hole is globally favored thermodynamically, whereas if $Q < Q_c$ we have that reference background solution is globally favored.

There are two more critical values of the charge that are of interest. These correspond to those black hole configurations with zero mass and those black hole configurations with vanishing temperature. The zero mass black holes have a charge that satisfies the following expression:
\begin{equation}
Q_{M=0}^2 = \frac{2}{g^2} \left[- \frac{(d - 1)}{{\kappa ^2 }}k{r_H}^{2}  -
c\left( {d - 1} \right)\left( {d - 3} \right)k^2   + \frac{\left( {d - 2}
\right) \Lambda }{d}{r_H}^{4} \right] {r_H}^{2\left( d - 3 \right)}.\label{eq:qm0}
\end{equation}
This bound is not physical in the sense that anti-de Sitter space allows for black holes with negative thermodynamic mass but positive horizon radius and entropy. However, we find it instructive to include this bound in our analysis of the thermodynamics, in order to explore regions of the parameter space containing negative and positive mass black holes with greater clarity.

The second bound, which is in fact physical, is the extremality bound which requires that our black holes have a non-negative temperature. For a given $r_H$ the charge of an extremal black hole is given by:
\begin{equation}
Q_{T_H=0}^2 = \frac{2}{g^2} \left[ \frac{(d - 1)}{{\kappa ^2
}}k{r_H}^{2}+c\frac{{\left( {d - 1} \right)\left( {d - 3} \right)\left( {d - 4}
\right)}}{{d - 2}}k^2 -\Lambda {r_H}^{4} \right] {r_H}^{2\left( d - 3 \right)}.\label{eq:qt0}
\end{equation}
So the physical bound on the black hole charges following from the above expression is that $Q\le Q_{T_H=0}$. Before we study the full global phase structure expressed in the $\phi$, $c$ plane, it will be convenient to consider the local structure first and build the global picture from there. In the next section we discuss the conditions for local thermodynamic and electric stability and study the various configurations arising.

\section{Local Stability}
\label{sec:localgb}

Having obtained the conditions for the global thermodynamic stability of a solution, we proceed to explore the local thermodynamic structure by computing the specific heat and electrical permittivity of our system. The specific heat informs us about the thermal stability of the black hole under temperature fluctuations, and the electrical permittivity informs us about the thermal stability of the black hole under electrical fluctuations.

\subsection{Specific Heat and Electrical Permittivity}

Even though a black hole configuration may be found to be in global thermodynamic preference to the reference background, it may still be locally unstable to some other globally favored configuration. It is not clear to what stable configurations such locally unstable black holes decay and our guess is that such configurations do not satisfy our simple ansatz. Additionally, we will also find configurations of multiple locally stable black hole solutions corresponding to a single temperature, for a fixed potential. Such configurations open the possibility for first order phase transitions from the metastable black hole to the globally favored one.

In order to analyze local stability, we will consider two notions of local stability as discussed in \cite{Chamblin:1999tk}. Firstly, we will consider the specific heat of our system in the grand canonical ensemble. This is given by the following expression:
\begin{equation}
C_\phi   = T_H \left( {\frac{{\partial S}}{{\partial T_H}}} \right)_\phi.
\end{equation}
For regions of the parameter space where the specific heat is positive, we have that the black holes are locally stable to thermal fluctuations. If the specific heat is positive, then an increase in temperature for fixed potential will result in an increase in the entropy resulting in a thermally stable situation. On the other hand, if temperature fluctuations lead to a decrease in the entropy of the black hole the system is locally unstable.

Suppose for a moment that we have a temperature with $n$ black hole solutions. Such a temperature will also have $(n-1)$ turning points as a function of the radius for fixed $\phi$. For large radii we have that the temperature and entropy are both increasing functions of $r_H$ and thus the largest black hole has $dS/dr_H>0$ and $(dT_H/dr_H)^{-1}>0$ for a fixed $\phi$. This condition leads to a positive specific heat and thus local stability under temperature fluctuations. Furthermore, the $dT_H/dr_H$ derivative changes sign consecutively with decreasing horizon radius, which in turn leads to a sequence of locally stable and locally unstable black holes for decreasing radius, so long as $dS/dr_H>0$ holds for fixed $\phi$. The only case where we might have $dS/dr_H<0$ is for $c k<0$; however, it is easy to show that such black holes will have a radius that violates the bound obtained in (\ref{eq:kcbound}) and are thus rendered unphysical. We notice this feature throughout our analysis in the situations where we have many black holes. Finally, we note that if our temperature has only one black hole solution with positive entropy then this black hole will be locally stable.

There is a second notion of local stability that we will study throughout the paper. This was introduced by Chamblin et al. in \cite{Chamblin:1999tk} for the case of charged black hole thermodynamics and it is known as the \emph{isothermal electrical permittivity} $\varepsilon_T$ of the black hole. It is given as follows,
\be
\varepsilon _T  = \left( {\frac{{\partial Q}}{{\partial \phi }}} \right)_{T_H}.
\ee

If $\varepsilon_T$ is non-negative our system is stable under electrical fluctuations. More specifically, if the electric potential at the surface of the black hole increases as a result of introducing an infinitesimal amount of charge to, the black hole the system is locally electrically stable. In contrast a black hole whose electric potential decreases when absorbing a small amount of charge is locally unstable\footnote{Let us clarify this with an analogy from classical thermodynamics. When studying the Gibbs free energy in a hydrodynamic system, our electric potential corresponds to the volume $V$ of the system and the charge of the black hole to the pressure $P$. Then a positive isothermal electrical permittivity will correspond to minus the isothermal compressibility. In particular, a negative $\varepsilon_T$ would be analogous to a thermodynamic system that decreases in size as a result of increasing the pressure, which is clearly locally unstable.}.

Having discussed our notions of thermodynamic stability, we can obtain their explicit expressions as functions of $r_H$ and $Q$. The specific heat for constant electric potential is given by:
\begin{multline}
\frac{C_\phi}{\Sigma_k}   =  - \frac{{4\pi \left( {d - 1} \right)r_H ^{d - 3} \left( {r_H ^2  +
\tilde k} \right)^2 \left[ {\left( {d - 1} \right)\frac{k}{{\kappa ^2 }}\left(
{2r_H ^2  + \frac{{d - 4}}{{d - 2}}\tilde k} \right) - 2\Lambda r_H ^4  -
\frac{{g^2 Q^2 }}{{r_H ^{2\left( {d - 3} \right)} }}} \right]}}{{\kappa ^2
\left( {A - \left( {r_H ^2  - \tilde k} \right)\frac{{g^2 Q^2 }}{{r_H ^{2\left(
{d - 3} \right)} }}} \right)}}.
\label{eq:spef}
\end{multline}
The isothermal electrical permittivity is given by:
\be
\frac{\varepsilon _T}{\Sigma_k}  = \frac{g}{{2\left( {d - 2} \right)r_H ^{d - 2} }}\left\{ {1 +
\frac{{2\left( {d - 2} \right)\left( {r_H ^2  + \tilde k} \right)\frac{{g^2 Q^2
}}{{r_H ^{2\left( {d - 3} \right)} }}}}{{A - \left[ {\left( {2d - 3} \right)r_H
^2  + \left( {2d - 5} \right)\tilde k} \right]\frac{{g^2 Q^2 }}{{r_H ^{2\left(
{d - 3} \right)} }}}}} \right\}.
\label{eq:perm}
\ee
We have defined the following quantities in the above expressions:
\begin{eqnarray}
A &\equiv& \left( {d - 1} \right)\frac{k}{{\kappa ^2 }}\left( {2r_H ^4  + \frac{{d -
8}}{{d - 2}}\tilde kr_H ^2  + \frac{{d - 4}}{{d - 2}}\tilde k^2 } \right) +
2\Lambda r_H ^4 \left( {r_H ^2  + 3\tilde k} \right),\\
\tilde k &\equiv& 2\left( {d - 3} \right)c k \kappa^2.
\end{eqnarray}
At this point we can proceed to study the local stability of our black hole solutions for the various regions of our parameter space. For the sake of completeness we have also included the corresponding figures of the following analysis in the $r_H$, $T_H$ parametrization for $d=4$ in appendix \ref{sec:rt}.

At this point, we should emphasize that the effects we uncover are qualitatively different from those obtain in the pure Einstein-Maxwell theory \cite{Chamblin:1999tk,Cvetic:1999ne}. In other words, effects due to one-loop corrections become as significant as tree level effects, which suggests higher order effects are also important. It was shown in \cite{Dey:2006ds,Dey:2007vt}, however, that for $d=4$ spatial dimensions the Gauss-Bonnet corrections can be matched with gauge coupling corrections in a phenomenological dual matrix model. This gives some evidence that the thermodynamic phase structure we uncover should be visible in a similarly constructed phenomenological dual matrix model.

\subsection{$k>0$}
\subsubsection*{$d=4$ Spatial Dimensions}
\begin{figure}[h]
\begin{center}
$\begin{array}{c@{\hspace{0.1in}}c}
\includegraphics[angle=0,width=0.5\textwidth]{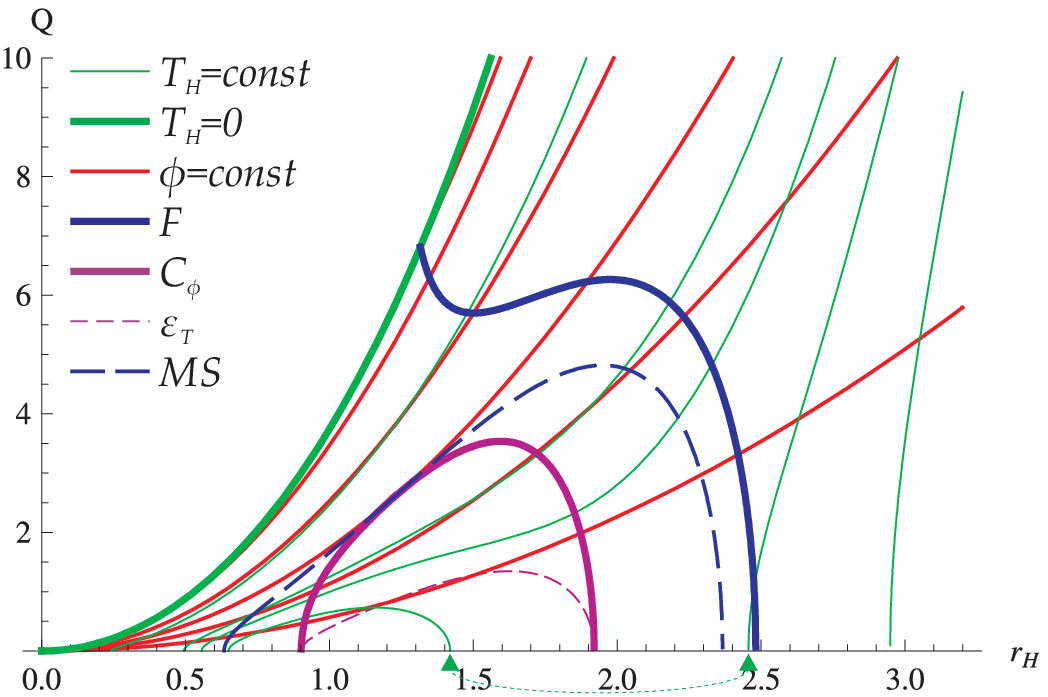} &
   \includegraphics[angle=0,width=0.5\textwidth]{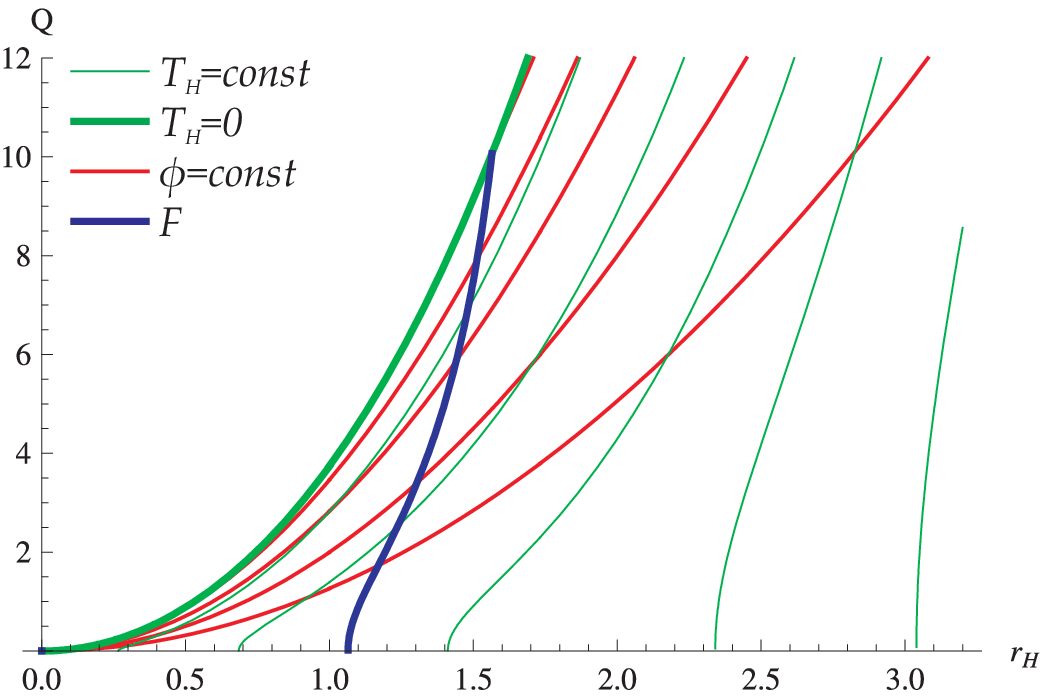} \\
   \text{(a)}&\text{(b)}
\end{array}$
\end{center}
\caption{Thermal structure for $k=2$, $d=4$ and (a) $c<\frac{1}{6\kappa^4|\Lambda|}$, (b) $c>\frac{1}{6\kappa^4|\Lambda|}$. }
\label{fig:k2c1-8}
\end{figure}

We begin by studying the $\bold{c=0}$ in five spacetime dimensions, which was discussed in \cite{Chamblin:1999tk,Chamblin:1999hg}. When $\phi^2<3k/8\kappa^2$, we find temperatures corresponding to two black hole solutions. The smaller of these always has $C_\phi<0$ and is thus rendered locally unstable under temperature fluctuations whereas the larger one is thermally and electrical locally stable. Only a subset of the thermally unstable are also electrically unstable; however, all thermally stable black holes are also electrically stable. When $\phi^2>3k/8\kappa^2$ only one black hole solution exists which is always locally stable both electrically and thermally as well as being always globally stable.
\newline
\newline
The case $\bold{c>0}$ with $d=4$ is depicted in figures \ref{fig:k2c1-8}(a) and (b) for $c<1/6\kappa^4|\Lambda|$ and $c>1/6\kappa^4|\Lambda|$ respectively \cite{Dey:2007vt}. By looking at the slices of constant $\phi$ in figure \ref{fig:k2c1-8}(a) we find slices which intersect the constant temperature slices once (for higher values of $\phi$) or three times (for lower values of $\phi$) corresponding to one or three black hole solutions. In the region with three black holes, only the largest and smallest ones are locally stable, whereas the intermediate sized black hole always resides within the region of local instability enclosed by the $C_\phi$-curve. These black holes may also live in the region of electrical instability which resides within the region of local instability for all the cases we consider. The smallest black hole lives within the region of global instability enclosed by the global stability ($F$) curve and will thus always have a decay channel to thermal anti-de Sitter space. The largest black hole may or may not be globally favored. A configuration where a black hole is globally unstable may have two decay channels: either the reference background or the black hole with lower free energy. Such metastable configurations reside within the region enclosed by the metastable ($MS$) curve. For lower values of the potential, the constant $\phi$ curves cross the $F$-curve, thus it can experience a Hawking-Page transition. For larger values of $\phi$, the constant $\phi$ slices cross the extremality curve, $T_H=0$, without ever reaching the $F$-curve. This means, in particular, that even at zero temperature the ground state of the boundary theory has non-vanishing thermodynamic entropy\footnote{Note, however, as discussed in \cite{Chamblin:1999hg} that this does not imply that there is no longer a confinement-deconfinement phase transition. The temporal Wilson lines have vanishing expectation value since they are unable to wrap around a thermal circle with infinite periodicity. Similarly the spatial Wilson loops are not obstructed by the horizon which is infinitely far down the throat for an extremal black hole and thus cannot exhibit an area law.}.

\begin{figure}[h]
\begin{center}
\includegraphics[angle=0,width=0.5\textwidth]{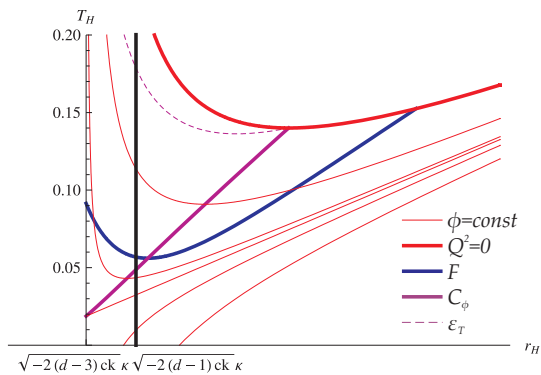}
\end{center}
\caption{Thermal structure for $k=2$, $d=4$ and $c=-\frac{1}{8\kappa^4|\Lambda|}$. }
\label{fig:k2cm1-8}
\end{figure}

As we increase the value of $c$, we no longer find temperatures with three black hole solutions over slices of constant $\phi$. Instead we obtain the situation in figure \ref{fig:k2c1-8}(b). The region of local instability has disappeared altogether and we find single black hole solutions that can undergo Hawking-Page transitions. Furthermore, for large enough $\phi$ some of the black holes in figure \ref{fig:k2c1-8}(b) are globally favored along the whole constant $\phi$ slice and experience no Hawking-Page transition.
\newline
\newline
Finally, we consider the case of $\bold{c<0}$ for $d=4$. When $c<-1/6|\Lambda| \kappa^4$ the behavior is identical to the $d>4$ case and it will be described below. Here we describe the case $c>-1/6|\Lambda| \kappa^4$, which is depicted in figure \ref{fig:k2cm1-8}. For slices of constant $\phi$ we find constant temperature slices that intersect them at either one point for larger values of $\phi$ or at two points for smaller values of $\phi$. For regions with two black holes, the smaller black hole is always within the region of local instability. There are two cases to be noted. The first case corresponds to slices of constant $\phi$, where the larger stable black hole can experience a Hawking-Page transition by moving along the $\phi$ slice. The second case corresponds to the larger black hole unable to experience a Hawking-Page transition because it encounters the region of negative entropy, to the left of the $S$-curve or $C_{\phi}$-curve, before reaching the $F$-curve. The latter case occurs for relatively large values of $\phi$. Finally, we find that for those slices of constant temperature intersecting the constant $\phi$ slices at a single point, and thus corresponding to a single black hole solution, there is no Hawking-Page transition accessible. More specifically the constant $\phi$ curve ends at the extremality bound without intersecting the global stability bound and thus remains globally stable as we continuously lower the temperature.
\subsubsection*{$d>4$ Spatial Dimensions}
\begin{figure}[h]
\begin{center}
\includegraphics[angle=0,width=0.5\textwidth]{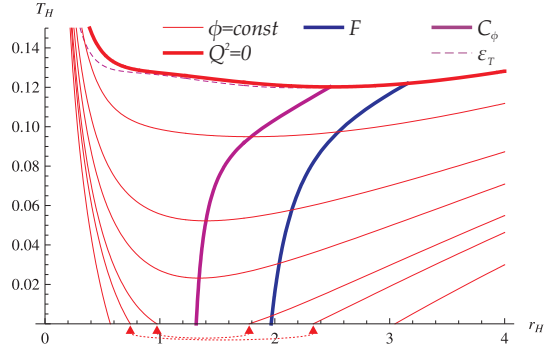}
\end{center}
\caption{Thermal structure for $k=d-2$, $d>4$ and $c>0$.}
\label{fig:k3c1-8}
\end{figure}

\begin{figure}[h]
\begin{center}
$\begin{array}{c@{\hspace{0.1in}}c}
\includegraphics[angle=0,width=0.5\textwidth]{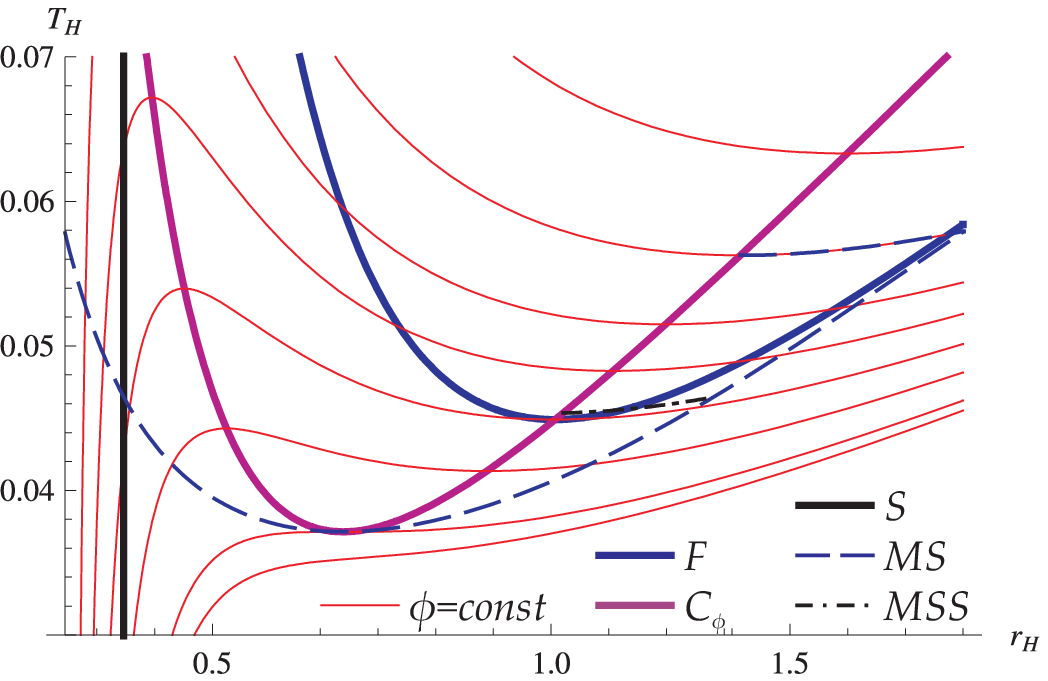}&
\includegraphics[angle=0,width=0.5\textwidth]{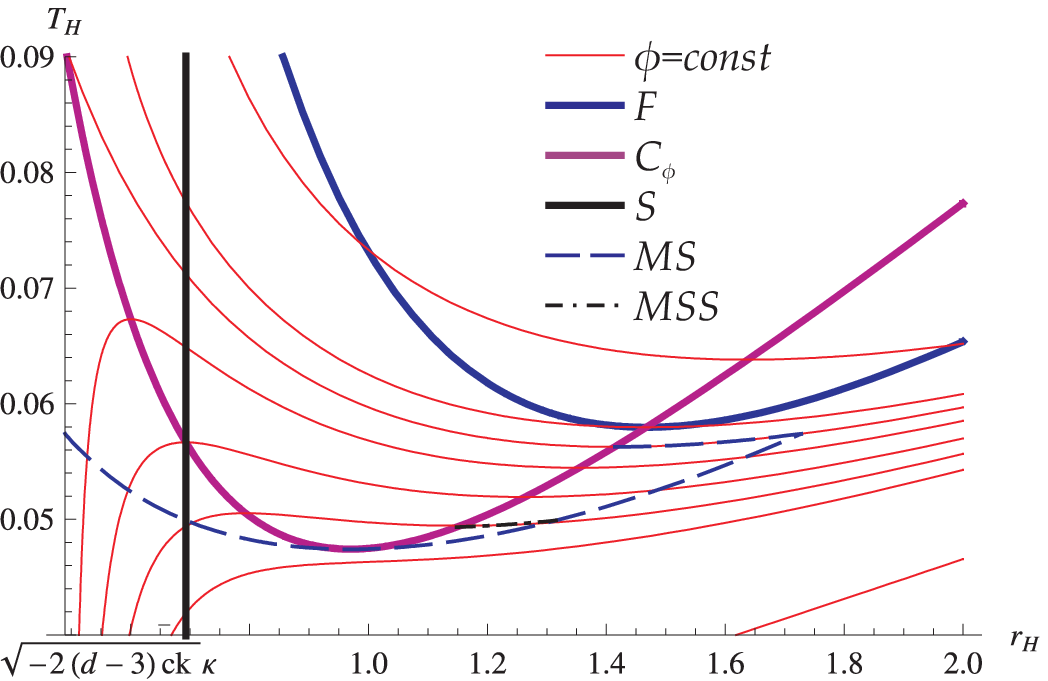}
\\
\text{(a)}&\text{(b)}
\end{array}$
\end{center}
\caption{Thermal structure for $k=d-2$, $d>4$. The two regions of $c$ displayed are:  (a) $\tilde c<c<0$ and (b) $-\frac{\left(d-3\right)\left(d-4\right)\left(2d-3\right)}{4\left(d-1\right)\left(d-2\right)\left(2d-5\right)\kappa^4|\Lambda|}<c<\tilde c$.}
\label{fig:k3cm1-50}
\end{figure}

For the higher dimensional case we find that the $\bold{c=0}$ case remains unchanged \cite{Chamblin:1999tk, Chamblin:1999hg}. Once again there is a critical value of the electric potential $\phi^2_c = \frac{(d-1)k}{2(d-2)^2\kappa^2}$. If $\phi^2 > \phi^2_c$ we have a black hole solution that is always globally and locally stable and thus undergoes no Hawking-Page transition. If $\phi^2<\phi^2_c$ we have two black hole solutions of which only the smaller one is locally unstable and the other may undergo Hawking-Page transitions as we vary the temperature.
\newline
\newline
When $\bold{c>0}$ in higher dimensions, we find temperatures with two black hole solutions of which the smaller one is always locally unstable. For small enough values of the potential we have that the larger black hole may undergo a Hawking-Page transition. For large values of the potential the constant $\phi$ slices end at the extremality bound, as is evident from figure \ref{fig:k3c1-8} and thus there is no Hawking-Page transition.
\newline
\newline
Finally, when $\bold{c<0}$ in higher dimensions we find various deviations from the case with five spacetime dimensions. Let us begin with small negative values of $c$ as exhibited in figure \ref{fig:k3cm1-50}(a). We find temperatures with three black hole solutions of which the smallest and largest are always locally stable, whereas the intermediate sized black hole is always locally unstable. For such three black hole configurations the smallest is always globally favored, but we find values of $\phi$ where \emph{all three} can be globally favored for a particular temperature. At such temperatures we can posit the existence of a metastable phase transition to the black hole with the lowest free energy provided that it has positive entropy. In the case that the globally stable black hole has negative entropy such transitions cannot occur\footnote{See \cite{Cvetic:2001bk} for a different interpretation about the negative entropy black holes}. This phenomenon happens for arbitrarily small negative values of the Gauss-Bonnet coefficient $c$ and should in principle be visible in the thermodynamic structure of the dual boundary theory. Once again the region of metastability is bounded by the $MS$-curve and the $C_\phi$-curve and we find that either the larger or smaller black holes can be metastable. Furthermore, the large metastable black holes that live within the metastable region but above the $MS,S$-curve would decay to small black holes with negative entropy. Thus such large black holes are not truly metastable modes.

It is perhaps amusing that we also find slices of constant $\phi$ that cross the global stability curve at two points. Naively this would seem like the possibility for two Hawking-Page transitions. It turns out that only one of the Hawking-Page transitions occurs outside the region of local instability; only the largest black hole configuration can experience a Hawking-Page transition as the temperature is varied. We also find constant temperature slices with two black hole solutions.  Only the larger one is locally stable and can always experience a Hawking-Page transition upon varying the temperature. For large values of $\phi$ we find single black hole solutions that are always globally and locally stable and never cross the $F$-curve as can be seen from \ref{fig:k3cm1-50}(a).

As we push $c$ to slightly more negative values, we find that for any value of $\phi$ with the three black holes these are always globally stable and thus undergo no Hawking-Page transition. Furthermore, there are regions where the smallest of the three black holes is in a region of negative entropy so that the metastable phase transition towards these cannot occur. These configurations are displayed in figure \ref{fig:k3cm1-50}(b).

\begin{figure}[h]
\begin{center}
$\begin{array}{c@{\hspace{0.1in}}c}
\includegraphics[angle=0,width=0.5\textwidth]{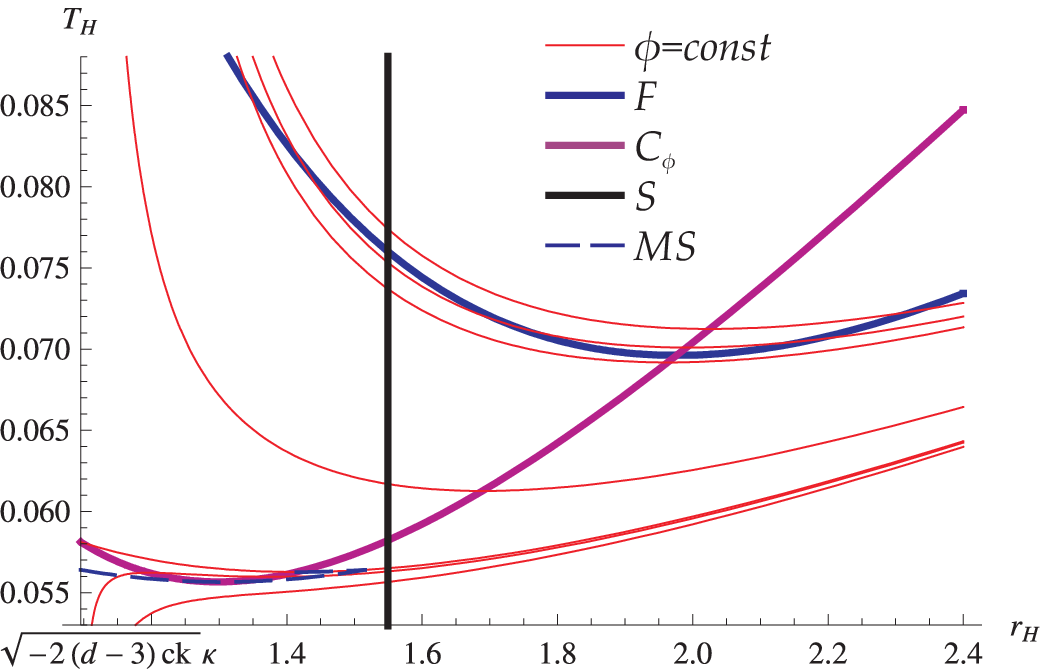}&
\includegraphics[angle=0,width=0.5\textwidth]{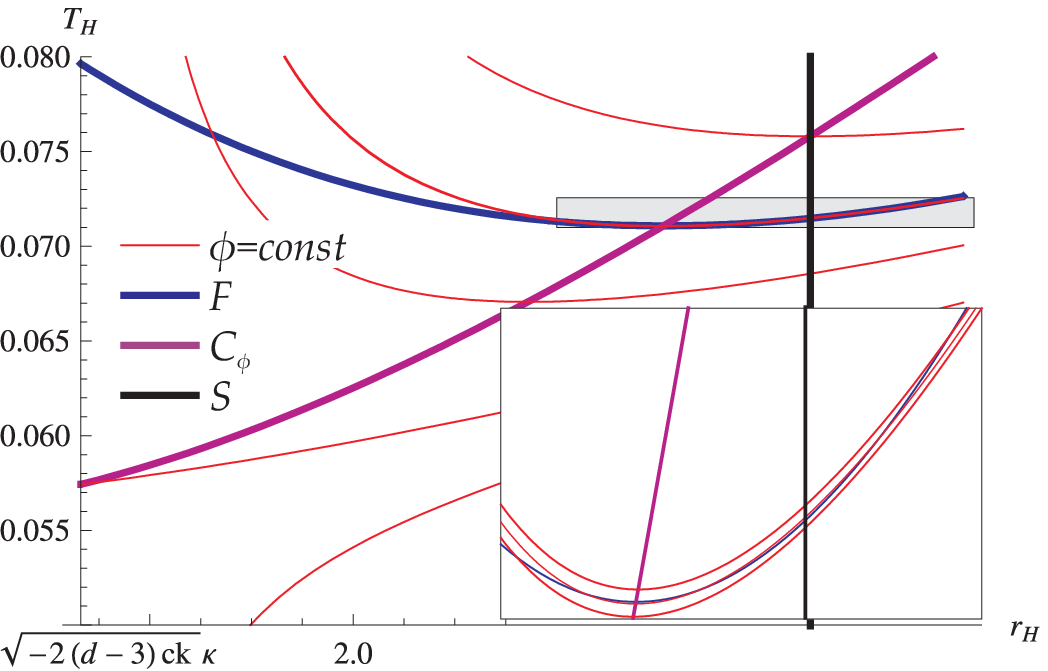}
\\
\text{(a)}&\text{(b)}
\end{array}$
\end{center}
\caption{Thermal structure for $k=d-2$, $d>4$. The two regions of $c$ displayed are: (a) $-\frac{\left(d-3\right)\left(d-4\right)\left(2d-3\right)}{4\left(d-1\right)\left(d-2\right)\left(2d-5\right)\kappa^4|\Lambda|}<c<-\frac{d\left(d-3\right)}{4\left(d-1\right)\left(d-2\right)\kappa^4|\Lambda|}$ and \newline (b) $-\frac{d\left(d-3\right)}{4\left(d-1\right)\left(d-2\right)\kappa^4|\Lambda|}<c<-\frac{d}{4\left(d-2\right)\kappa^4|\Lambda|}$. }
\label{fig:k3cm1-4}
\end{figure}

As $c$ is cranked down to even lower values we find that the two smaller black holes occur in regions of negative entropy and the metastable transitions are lost altogether as is observed in \ref{fig:k3cm1-4}(a). We also find slices of constant $\phi$ intersected twice by the same temperature, corresponding to two black hole solutions. A small subset of these slices of constant $\phi$ encounter the global stability curve at three points. Once again, only the largest black hole configuration is physical and can experience a Hawking-Page transition. The other two Hawking-Page transitions would happen in regions where the black hole is either locally unstable or has a negative entropy and hence are rendered unphysical. These features are shown in figure \ref{fig:k3cm1-4}(a) where the two unphysical Hawking-Page transitions occur in a region of local instability and figure \ref{fig:k3cm1-4}(b) where one of the unphysical Hawking-Page transitions occurs in a region of negative entropy yet local stability and the other unphysical Hawking-Page transition occurs in a region of local instability.
\begin{figure}[h]
\begin{center}
$\begin{array}{c@{\hspace{0.1in}}c}
\includegraphics[angle=0,width=0.5\textwidth]{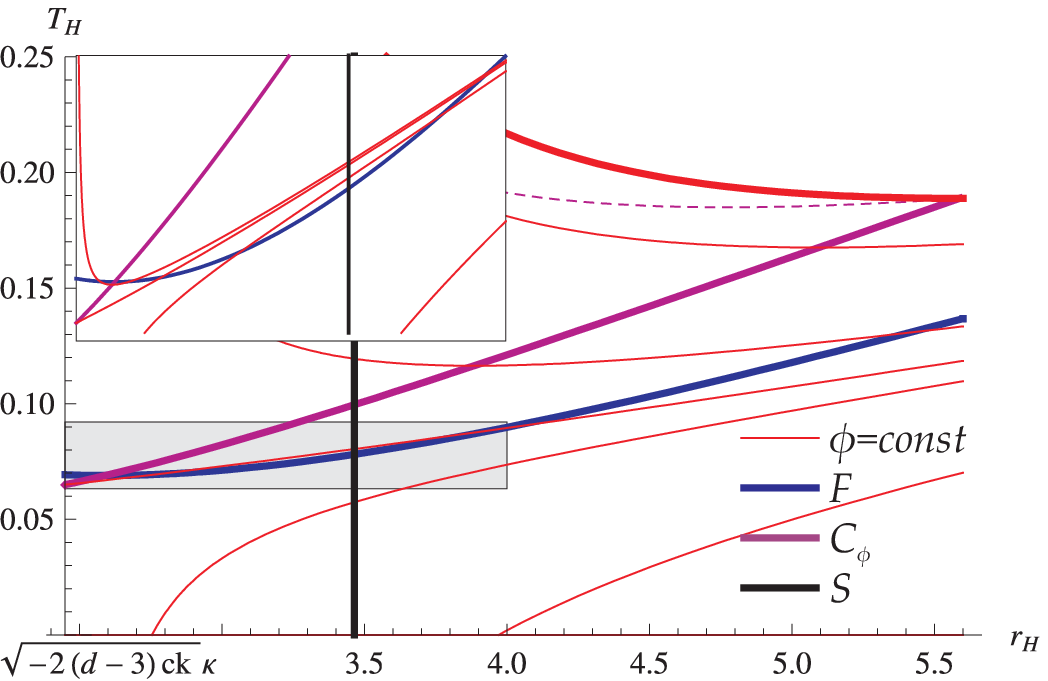}&
\includegraphics[angle=0,width=0.5\textwidth]{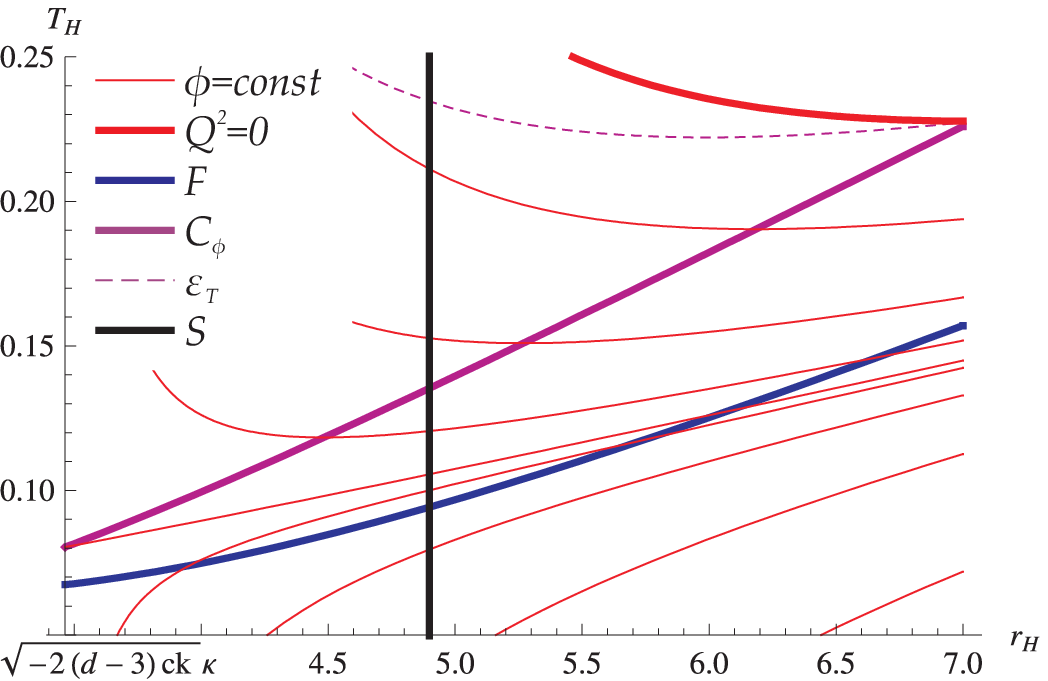}
\\
\text{(a)}&\text{(b)}
\end{array}$
\end{center}
\caption{Thermal structure for $k=d-2$, $d>4$ The two regions of $c$ displayed are: (a) $-\frac{d\left(d-1\right)\left(2d-5\right)}{4\left(d-2\right)\left(d-3\right)\left(2d-3\right)\kappa^4|\Lambda|}<c<-\frac{d}{4\left(d-2\right)\kappa^4|\Lambda|}$ and (b) $c<-\frac{d\left(d-1\right)\left(2d-5\right)}{4\left(d-2\right)\left(d-3\right)\left(2d-3\right)\kappa^4|\Lambda|}$. }
\label{fig:k3cm1}
\end{figure}

We encounter values of $c$ where there are either one or two black hole solutions as seen in figures \ref{fig:k3cm1-4}(b) and \ref{fig:k3cm1}(a) and \ref{fig:k3cm1}(b). In figure \ref{fig:k3cm1}(a) we find that the constant $\phi$ slices with two black holes can cross the global stability curve three times for larger values of $\phi$ or once for smaller values of $\phi$. Once again only the Hawking-Page transition of the largest black hole is physical. In figure \ref{fig:k3cm1}(b) there is no unphysical Hawking-Page transition for constant $\phi$ slices with two black holes. The constant $\phi$ slices with single black hole solutions either cross the global stability curve twice for smaller values of $\phi$ or never for larger values of $\phi$. In the case that they cross it twice, only the Hawking-Page transition between the reference background and the largest black hole configuration is physical.

\subsection{$k<0$}
\begin{figure}[h]
\begin{center}
$\begin{array}{c@{\hspace{0.1in}}c}
\includegraphics[angle=0,width=0.5\textwidth]{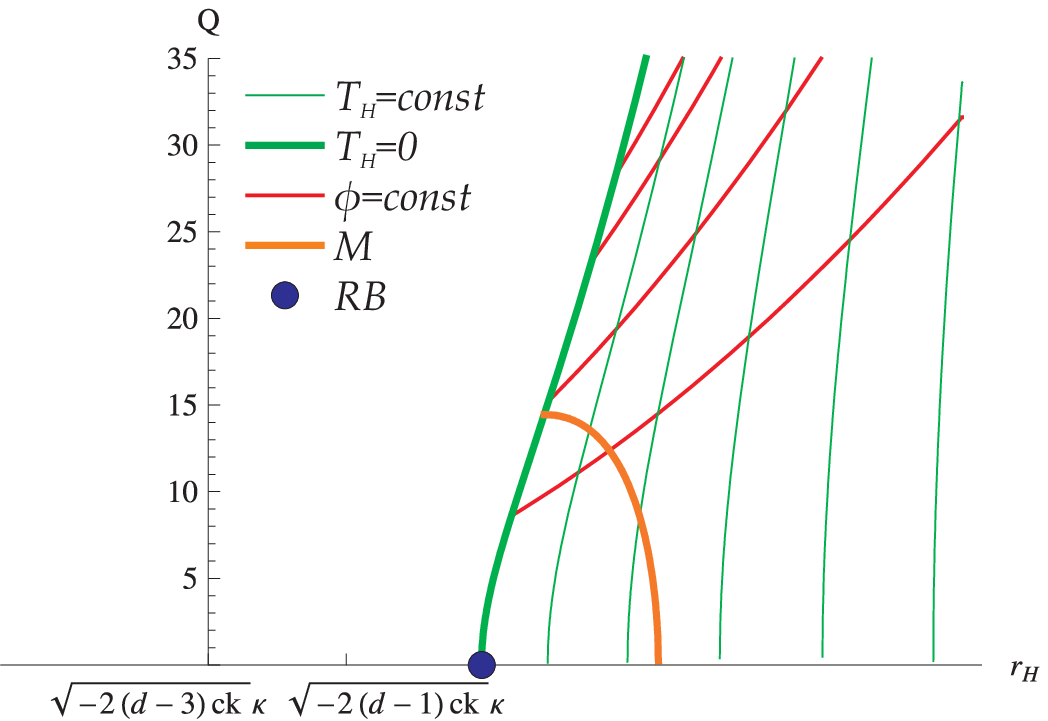} &
   \includegraphics[angle=0,width=0.5\textwidth]{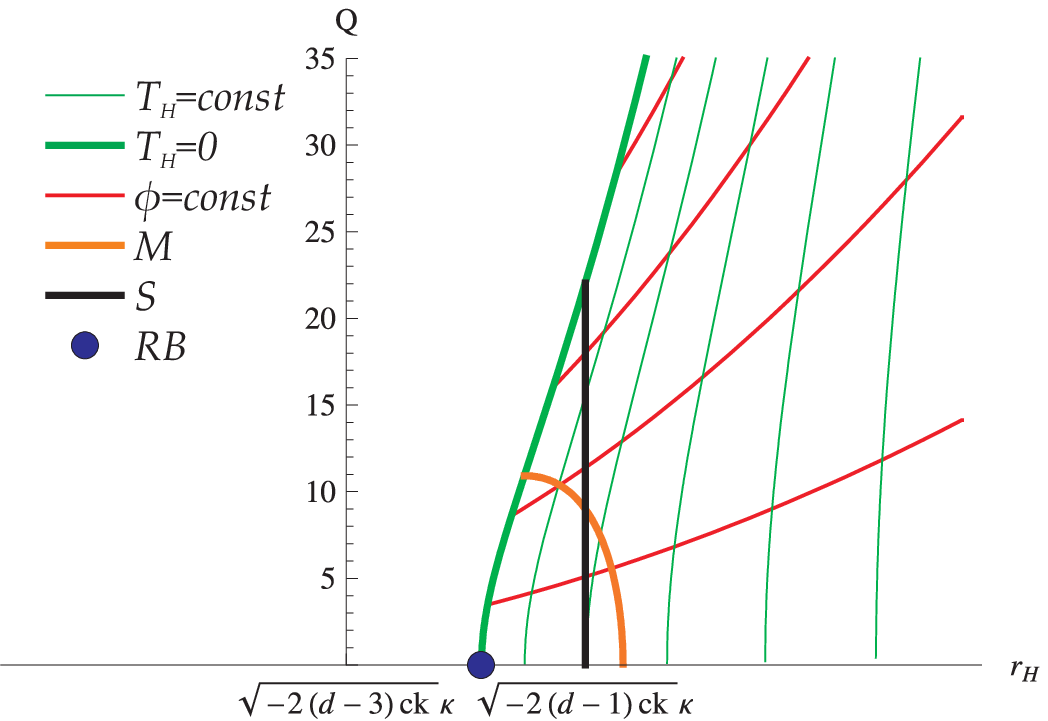} \\
   \text{(a)}&\text{(b)}
\end{array}$
\end{center}
\caption{Thermal structure for $k=-\left(d-2\right)$, $d\ge 4$. The two regions of $c$ displayed are: (a) $c<\frac{d^2+d-8}{4\left(d-1\right)\left(d-2\right)\kappa^4|\Lambda|}$ and (b) $\frac{d^2+d-8}{4\left(d-1\right)\left(d-2\right)\kappa^4|\Lambda|}<c<\frac{d\left(d+1\right)}{4\left(d-1\right)\left(d-2\right)\kappa^4|\Lambda|}$. }
\label{fig:km2c1-4}
\end{figure}

\begin{figure}[h]
\begin{center}
$\begin{array}{c@{\hspace{0.1in}}c}
\includegraphics[angle=0,width=0.5\textwidth]{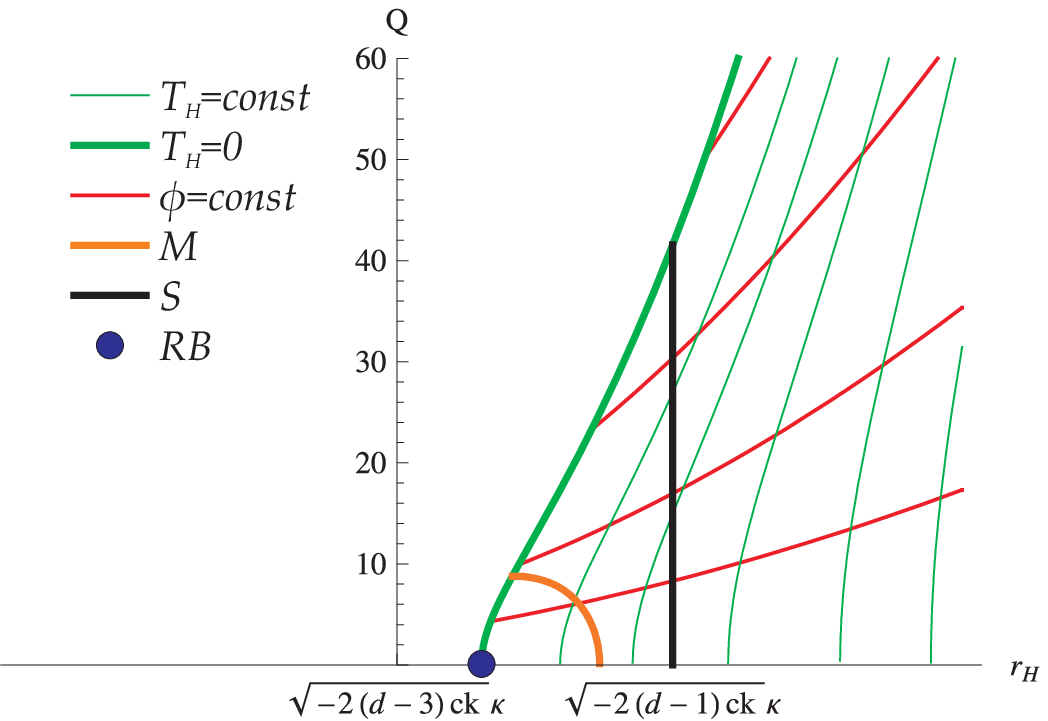} &
\includegraphics[angle=0,width=0.5\textwidth]{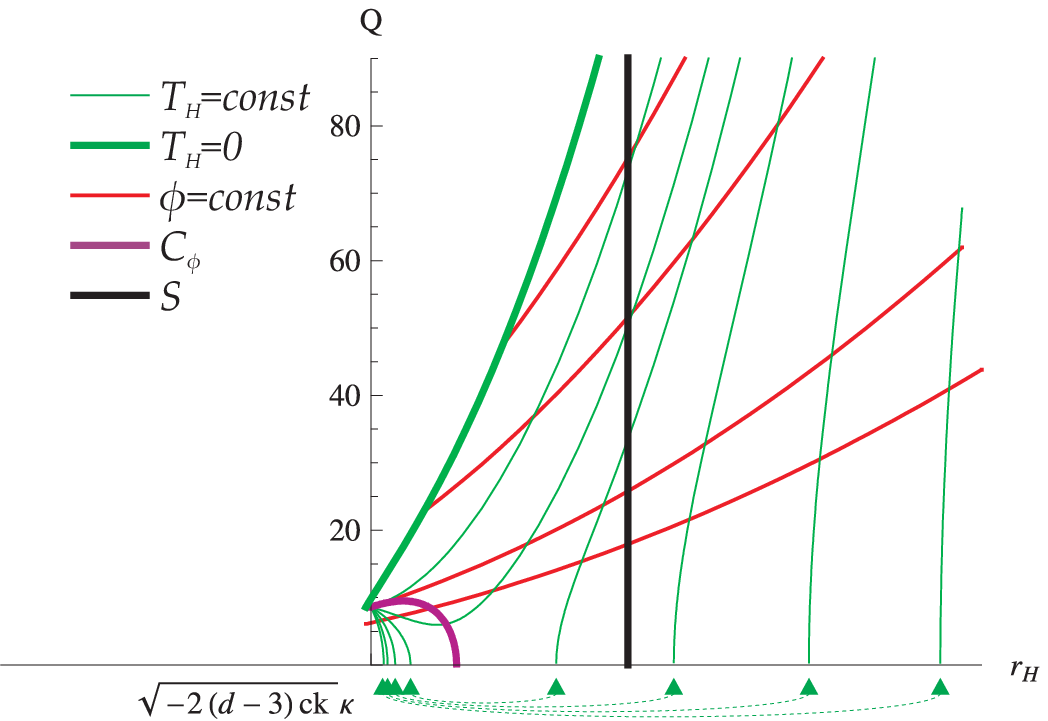} \\
\text{(a)}&\text{(b)}
\end{array}$
\end{center}
\caption{Thermal structure for $k=-\left(d-2\right)$, $d\ge 4$. The two regions of $c$ displayed are: (a) $\frac{d\left(d+1\right)}{4\left(d-1\right)\left(d-2\right)\kappa^4|\Lambda|}<c<\frac{d\left(d-1\right)}{4\left(d-2\right)\left(d-3\right)\kappa^4|\Lambda|}$ and (b) $c>\frac{d\left(d-1\right)}{4\left(d-2\right)\left(d-3\right)\kappa^4|\Lambda|}$. }
\label{fig:km2c17}
\end{figure}

We now turn our attention to those black holes with hyperbolic horizons. The situation with $\bold{c=0}$ was studied in \cite{Cai:2004pz}, where it was found that the black holes are both globally and locally favored and experience no Hawking-Page phase transition in the grand canonical ensemble, as we discussed in section \ref{sec:free}. This result holds for all $d\ge4$.
\newline
\newline
The case of $\bold{c>0}$ is depicted in figures \ref{fig:km2c1-4} and \ref{fig:km2c17}. The reference background, which we have chosen to be the neutral extremal black hole, resides at the point $RB$. The first situation we encounter is shown in figure \ref{fig:km2c1-4}(a), where we have temperatures with single black hole solutions. Such black holes can have negative or zero mass only if they reside within the region bounded by the $M$-curve.
The next situation which is closely related is given figure \ref{fig:km2c1-4}(b). In this case we still have temperatures corresponding to single black hole solutions; however, we also encounter the $S$-curve discriminating those solutions with negative and positive entropy. This entropy bound, however, does not enclose all the massless and negative mass modes.

As we increase the Gauss-Bonnet coefficient $c$ we find that the negative entropy bound has engulfed all the massless and negative mass modes, as seen in figure \ref{fig:km2c17}(a). We also arrive at a similar situation in figure \ref{fig:km2c17}(b), where we have temperatures with two black hole solutions as well as temperatures with single black hole solutions. As is usual the smaller of the two black holes is always locally unstable as well as having negative entropy, whereas the larger one is always locally stable. There are no massless modes present with or without negative entropy in this situation\footnote{One may be interested in possible decays between black holes and the charged extremal black hole with the same external potential. We find that such decays are only possible when the extremal black hole possesses negative entropy.}.
\newline
For the case $\bold{c<0}$ we find that there is again a critical value of $\phi$ that separates the phase structure into two regions. The region with small values of $\phi$ has temperatures corresponding to black hole solutions that can have zero, negative or positive mass and the region with large values of $\phi$ where the black holes have positive mass. This situation is qualitatively equivalent to the one in figure \ref{fig:km2c1-4}(a).

\section{Global Thermal Phase Structure for $c\neq0$ \& $\varepsilon=0$}

\subsection{$k>0$}
\subsubsection*{$d=4$ Spatial Dimensions}
We begin by summarizing the thermal phase structure in the $\phi$ vs. $c$ plane for $d=4$ spatial dimensions. The situation is depicted in figure \ref{fig:k2}(a) and a derivation of the curves bounding the various regions is provided in appendix \ref{sec:figs}. There are three main regions\footnote{For convenience we have also included the definitions of the various regions in Table 1.} denoted by the Roman numerals $I$, $II$ and $III$ containing one, two and three black holes respectively. These regions are identified by mapping out the regions of local instability given by the region within the green curve.
\begin{figure}[h]
\begin{center}
\includegraphics[angle=0,width=0.8\textwidth]{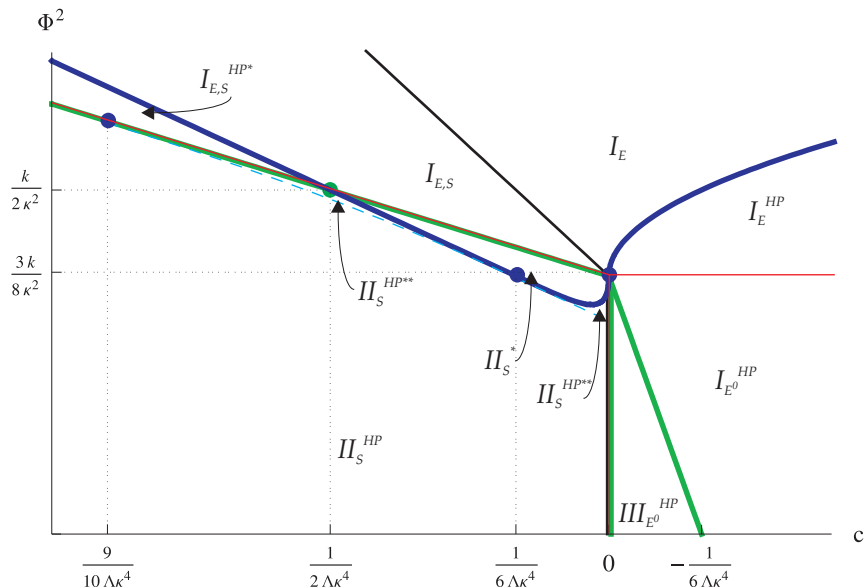}
\end{center}
\caption{Thermal phase structure for $k>0$ and $d=4$.}
\label{fig:k2}
\end{figure}Once we have identified the three main regions we can identify subregions given by the particular properties of the configurations. These are labeled by the subscripts and superscripts. The subscript $E$ denotes regions where there are extremal black hole configurations.
The subscript $E^{0}$ denotes regions where the only extremal solutions are ones with vanishing horizon radius. The subscript $S$ denote regions where there exist configurations with negative entropy. In $S$ regions, possible extremal black holes always have negative entropy. Finally, the superscript $HP$ denotes configurations that experience a Hawking-Page transition. The superscript $HP^*$ denotes the region where the Hawking-Page transition for the large black hole happens in a region of negative entropy. Note that we have metastable black holes in region $III$ where there are three black hole solutions. Of particular interest is region $III$ where we have three black hole solutions of which the intermediate one is locally unstable. The metastable configuration can decay to either the more stable black hole hole configuration or the reference background.

\subsubsection*{$d>4$ Spatial Dimensions}
We now proceed to the $d>4$ case depicted in figure \ref{fig:k3}. Once again, the regions of the diagram are obtained explicitly in appendix \ref{sec:figs}. The region within the shaded box in figure \ref{fig:k3} is expanded. \begin{figure}[h]
\begin{center}
\includegraphics[angle=0,width=0.8\textwidth]{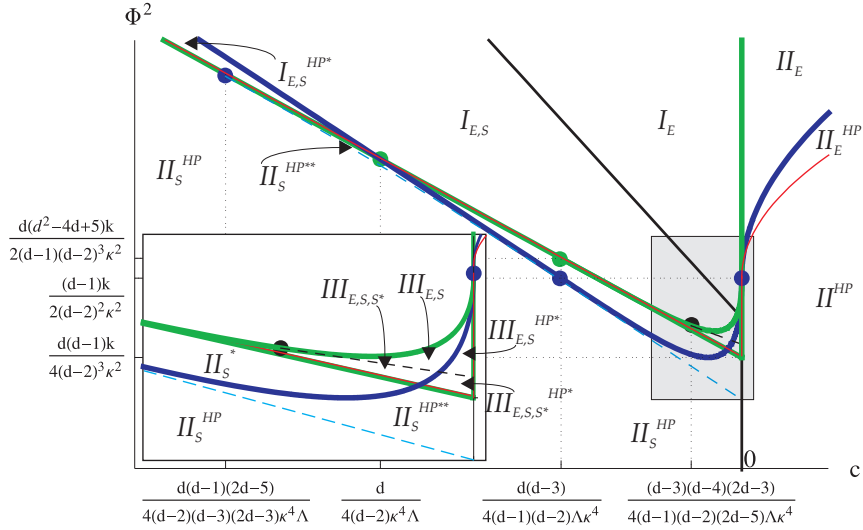}
\end{center}
\caption{Thermal phase structure for $k>0$ and $d>4$. }
\label{fig:k3}
\end{figure}Once again the phase diagram is divided into three main regions $I$, $II$ and $III$ denoting the number of black hole solutions. These regions are delineated by the curves of local instability. As before the metastable black holes are found in region $III$. Within each main region there are various subregions as before. We have encountered, however, two new types of subregions. The subscript $S^*$ denotes a subregion of $III$ where the smallest black holes always have negative entropy. The subscript $HP^{**}$ denotes subregions where we find the lines of constant $\phi$ cross the global stability three times and thus there are two unphysical Hawking-Page transitions. Of particular interest is the region $III_{E,S}$, where we encounter three globally stable black hole solutions at a constant temperature. The intermediate sized one is locally unstable. If we are in the metastable configuration we can have a phase transition to the globally stable configuration without decaying to the reference background.

\subsection{$k<0$}
We proceed to study the case with $k<0$ which is depicted in figure \ref{fig:km2}. As before the explicit regions given in these diagrams are derived in appendix \ref{sec:figs}. There are only two main regions denoted by $I$ and $II$, where we have one or two black hole solutions. As usual, these regions are split into various subregions. A new subregion that we have not yet encountered is denoted by the subscripts $M$ and $M^*$, where we find black holes with vanishing or negative thermodynamic mass.
\begin{figure}[h]
\begin{center}
\includegraphics[angle=0,width=0.8\textwidth]{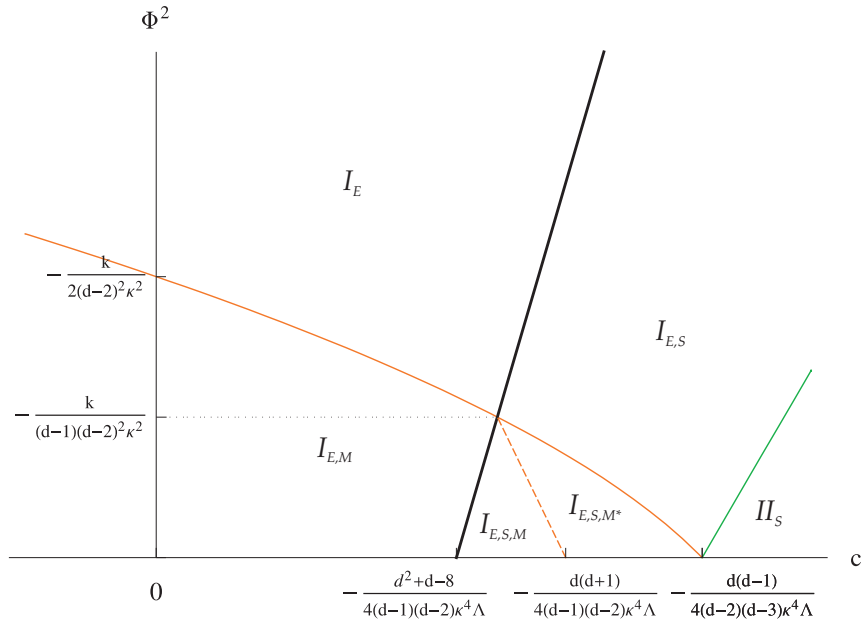}
\end{center}
\caption{Thermal phase structure for $k<0$.}
\label{fig:km2}
\end{figure}The asterisk in $M^*$ refers to regions where the negative or vanishing mass black hole has negative entropy. Finally, we mention that black holes that are unstable locally occur only in regions of parameter space that are hidden by the negative entropy bound.

\subsection{$k=0$}
Finally, we visit the $k=0$ case which has the simplest thermal structure. Particularly, for the case of a flat asymptotic geometry in the grand canonical ensemble, we find again the absence of a Hawking-Page phase transition. This is the same result as that for the $c=0$ case and we find that it holds for all values of $c$.

\begin{center}
\begin{tabular}{|c|l|}
\hline
\multicolumn{2}{|c|}{Table 1: Definitions of Subregions} \\
\hline
$I$ & Single black hole for any given temperature. \\
$II$ & Two black holes for any given temperature. \\
$III$ & Three black holes for some temperatures. \\
\hline
${HP}$ & Physical Hawking-Page transition.\\
$*$ & Single unphysical Hawking-Page transition.\\
$**$ & Two unphysical Hawking-Page transitions.\\
\hline
$E$ & Extremal black hole present.\\
$E^0$ & Extremal black hole has vanishing radius.\\
\hline
$S$ & Negative entropy modes present.\\
$S^*$ & Smallest black holes have negative entropy.\\
\hline
$M$ & Negative mass modes present.\\
$M^*$ & All negative mass modes have negative entropy.\\
\hline
\end{tabular}
\end{center}

As a final remark we reiterate here that there is a vastly richer thermal phase structure for the case of charged black holes with a Gauss-Bonnet term than the usual Hawking-Page transition between neutral black holes and thermal anti-de Sitter space. There are configurations with locally unstable black holes which are globally stable as those discovered in \cite{Chamblin:1999tk,Cvetic:1999ne}, as well as globally unstable yet locally stable configurations. The black holes we are considering have one more free parameter compared to the neutral black holes considered in the original work \cite{Hawking:1982dh}. Thus the free energy is a function of two varying parameters constrained along slices of constant $\phi$. A metastable black hole can thermally evolve to some $r^*_H,Q^*$ configuration that is locally and globally stable. This is particularly true in the grand canonical ensemble, where the electric potential is kept fixed but not the electric charge so that the black holes can emit and absorb charged radiation.

\newpage
\begin{center}
\section*{PART III - $F^4$ Thermodynamics}
\end{center}
\addcontentsline{toc}{part}{PART III - $F^4$ Thermodynamics}
\section{Global Stability for $\varepsilon \neq 0$ and $c=0$}
\label{sec:eps}
We can now direct our attention to the case of $\varepsilon \neq 0$. Once we obtain the general expressions for the free energy, specific heats and thermodynamic quantities of the theory, we proceed to explore thermal phase structure of the $F^4$ theory with no Gauss-Bonnet term in the action. Finally, we study the deformations to the thermodynamic structure of the Einstein-Maxwell-Gauss-Bonnet theory caused by $F^4$ corrections.

\subsection{Computing the Free Energy}
In order to compute the free energy of the $\varepsilon \neq 0$ theory we need to slightly extend the discussion in section \ref{sec:free}. In particular, we have to evaluate the full Euclidean action given by the sum of (\ref{eq:I1}) and (\ref{eq:I2}). Once again we will be working in the $r_H,Q$ parametrization due to the non-trivial $Q$ dependence of the Euclidean action. The first term in the Euclidean action becomes:
\be
 - \frac{{F^\varepsilon_1 -F^\varepsilon_{vac}}}{{\Sigma_k }} =  - \frac{2}{{d - 3}}\left[ {\frac{1}{{\kappa ^2 }}\left( { 4\pi {r_H} ^{d - 1} T_H  - \frac{{d - 1}}{{d - 2}}k{r_H} ^{d - 2} } \right) - \frac{\mu}{2} + \frac{2}{d}\Lambda {r_H} ^d }\right],
  \label{eq:epsfree1}
\ee
where now $T_H$ and $\mu$ are considered as functions of $r_H$ and $Q$ and furthermore depend on $\varepsilon$ as dictated by (\ref{eq:htrq}) and (\ref{eq:murq}). $F^\varepsilon_{vac}$ was defined in (\ref{eq:fvac}). The second term in the Euclidean action becomes:
\be
-\frac{F^\varepsilon_2}{\Sigma_k} = \frac{8\varepsilon}{d-3}  \int^\infty_{r_H} dr r^{d-1} f(r)^4.
\ee
The above expressions are for general $\varepsilon$ and no approximations have been made. We will perform our analysis numerically using the numerical integration packages contained in Mathematica \cite{mathematica}. Thus the basic object we will be working with is given by the full Euclidean action:
\be
I^\varepsilon_{total} = I^\varepsilon_1 + I^\varepsilon_2 = -\beta(F^\varepsilon_1 + F^\varepsilon_2).
\ee
The reference vacua used to test the global stability of a given black hole solution are once again those introduced in section \ref{sec:free}, since once we set $Q$ to zero all $\varepsilon$ effects vanish. For the sake of completeness we include the $\chi \ll 1$ expansion of the total free energy below:
\begin{multline}
F^{\varepsilon}_{1} + F^{\varepsilon}_{2} = \\ \frac{{kr_H ^{d - 2} }}{{\left( {d - 2} \right)\kappa ^2 }} + \frac{{\Lambda
r_H ^d }}{{d\left( {d - 1} \right)}} - \frac{{g^2 Q^2 }}{{2\left( {d - 1}
\right)r_H ^{d - 2} }}\left[\frac{1}{(d-2)} - \frac{{ \chi}}{{4\left( {3d - 4} \right) }} + \mathcal{O}(\chi^2)\right].
\end{multline}
Furthermore, we include the Hawking temperature expanded to first order in $\chi$:
\be
T_H^\varepsilon = \frac{1}{{8\pi \left( {d - 1} \right)}}\left[ \frac{{2\left( {d - 1}
\right)k}}{{r_H }} - 2\kappa ^2 \Lambda r_H  - \frac{{\kappa ^2 g^2 Q^2 }}{{r_H
^{2d - 3} }}\left(1 - \frac{\chi}{2} + \mathcal{O}(\chi^2) \right)
\right].
\label{eq:Teps}
\ee

\subsection{Thermodynamic Quantities}
Having obtained the Euclidean action we can in principle obtain all the thermodynamic quantities. We must compute the appropriate thermodynamic derivatives given in appendix \ref{sec:thder} for the $r_H$, $Q$ parametrization. We note, however, that the fixed external electric potential is modified to the following:
\be
\phi^\varepsilon(r_H,Q) = \frac{1}{2g} \int^\infty_{r_H} dr f(r),
\label{eq:phieps}
\ee
which in the limit $\chi \ll 1$ becomes
\be
\phi^\varepsilon  = \frac{{g Q}}{{2r_H ^{d - 2} }}\left[ \frac{1}{\left( {d - 2} \right)} - \frac{\chi}{{\left( {3d - 4} \right) }} + \mathcal{O}(\chi^2)\right].
\label{eq:phichi}
\ee
Even though we are barred from obtaining full analytic expressions, by direct application of Wald's formula \cite{Wald:1993nt}, we expect that the entropy as a function of $r_H$ and $Q$ is not affected by a contribution to the matter Lagrangian. Indeed we can verify this to third order in $\chi$ and numerically for arbitrary values of $\varepsilon$. In fact one can explicitly check that the thermodynamic electric charge also remains unchanged for non-zero $\varepsilon$. Thus we obtain:
\begin{eqnarray}
\mathcal{Q}^\varepsilon &=& -\frac{1}{\beta_H}\left( \frac{\partial I^\varepsilon_{total}}{\partial {\phi^\varepsilon}} \right)_{\beta_H} = \Sigma_k \times 2 g Q, \\
S^\varepsilon &=& \beta_H \left(\frac{\partial I^\varepsilon_{total}}{\partial \beta_H}\right)_{\phi^\varepsilon} - I^\varepsilon_{total}=
\Sigma_k \times \left(\frac{{4\pi r_H^{d - 1} }}{{\kappa ^2 }} + 8\pi c k\left( {d - 1} \right)r_H^{d - 3} \right),
\end{eqnarray}
We can then use the first law to obtain the thermodynamic energy
\be
E^\varepsilon = \frac{I^\varepsilon_{total}+S^\varepsilon}{\beta_H}+\mathcal{Q}^\varepsilon \phi^\varepsilon = -\Sigma_k \times \mu^\varepsilon(r_H,Q).
\ee
Using the expansion in (\ref{eq:intexp}) we expand $\mu^{\varepsilon}$ in $\chi \ll 1$:
\begin{multline}
\mu^\varepsilon = - \frac{1}{{\kappa ^2 }}\frac{{d - 1}}{{d - 2}}k{r_H} ^{d - 2}  -
c\frac{{\left( {d - 1} \right)\left( {d - 3} \right)}}{{d - 2}}k^2 {r_H} ^{d -
4}  \\+ \frac{\Lambda }{d}{r_H} ^d  - \frac{g^2Q^2}{2(d-2)r^{d-2}}\left[1+\frac{(d-2)}{2(4-3d)}\chi + \mathcal{O}(\chi^2)\right].\label{eq:mueps}
\end{multline}

\subsection{Hawking-Page Transitions}

In principle we would like to obtain the analytic quantities for the critical parameters that dictate the Hawking-Page transitions. We can achieve this by expressing $I^\varepsilon_{total}$ in terms of our two free parameters $r_H$ and $Q$ and solving $I^\varepsilon_{total} = 0$. This is only possible numerically for arbitrary values of $\varepsilon$ although one can obtain analytic expressions upon expanding in $\chi$, which are given below:
\begin{eqnarray}
(Q_c^\varepsilon)^2  &=& Q_{c} ^2 \left[ 1 + \frac{{\left( {d - 2} \right) }}{ 2{(3d - 4)}} \chi_c + \mathcal{O}\left( {\chi_c^2 } \right) \right],\\
(Q_{M = 0}^\varepsilon)^2  &=& Q_{{\rm M} = 0}^2 \left[ 1 + \frac{{\left( {d - 2} \right) }}{{ 2{(3d - 4)} }}\chi_{M=0} + \mathcal{O}\left( {\chi_{M=0}^2 } \right)\right],\\
(Q_{T_H  = 0}^\varepsilon)^2  &=& Q_{T_H  = 0} ^2 \left[ {1 + \frac{\chi_{T_H=0}}{{2}} + \mathcal{O}\left( {\chi^2_{T_H=0} } \right)} \right]
\end{eqnarray}
where $Q^2_c$, $Q^2_{M=0}$ and $Q^2_{T_H=0}$ are given in (\ref{eq:qcgc}), (\ref{eq:qm0}) and (\ref{eq:qt0}) respectively where we set $c=0$. We have also defined the unitless parameter $\chi_x \equiv (8g^6Q^2_x\varepsilon)/r_H^{2(d-1)}$, where $x=\{c,T_H=0,M=0\}$. Black holes with $Q>Q_c^\varepsilon$ are thermodynamically favored, massless black holes have $Q = Q_{M = 0}^\varepsilon$ and positive temperature black holes satisfy $Q\le Q^\varepsilon_{T_H  = 0}$.

\section{Local Stability}
To test the local thermodynamic stability for non-zero $\varepsilon$ we must obtain the specific heat, $C_\phi$, and the isothermal electrical permittivity, $\varepsilon_T$, of our solutions. These quantities are defined by equations (\ref{eq:spef}) and (\ref{eq:perm}). Analytic expressions for $C_\phi$ and $\varepsilon_T$ can be obtained only upon expansion in $\varepsilon$ and we will treat them on the same numerical footing as everything else. For completeness, however, we include the expansions for the local equilibrium parameters below:
\begin{multline}
\frac{C_\phi}{\Sigma_k}   =  - \frac{{4\pi \frac{{d - 1}}{{\kappa ^2 }}\left( {2\frac{{d - 1}}{{\kappa ^2 }}kr_H ^2  - 2\Lambda r_H ^4  - \frac{{g^2 Q^2 }}{{r_H ^{2\left( {d - 3} \right)} }}} \right)r_H ^{d - 1} }}{{2\frac{{d - 1}}{{\kappa ^2 }}kr_H ^2  + 2\Lambda r_H ^4  - \frac{{g^2 Q^2 }}{{r_H ^{2\left( {d - 3} \right)} }}}}\\ - \frac{{4\pi \frac{{\left( {d - 1} \right)\left( {d - 4} \right)}}{{\kappa ^2 }}\frac{{g^2 Q^2 }}{{r_H ^{2\left( {d - 3} \right)} }}\left( {2\frac{{d - 1}}{{\kappa ^2 }}kr_H ^2  + 4\frac{d}{{d - 4}}\Lambda r_H ^4  - \frac{{g^2 Q^2 }}{{r_H ^{2\left( {d - 3} \right)} }}} \right)r_H ^{d - 1} }}{{\left( {3d - 4} \right)\left( {2\frac{{d - 1}}{{\kappa ^2 }}kr_H ^2  + 2\Lambda r_H ^4  - \frac{{g^2 Q^2 }}{{r_H ^{2\left( {d - 3} \right)} }}} \right)^2 }}\chi+\mathcal{O}(\chi^2).
\end{multline}
Having obtained all the basic global and local thermodynamic quantities, we can begin exploring the various regions of the parameter space. Our approach will consist of first exploring the thermal phase structure for the theory with only an $F^4$ correction and no Gauss-Bonnet term. Once we identify the relevant qualitative regions of the $\varepsilon - \phi^\varepsilon$ plane with $c=0$, we will proceed to explore how the $F^4$ term deforms the various qualitative regions we found in the previous section for the $\varepsilon = 0$ case.

\subsection{$k>0$}

\begin{figure}[h]
\begin{center}
$\begin{array}{c@{\hspace{0.1in}}c}
\includegraphics[angle=0,width=0.5\textwidth]{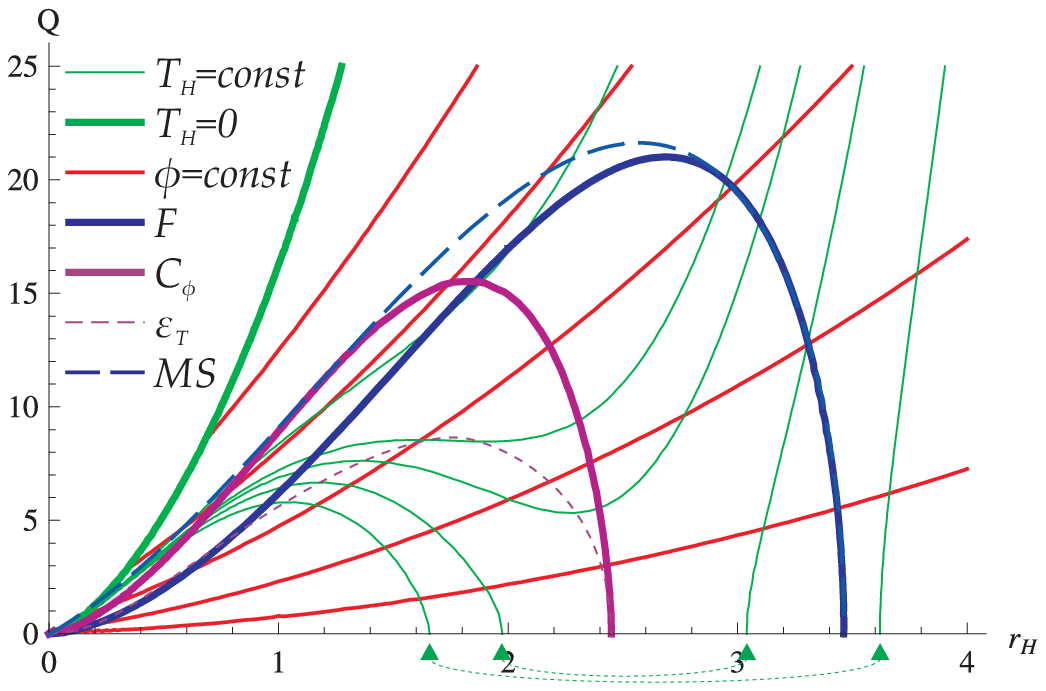} &
\includegraphics[angle=0,width=0.5\textwidth]{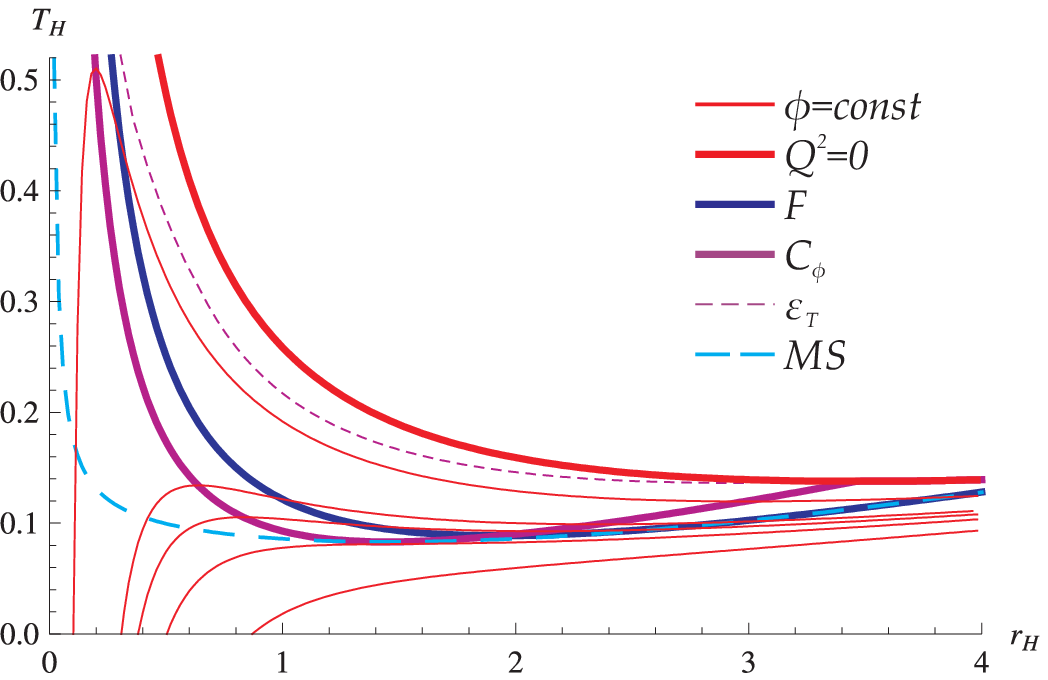} \\
\text{(a)}&\text{(b)}
\end{array}$
\end{center}
\caption{Thermal structure for $k>0$, $d\ge 3$ and  $\varepsilon > \varepsilon^+_{2}$ in both the $r_H$, $Q$ plane (a) and the $r_H$, $T_H$ plane (b). }
\label{fig:e01k2}
\end{figure}

\begin{figure}[h]
\begin{center}
$\begin{array}{c@{\hspace{0.1in}}c}
\includegraphics[angle=0,width=0.5\textwidth]{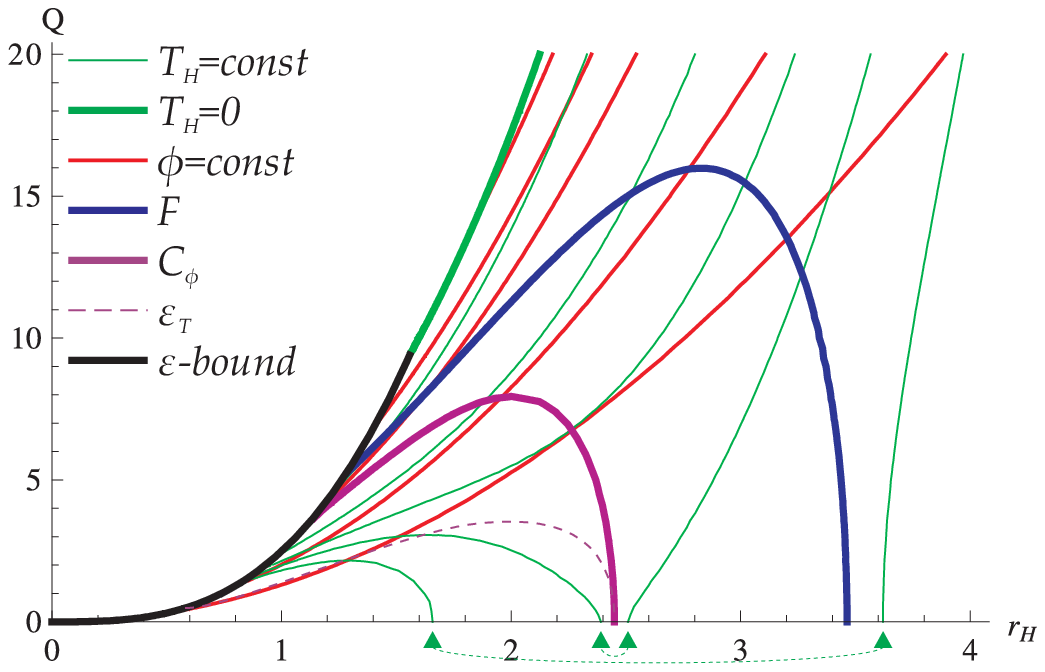} &
\includegraphics[angle=0,width=0.5\textwidth]{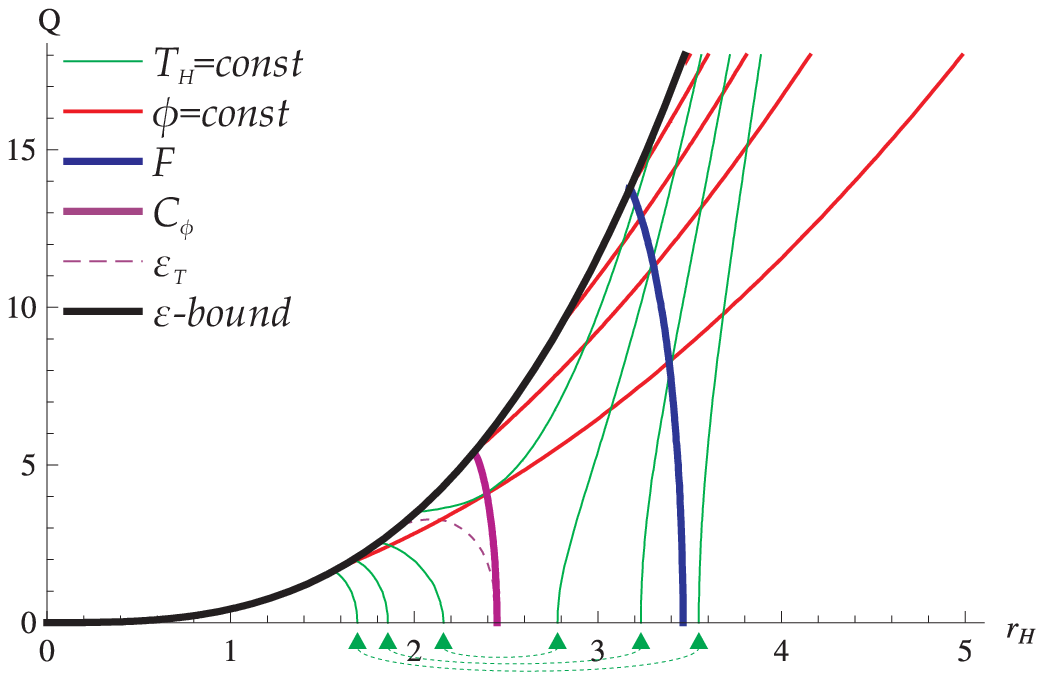} \\
\text{(a)}&\text{(b)}
\end{array}$
\end{center}
\caption{Thermal structure for $k>0$, $d\ge 3$ and (a) $\varepsilon^+_{1} < \varepsilon < \varepsilon^+_{2}$, (b) $\varepsilon < \varepsilon^+_{1}$. }
\label{fig:em01k2}
\end{figure}

We begin by exploring the $\bold{\varepsilon>0}$ case, for $d\ge 3$, depicted in figure \ref{fig:e01k2}. For large enough values of $\phi^\varepsilon$, we find that there is only one black hole which is always globally and locally favored and undergoes no Hawking-Page transition. As we decrease the value of $\phi^\varepsilon$, we encounter constant $\phi^\varepsilon$ slices with three black hole solutions. Of the three black holes the medium sized one is always locally unstable, whereas the largest one can experience a Hawking-Page transition.
The smallest black hole is always globally favored and it cannot reach the global stability curve via a continuous change in the temperature without first crossing the local instability curve. Interestingly we find temperatures where two of the three black hole solutions are globally favored and separated by an intermediate black hole which is both locally and globally unstable. For such situations we can conceive of a phase transition from the least to most globally stable configuration. We also find temperatures where all three black hole solutions are globally favored although as always the intermediate one is locally unstable. A similar phase transition should happen for such situations. The metastable black hole is qualitatively similar to the one we found in the $c<0$, $k>0$ case for $d>4$ with $\varepsilon=0$. The local electrical stability curve is always hidden within the local thermal instability region, as was the case for the $c\neq0$, $\varepsilon = 0$ black hole. This is in fact a common feature for all values of $\varepsilon$ and $k$ both positive and negative.
\newline
\newline
For the $\bold{\varepsilon<0}$ case, with $d\ge 3$, depicted in figure \ref{fig:em01k2}(a) and (b), we find constant $\phi^\varepsilon$ slices containing one or two black hole solutions. For large enough values of $\phi^\varepsilon$ we find a single black hole solution that is globally favored and undergoes no Hawking-Page transition. Lowering the value of $\phi^\varepsilon$ we find that the single black hole solution has access to Hawking-Page transitions. Finally, for small enough values of $\phi^\varepsilon$ we have two black hole solutions for which the larger one experiences Hawking-Page transitions and the smaller is always locally unstable. We find that all extremal black hole solutions disappear after a critical $\varepsilon<0$ due to the bound on the minimum allowed horizon radius (\ref{eq:epsbound}) depicted by the $\varepsilon$-bound-curve in figure \ref{fig:em01k2}(b). This bound also reduces the region of allowed Hawking-Page transitions as $\varepsilon \to -\infty$.

\subsection{$k\leq 0$}

\begin{figure}[h]
\begin{center}
\includegraphics[angle=0,width=0.5\textwidth]{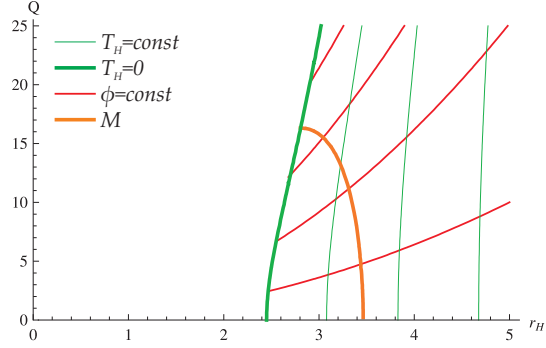}
\end{center}
\caption{Thermal structure for $k<0$, $d\ge3$ and $\varepsilon > \varepsilon^-_{2}.$}
\label{fig:em01km2}
\end{figure}

\begin{figure}[h]
\begin{center}
$\begin{array}{c@{\hspace{0.1in}}c}
\includegraphics[angle=0,width=0.5\textwidth]{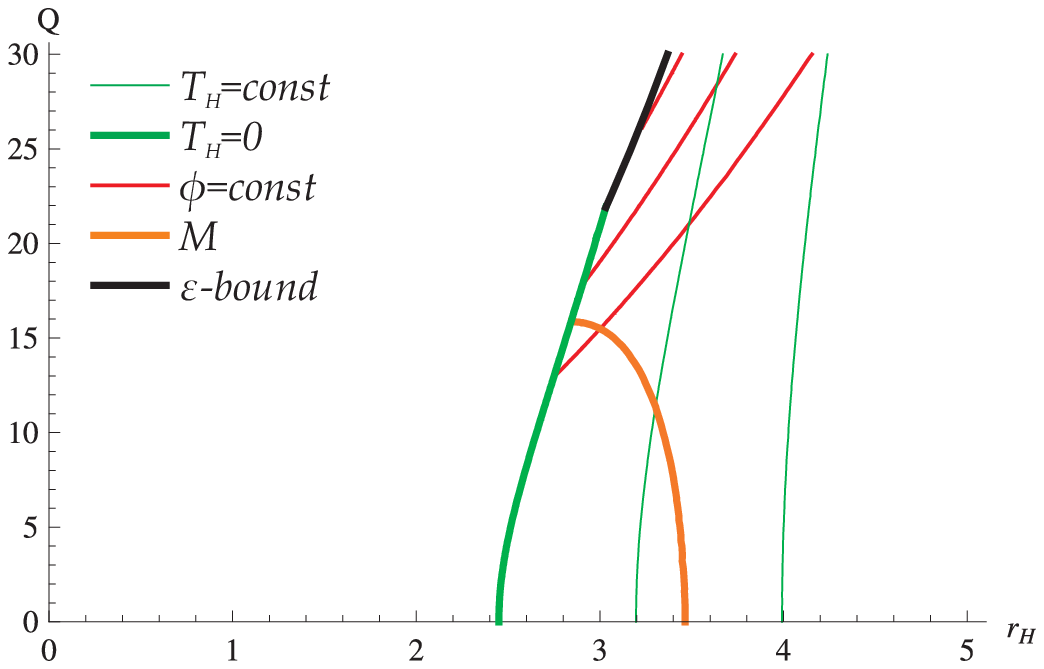} &
\includegraphics[angle=0,width=0.5\textwidth]{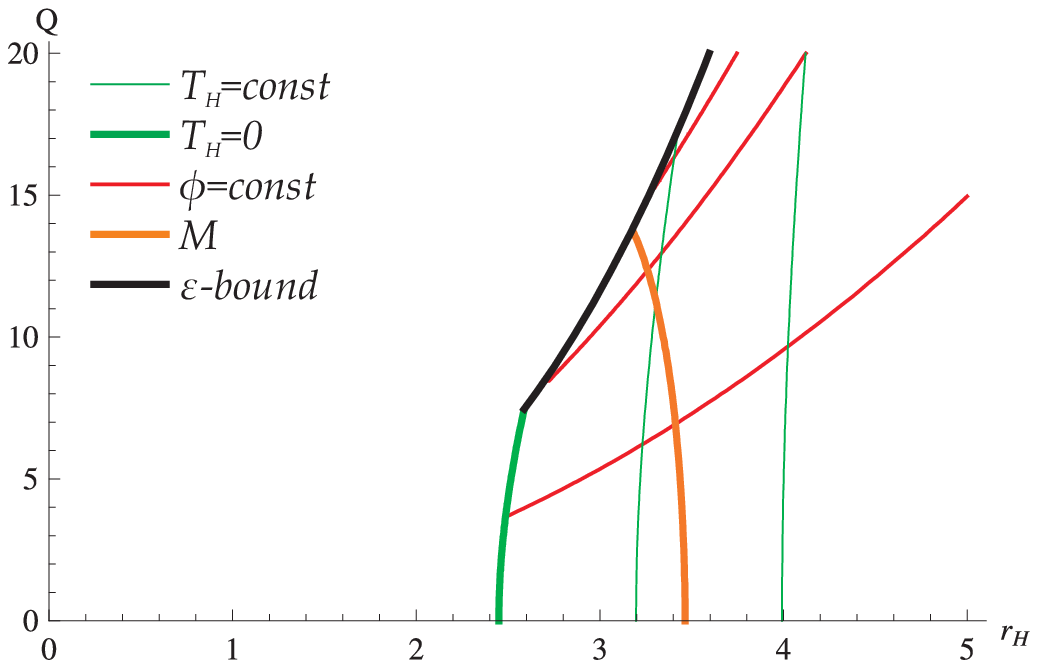} \\
\text{(a)}&\text{(b)}
\end{array}$
\end{center}
\caption{Thermal structure for $k<0$, $d\ge3$ and (a) $\varepsilon^-_1 < \varepsilon < \varepsilon^-_2$, (b) $\varepsilon < \varepsilon^-_1$. }
\label{fig:em003km2}
\end{figure}

The case for $\bold{\varepsilon}>0$, with $d \ge 3$, shown in figure \ref{fig:em01km2} constitutes of a single black hole solution which may be extremal, that is always globally and locally stable and never undergoes a Hawking-Page transition. A similar situation holds for $\bold{\varepsilon}<0$ displayed in figures \ref{fig:em003km2}(a) and (b), except that there is a further constraint due to the bound (\ref{eq:epsbound}) that eliminates extremal solutions for large enough $\phi^\varepsilon$. Finally, for the $k=0$ case there is as usual no rich thermodynamic structure.

\section{Global Thermal Phase Structure for $\varepsilon \neq 0$ \& $c=0$}
Let us summarize the results for $\varepsilon \neq 0$ and $c=0$ with the global thermal phase diagrams in the $\varepsilon-(\phi^\varepsilon)^{-2}$ plane.
\subsection{$k>0$}
\begin{figure}[h]
\begin{center}
\includegraphics[angle=0,width=0.8\textwidth]{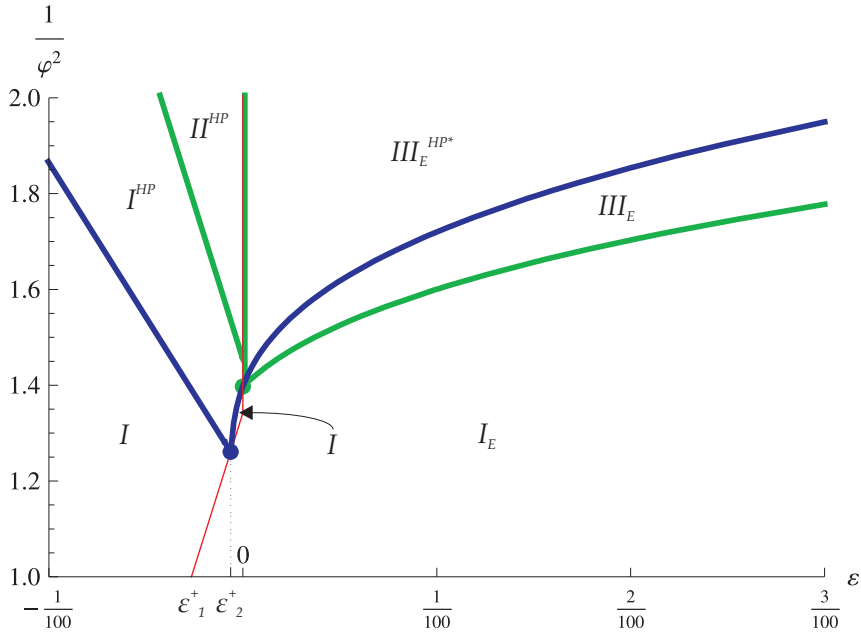}
\end{center}
\caption{Thermal phase structure for $k>0$.}
\label{fig:ek2}
\end{figure}

We begin by examining the case with $k>0$ depicted in figure \ref{fig:ek2}. Notice that we have expressed the thermal phase structure in the $\varepsilon-(\phi^\varepsilon)^{-2}$ plane, so that large potentials are in the lower part of the diagram. The phase diagram is separated into three main regions with one or three black hole solutions and denoted by $I$, $II$ and $III$ respectively. The expressions for the curves bounding the various regions are provided in appendix \ref{sec:figs}. Of particular interest is region $III$ where we encounter the possibility of three globally favored black holes leading to phase transitions between the least thermally favored to the most thermally favored. This situation is qualitatively the same as the one we studied in the $c<0$, $\varepsilon=0$, $k>0$ case for $d>4$ spatial dimensions. However, it is a new phenomenon in $d=4$ spatial dimensions, since in the $c>0$, $\varepsilon=0$, $k>0$ case the smallest black hole always has a higher free energy than the reference background.

\subsection{$k<0$}
\begin{figure}[h]
\begin{center}
\includegraphics[angle=0,width=0.8\textwidth]{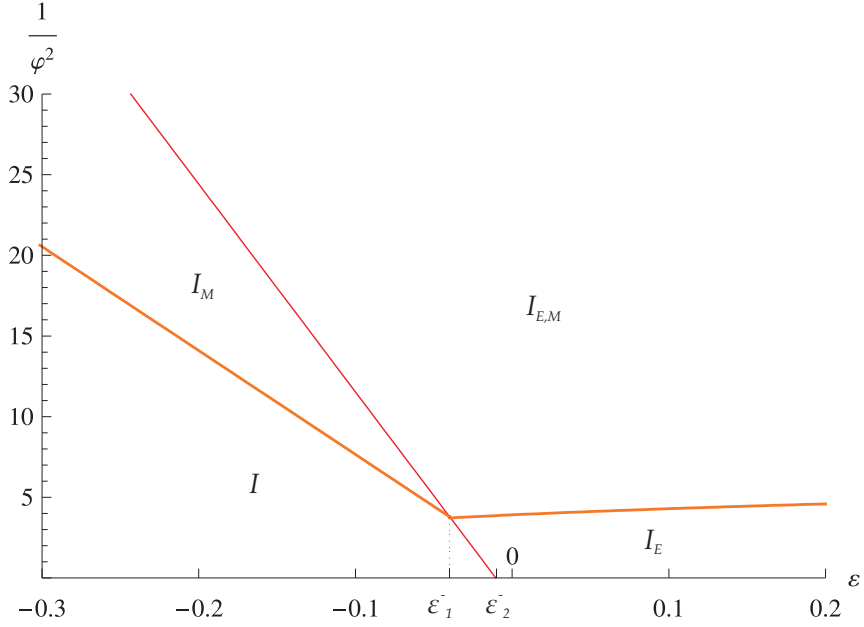}
\end{center}
\caption{Thermal phase structure for $k<0$.}
\label{fig:ekm2}
\end{figure}

The case with $k<0$ given in figure \ref{fig:ekm2} is much simpler. There is only one main region containing single black holes and denoted by $I$. We find subregions with extremal black holes and or massless and negative mass black holes.

\section{Metastability}
\subsection{Gauss-Bonnet}
We now discuss the various metastable configurations we have encountered throughout the analysis. We begin with the $\bold{\varepsilon = 0}$ case depicted in figure \ref{fig:msgb}. The results are expressed as constant $\phi$ slices in the $F - T_H$ plane. The locally unstable black hole is always depicted by the dashed curve connecting the small and large black hole curves. The $d=4$ case, shown in figure \ref{fig:msgb}(a), clearly shows that one of the black holes is always globally unstable and corresponds to region $III_{E^0}^{HP}$ in figure \ref{fig:k2}(a). Thus the metastable black hole has two decay channels; the reference background or the thermally favored black hole. Figures \ref{fig:msgb}(b), (c) and (d) are the various situations we encounter in the $d>4$ case. The black holes have negative entropy to the left of the black dot. In figure \ref{fig:msgb}(b) we find configurations with two decay channels as well as a single decay channel corresponding to regions $III^{HP^*}_{E,S}$ and $III^{HP^*}_{E,S,S^*}$ of figure \ref{fig:k2}(b). In figure \ref{fig:msgb} (c) and (d) we have a single decay channel since the metastable black hole has lower free energy than the reference background. Thus these metastable configurations are found in regions $III_{E,S}$ and $III_{E,S,S^*}$ of figure \ref{fig:k2}(b).
\begin{figure}[h]
\begin{center}
$\begin{array}{c@{\hspace{0.1in}}c}
\includegraphics[angle=0,width=0.5\textwidth]{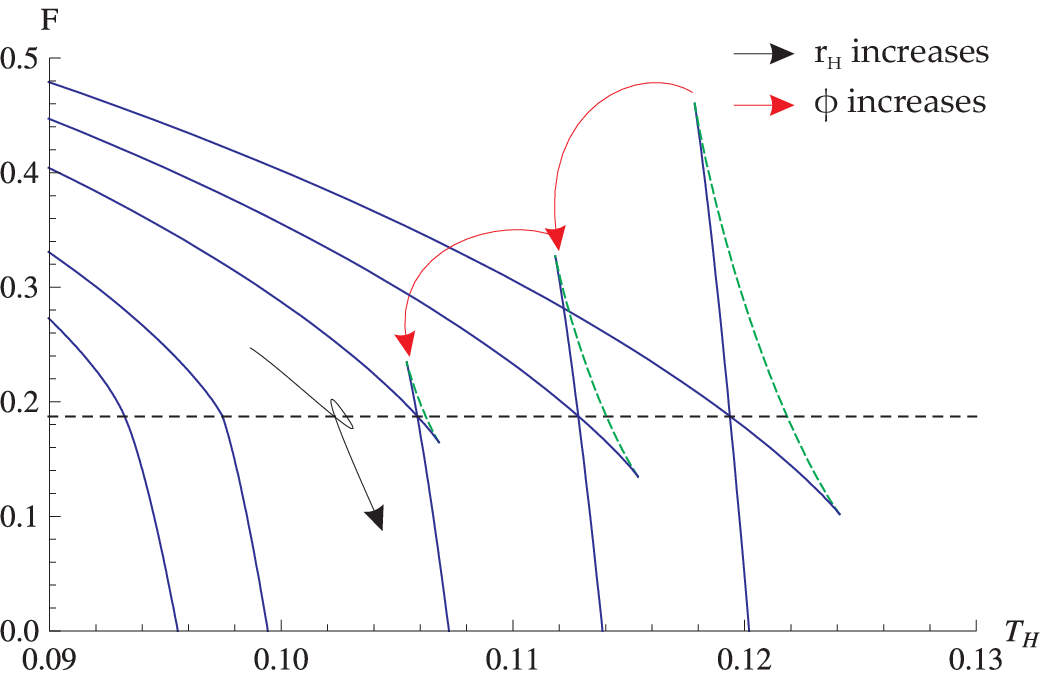} &
\includegraphics[angle=0,width=0.5\textwidth]{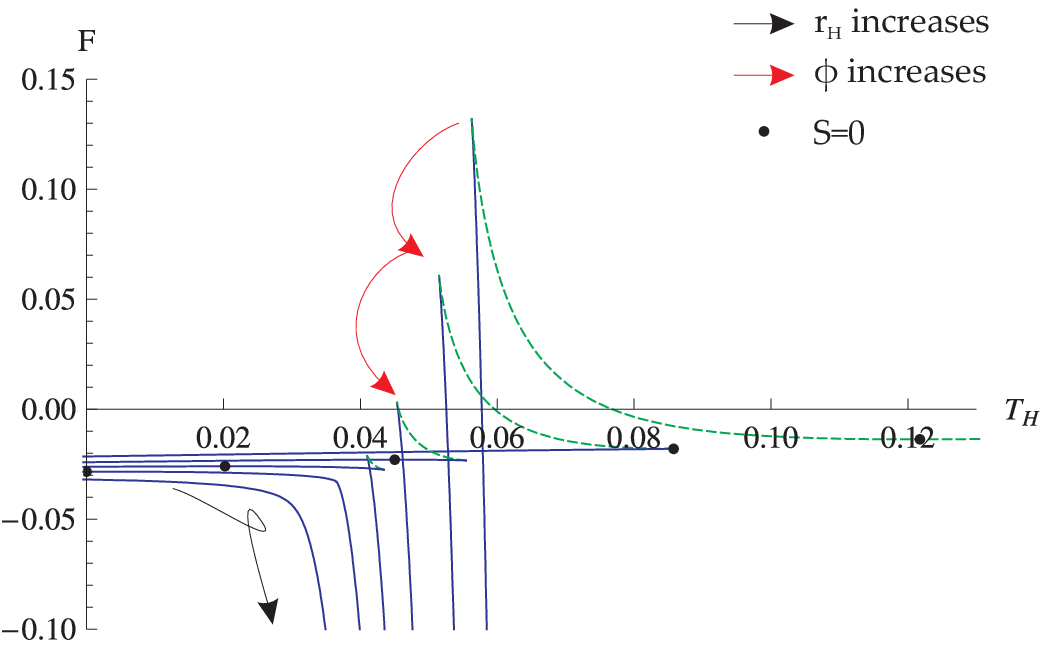}\\
\text{(a)} & \text{(b)}
\end{array}$
\end{center}
\label{fig:msgb}
\end{figure}
\begin{figure}[h]
\begin{center}
$\begin{array}{c@{\hspace{0.1in}}c}
\includegraphics[angle=0,width=0.5\textwidth]{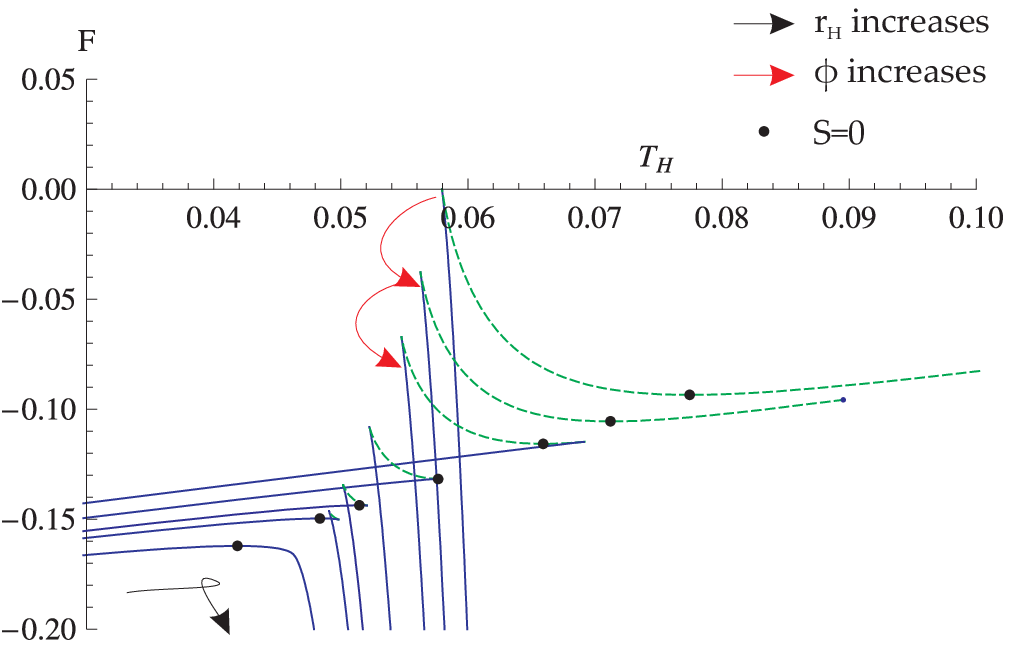} &
\includegraphics[angle=0,width=0.5\textwidth]{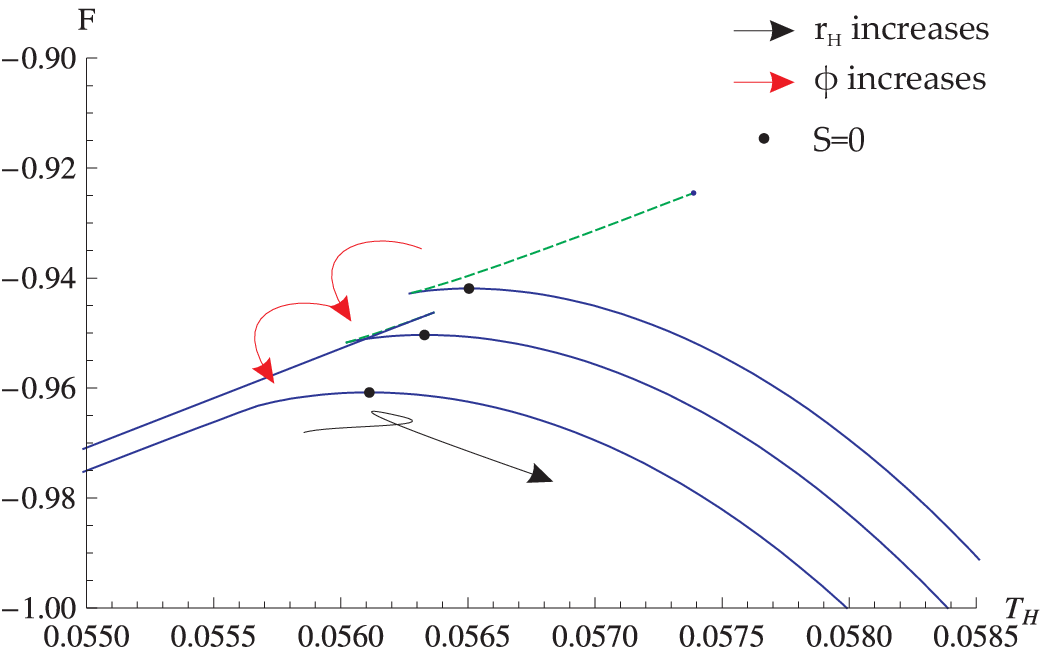}\\
\text{(c)} & \text{(d)}
\end{array}$
\end{center}
\caption{Metastability for $c\neq 0$, $\varepsilon = 0$ with (a) $d=4$ and (b), (c) and (d) $d>4$.}
\label{fig:msgb}
\end{figure}
\subsection{$F^4$}
\begin{figure}[h]
\begin{center}
\includegraphics[angle=0,width=0.5\textwidth]{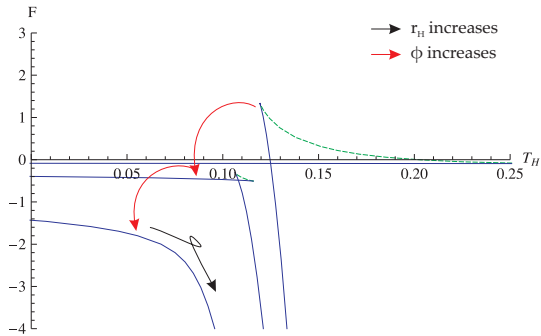}
\end{center}
\caption{Metastability for $\varepsilon \neq 0$, $c = 0$. }
\label{fig:mse}
\end{figure}

The $\bold{c=0}$ situation is depicted in figure \ref{fig:mse}. We find configurations where the metastable black hole can decay to either the reference background or the favored black hole corresponding to region $III^{HP^*}_E$ in figure \ref{fig:ek2}, or configurations where the metastable black hole has a single decay channel to the globally favored black hole corresponding to region $III_E$ in figure \ref{fig:ek2}.
\section{$F^4$ Deformations of Gauss-Bonnet Thermodynamics}
\label{sec:deformations}
Our remaining task is to explore how the interesting phenomena we found in the thermal phase structure of the Gauss-Bonnet black hole are deformed by turning on the $\varepsilon$ coefficient. Our analysis is by no means exhaustive of all possible cases; however, various new phenomena are uncovered when the $\varepsilon$ coupling is of comparable order to the $c$ coupling.

\subsection{$d=4$ Spatial Dimensions with $\varepsilon>0$ and $c>0$}
We begin with the $\bold{c,\varepsilon>0}$ case with four spatial dimensions. Upon deforming the regions of $c$ with three black hole solutions by turning on the $\varepsilon$, we find that temperatures for which the two locally stable modes can in fact become globally stable.
\begin{figure}[h]
\begin{center}
$\begin{array}{c@{\hspace{0.1in}}c}
\includegraphics[angle=0,width=0.5\textwidth]{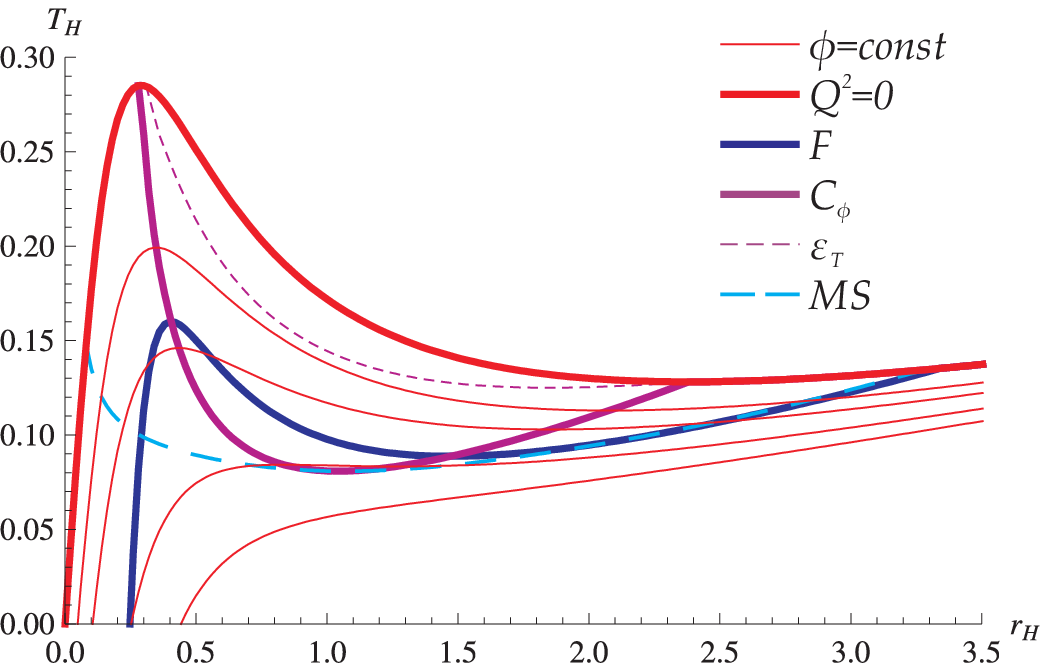} &
\includegraphics[angle=0,width=0.5\textwidth]{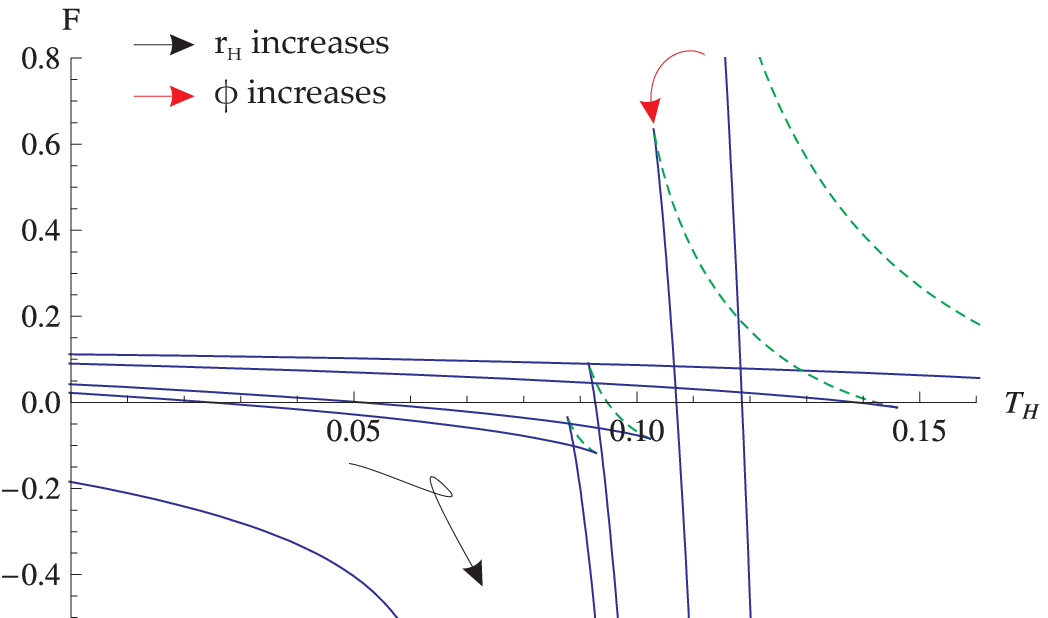} \\
\text{(a)}&\text{(b)}
\end{array}$
\end{center}
\caption{Thermal structure for $k=2$, $d=4$ and $c=\frac{1}{50\kappa^4|\Lambda|}$, $\varepsilon=\frac{1}{5g^4|\Lambda|}$ (a) and the relevant metastability curves (b).}
\label{fig:cposeposrt}
\end{figure}
This situation differs from the $\varepsilon=0$ case where the smaller locally stable black hole was always thermally unfavored with respect to the thermal anti-de Sitter vacuum.

We also find constant $\phi$ slices that intersect the global stability curve at three points. Of the three intersection points, one is always within the region of local instability. Interestingly we find that for the other two points the Hawking-Page transition occurs in a physical region of parameter space. This differs from our previous encounter of such multiple Hawking-Page transitions, which always occurred in unphysical regions. We depict the situation in figures \ref{fig:cposeposrt}(a) and (b). Notice that we if we begin in a globally stable black hole configuration and continuously decrease the temperature the system will only ever experience a single Hawking-Page transition, since the two regions where Hawking-Page transitions are available are separated by a region of local instability. The constant $\phi$ slices in the $F-T_H$ plane are observed in figure \ref{fig:cposeposrt} it is clear that for some slices both the large and small black hole can experience a Hawking-Page transition.

\subsection{$d>4$ Spatial Dimensions with $\varepsilon>0$ and $c>0$}
\begin{figure}[h]
\begin{center}
\includegraphics[angle=0,width=0.5\textwidth]{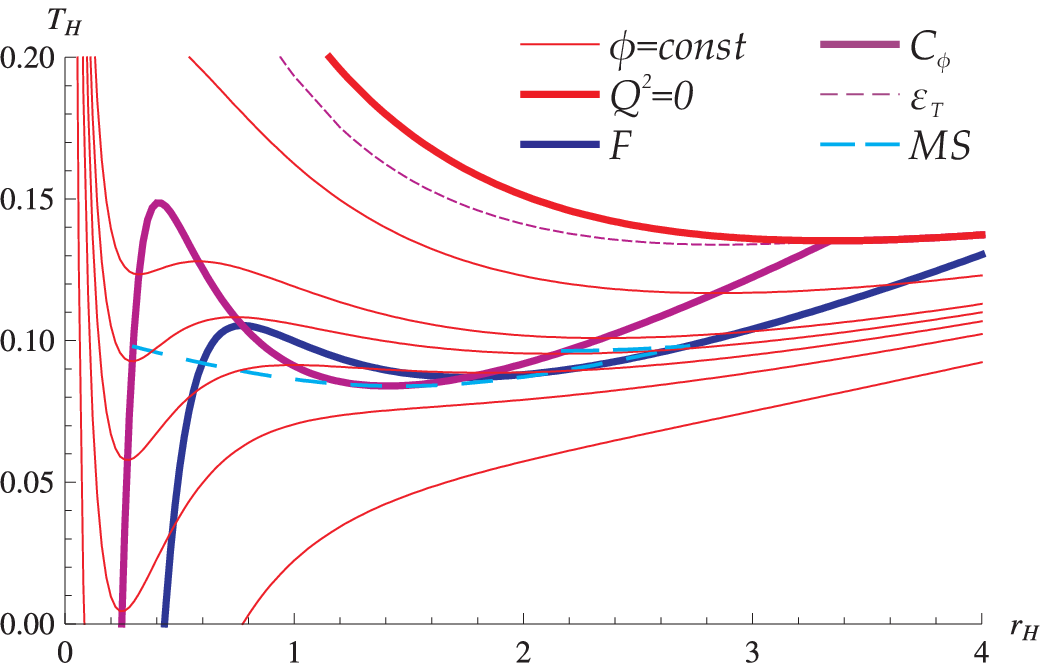}
\end{center}
\caption{Thermal structure for $k=3$, $d=5$ and $c=\frac{1}{50\kappa^4|\Lambda|}$, $\varepsilon=\frac{1}{10 g^4|\Lambda|}$.}
\label{fig:cposeposd5}
\end{figure}

For the $c>0$, $\varepsilon>0$ case with five or more spatial dimensions, depicted in figure \ref{fig:cposeposd5}, we find temperatures intersecting the constant $\phi$ slices four times, corresponding to four black hole solutions at a given temperature. Though novel, the smallest black hole has negative specific heat in accordance to our general argumentation.
Furthermore, constant $\phi$ slices that have two regions with physical Hawking-Page transitions. These regions are separated by a region of local instability so one cannot have two consecutive phase transitions as we continuously lower the temperature from the globally favored large black hole configuration. We see that phase transitions at constant temperatures between two globally favored black hole states occur for $c>0$ in $d\ge5$ dimensions with positive $\varepsilon$ switched on. Such phase transitions were only visible for $c<0$ in the pure Gauss-Bonnet case.

\subsection{$d = 4$ Spatial Dimensions with $\varepsilon>0$ and $c<0$}
\begin{figure}[h]
\begin{center}
\includegraphics[angle=0,width=0.5\textwidth]{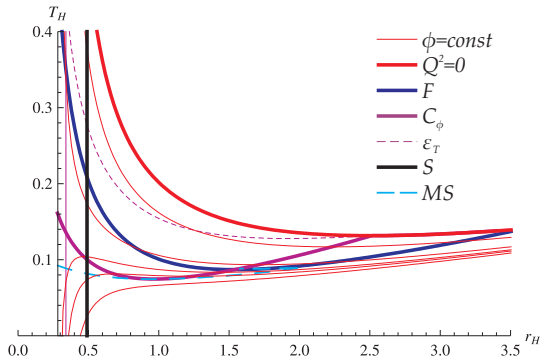}
\end{center}
\caption{Thermal structure for $k=2$, $d=4$ and $c=-\frac{1}{50\kappa^4|\Lambda|}$, $\varepsilon=\frac{1}{5g^4|\Lambda|}$.}
\label{fig:cnegepos}
\end{figure}

The $c<0$, $\varepsilon>0$ situation for $d = 4$ spatial dimensions is given in figure \ref{fig:cnegepos}. The situation is qualitatively identical to the $c<0$ case with $d\ge5$ uncovered in the pure Gauss-Bonnet theory and shown in figures \ref{fig:k3cm1-50}(a) and (b). There are constant $\phi$ slices with three black hole solutions at a given temperature of which the smaller and larger may be globally favored over the thermal anti-de Sitter vacuum. We simply note that this situation had not been observed for the pure Gauss-Bonnet case with $c<0$ and $d=4$. No secondary Hawking-Page transitions are observed in this case.
\section{Discussion and Summary of Results}
\begin{figure}[h]
\begin{center}
\includegraphics[angle=0,width=0.6\textwidth]{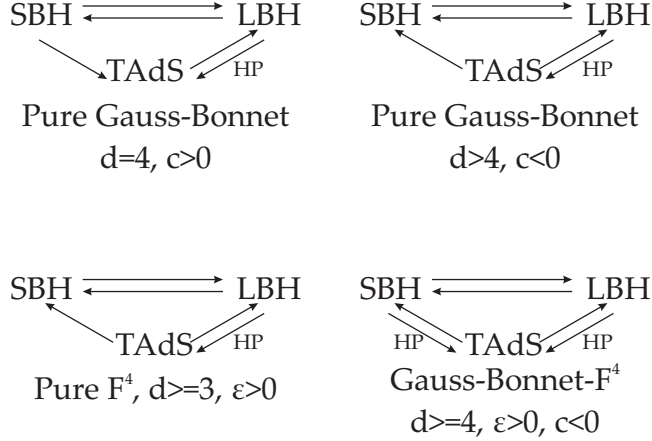}
\end{center}
\caption{All possible physical phase transitions in the Gauss-Bonnet-$F^4$ theory. SBH denotes the small locally stable black hole, BBH denotes the large locally stable black hole, TAdS denotes thermal anti-de Sitter space, and HP denotes an accessible Hawking-Page transition.}
\label{fig:transitions}
\end{figure}

We have uncovered various novel features in the thermal phase structure of both the Gauss-Bonnet black hole with no $F^4$ corrections and the Gauss-Bonnet-$F^4$ black hole. Our results are contained in the global phase structure figures \ref{fig:k2}, \ref{fig:k3} and \ref{fig:km2} for pure Gauss-Bonnet corrections, and in figures \ref{fig:ek2} and \ref{fig:ekm2} for pure $F^4$ corrections. In section \ref{sec:deformations} we discus the case with both couplings turned on. We have mapped out various regions of physical interest within the phase diagrams, such as regions containing extremal black holes, regions including massless and negative mass modes, and regions with negative entropy modes. We depict all possible phase transitions between locally stable  black holes and/or thermal anti-de Sitter space in figure \ref{fig:transitions}.
\newline
\newline
For the \emph{pure Gauss-Bonnet} case we have discovered new metastable modes for $k>0$ and $c<0$ in $d \ge 5$ spatial dimensions, as depicted in figures \ref{fig:msgb}(b), (c) and (d). More specifically, we find temperatures at a constant electric potential with three black hole solutions, of which the intermediate sized one is locally unstable. The smaller black hole is an $\mathcal{O}(\alpha')$ effect and thus its size is also of order $\sqrt{\alpha'}$. The larger black hole has a size of the order of the anti-de Sitter length. These modes have the interesting property that for certain values of the electric potential, they may both be thermally favored over the thermal anti-de Sitter vacuum at a fixed temperature. This differs from the situation in $d=4$ spacetime dimensions \cite{Dey:2007vt}, shown in figure \ref{fig:msgb}(a), where one of the metastable modes is always thermally unfavored with respect to the thermal anti-de Sitter vacuum. It may be somewhat alarming that the effect occurs for $c<0$; however, from a purely effective field theory point of view, we must take these effects seriously, unless there is evidence that this is an unphysical region of parameter space (see \cite{Hirayama:2008pf,Myung:2008af} for a relevant discussion for the neutral case). It would be very interesting to understand the dual field theory interpretation of these metastable modes given that the phase transition occurs between two states in the deconfined phase.

Another curious feature we uncover is the existence of constant electric potential slices that intersect the free energy curve more than once. Such a scenario would imply that as one lowers the temperature, the large hot black hole decays into thermal anti-de Sitter space, and the thermal anti-de Sitter space itself eventually collapses into a black hole at even lower temperatures. This, however, is not the case, since the secondary Hawking-Page transitions always occur in either a region of local instability or a region of negative entropy, and are thus unphysical.
\newline
\newline
For the \emph{pure $F^4$} case, which has not been previously explored, we uncover additional metastable modes for $d\ge3$, $k>0$ and $\varepsilon>0$. Once again, they consist of three black hole solutions, of which the intermediate sized one is locally unstable. Furthermore, we find that for certain values of the potential, both locally stable modes are thermally favored over the thermal anti-de Sitter at a fixed temperature, as depicted in figure \ref{fig:mse}, and thus they are both states in the deconfined phase of the dual theory. This feature persists for $d=3$ spatial dimensions when there is no Gauss-Bonnet correction present. Once again, any secondary Hawking-Page transitions observed occur in unphysical regions of parameter space.
\newline
\newline
Finally, we study the thermodynamics of our theory with \emph{both couplings} turned on in section \ref{sec:deformations}. For $d=4$ and both couplings positive, we uncover constant $\phi$ slices allowing for three black holes at a constant temperature, of which the smallest and largest may be globally favored. Furthermore, the secondary Hawking-Page transition becomes physical; thus, for a slice of constant $\phi$ there are two disjoint Hawking-Page transitions that can occur - starting from either the small or large black hole.

For $d\ge 5$ there are temperatures with four black hole solutions for a constant $\phi$, indicating the existence of a fourth locally unstable black. Furthermore, the secondary Hawking-Page transition for the small black hole is still present. It is of great interest to find a clear interpretation of the additional Hawking-Page transition from the point of view of the dual theory. Naively, it would seem that there are two disconnected deconfined phases at different temperatures that can experience a Hawking-Page transition to a confined phase as we continuously lower the temperature. However, such phenomena deserve a complete treatment in their own right.
\newline
\newline
We reiterate that a qualitative change in the tree level thermal phase structure arising from $\mathcal{O}(\alpha') $ corrections may imply that we must take into account all higher order corrections to get the correct physical picture. However, it has been possible to match the Gauss-Bonnet effects in $d=4$ with the thermal phase structure of a phenomenological dual matrix model \cite{Dey:2006ds,Dey:2007vt} with no $F^4$
corrections. This observation leads us to expect that a similar situation occurs for the thermal phase structure we have uncovered. It would be worth further studying this issue by obtaining the thermal phase structure of effective gravitational actions with even higher order corrections, even though it may be a rather challenging task.
\newline
\newline
One important question that we have left for future work is to study these effects from the point of view of the dual theory. Of course, this theory will be strongly coupled, and for general dimensions it is not clear what the precise dual theory even is. We can, however, construct phenomenological dual matrix models along the lines of \cite{Liu:2004vy,AlvarezGaume:2005fv,Dey:2006ds,Dey:2007vt} and compare the various phase structures of the bulk theories that we have uncovered with those of the dual matrix models. It is also of interest to perform our analysis in the canonical ensemble where the charge, rather than the electric potential, is held fixed. In this case, the reference background will no longer be thermal anti-de Sitter space, but an extremal black hole. Since we have uncovered new black hole solutions, it is important to study their dynamical stability with respect to small perturbations and see whether the regions of negative specific heat and dynamical instability coincide \cite{Hirayama:2008pf,Prestidge:1999uq,Reall:2001ag,Gubser:2000mm,Konoplya:2008ix}. Finally, there has been a recent interest in higher derivative corrections to the shear viscosity bound and the Gauss-Bonnet-$F^4$ black hole, and further higher derivative corrections thereof (particularly $R F^2$ terms) may be of interest to this story \cite{Policastro:2001yc,Kovtun:2003wp,Kats:2007mq,Brigante:2008gz}.

\section*{Acknowledgements}
We wish to thank S. Lahiri, K. Papadodimas and P. Petrov for useful discussions. D.A.'s work was partially funded by the DOE grant DE-FG02-91ER40654.

\appendix
\setcounter{equation}{0} 
\section{Equations of Motion} 
\label{section:appA}

\subsection*{Gauss-Bonnet Gravity}

The general action of $R^2$ gravity with matter is given by:
\begin{equation}
I = \int {d^{d + 1} x\sqrt { - g} \left( {\frac{1}{{\kappa ^2 }}R - \Lambda + aR^2 + bR_{\mu \nu } R^{\mu \nu } + cR_{\mu \nu \rho \sigma } R^{\mu \nu \rho \sigma } } \right)} + I_{matter}.
\end{equation}
We are particularly interested in the Gauss-Bonnet combination,
\begin{equation}
a = c,\quad b = - 4c.
\end{equation}
The equations of motion for the Gauss-Bonnet combination are given by:
\begin{multline}
T_{\mu \nu } = - \frac{1}{2}g_{\mu \nu } \left( {\frac{1}{{\kappa ^2 }}R - \Lambda + c \left( R^2 -4 R_{\mu \nu } R^{\mu \nu } + R_{\mu \nu \rho \sigma } R^{\mu \nu \rho \sigma } \right) } \right) + \frac{1}{{\kappa ^2 }}R_{\mu \nu } \\
+ c \left( { 2RR_{\mu \nu } - 4R_{\mu \rho } R_\nu ^\rho - 4R_{\mu \rho \nu \sigma } R^{\rho \sigma } + 2R_\mu ^{\lambda \rho \sigma } R_{\nu \lambda \rho \sigma } } \right).
\end{multline}

\section{Useful Intermediate Results}
\label{sec:intermediate}

In order to make life simpler for one to verify the equations of motion with our ansatz, we give some of the intermediate steps of the calculation.

\subsection*{Christoffel Symbols}
\label{subsec:christoffel}

The nonzero Christoffel symbols obtained from our ansatz are:
\be
\begin{aligned}
\Gamma _{rt}^t &= \nu ',&\quad\quad
\Gamma _{tt}^r &= - \nu 'e^{2\left( {\nu - \lambda } \right)}, \\
\Gamma _{rr}^r &= \lambda ', &\quad\quad
\Gamma _{ij}^r &= - re^{ - 2\lambda } \tilde g_{ij}, \\
\Gamma _{jr}^i &= \frac{{\delta _j^i }}{r}, &\quad\quad
\Gamma _{jk}^i &= \tilde \Gamma _{jk}^i.
\end{aligned}
\ee
\subsection*{Curvature Tensors}
The nonzero elements of the Riemann tensor:
\be
\begin{aligned}
R_{trtr} &= - e^{2\nu } R^t _{rtr} = e^{2\nu } \left( {\nu '' - \lambda '\nu ' + \nu '^2 } \right) \\
R_{titj} &= - e^{2\nu } R^t _{itj} = r\nu 'e^{2\left( {\nu - \lambda } \right)} \tilde g_{ij} \\
R_{rirj} &= r\lambda '\tilde g_{ij} \\
R_{ijkl} &= r^2 \tilde R_{ijkl} - r^2 e^{ - 2\lambda } \left( {\tilde g_{ik} \tilde g_{jl} - \tilde g_{il} \tilde g_{jk} } \right).
\end{aligned}
\ee
The elements of the Ricci tensor are:
\be
\begin{aligned}
R_{rr} &= - \left( {\nu '' - \lambda '\nu ' + \nu '^2 } \right) + \frac{{\left( {d - 1} \right)\lambda '}}{r} \\
R_{tt} &= e^{2\left( {\nu - \lambda } \right)} \left( {\nu '' - \lambda '\nu ' + \nu '^2 + \frac{{\left( {d - 1} \right)\nu '}}{r}} \right) \\
R_{ij} &= \left\{ {k - e^{ - 2\lambda } \left[ {r\left( {\nu ' - \lambda '} \right) + d - 2} \right]} \right\}\tilde g_{ij}.
\end{aligned}
\ee
Finally, the Ricci scalar is given by:
\begin{multline}
R = - e^{ - 2\lambda } \left( {2\left( {\nu '' - \lambda '\nu ' + \nu '^2 } \right) + \frac{{2\left( {d - 1} \right)\left( {\nu ' - \lambda '} \right)}}{r} + \frac{{\left( {d - 1} \right)\left( {d - 2} \right)}}{{r^2 }}} \right) \\
+ \frac{{\left( {d - 1} \right)k}}{{r^2 }}.
\end{multline}

\subsection*{Gauss-Bonnet terms}
We list various relevant contractions for the various derivations.\newline
Quantities in the Action:
\begin{multline}
R_{\mu \nu } R^{\mu \nu }
= e^{ - 4\lambda } \Biggl[ {2\left( {\nu '' - \lambda '\nu ' + \nu '^2 } \right)^2 + 2\left( {d - 1} \right)\left( {\nu '' - \lambda '\nu ' + \nu '^2 } \right)\frac{{\left( {\nu ' - \lambda '} \right)}}{r} } \\
{+ \frac{{\left( {d - 1} \right)^2 \left( {\nu '^2 + \lambda '^2 } \right)}}{{r^2 }}} \Biggr]
+ \frac{\left( {d - 1} \right)}{{r^4 }}\left( {k - e^{ - 2\lambda } \left[ {r\left( {\nu ' - \lambda '} \right) + \left(d - 2\right)} \right]} \right)^2
\end{multline}
\begin{multline}
R_{\mu \nu \lambda \rho} R^{\mu \nu \lambda \rho}
= 4e^{ - 4\lambda } \left( {\nu '' - \lambda '\nu ' + \nu '^2 } \right)^2 + 4\left( {d - 1} \right)e^{ - 4\lambda } \frac{{\lambda '^2 + \nu '^2 }}{{r^2 }} \\
+ \frac{2\left( {d - 1} \right)\left( {d - 2} \right)}{{r^4 }}\left( {\frac{k}{{d - 2}} - e^{ - 2\lambda } } \right)^2
\end{multline}
\begin{multline}
R^2 - 4R_{\mu \nu } R^{\mu \nu } + R_{\mu \nu \lambda \rho} R^{\mu \nu \lambda \rho}
= \left( {d - 1} \right) \Biggl\{ \frac{{\left( {d - 3} \right)\left( {d - 4} \right)k^2 }}{{\left( {d - 2} \right)r^4 }} \\
+ e^{ - 2\lambda } \left[ { - 4\frac{{k\left( {\nu '' - \lambda '\nu ' + \nu '^2 } \right)}}{{r^2 }} - 4\frac{{\left( {d - 3} \right)k\left( {\nu ' - \lambda '} \right)}}{{r^3 }} - 2\frac{{\left( {d - 3} \right)\left( {d - 4} \right)k}}{{r^4 }}} \right] \\
+ \left( {d - 2} \right) e^{ - 4\lambda } \left[ {\frac{{4\left( {\nu '' + \nu '^2 - 3\lambda '\nu '} \right)}}{{r^2 }} } { + 4\frac{{\left( {d - 3} \right)\left( {\nu ' - \lambda '} \right)}}{{r^3 }} + \frac{{\left( {d - 3} \right)\left( {d - 4} \right)}}{{r^4 }}} \right] \Biggr\}.
\end{multline}
${R^\mu} _\rho R^{\nu \rho }$ non-vanishing terms:
\begin{equation}
{R^t} _\rho R^{t\rho } = - e^{ - 2\nu - 4\lambda } \left( {\nu '' - \lambda '\nu ' + \nu '^2 + \frac{{\left( {d - 1} \right)\nu '}}{r}} \right)^2
\end{equation}
\begin{equation}
{R^r} _\rho R^{r\rho } = e^{ - 6\lambda } \left( { - \left( {\nu '' - \lambda '\nu ' + \nu '^2 } \right) + \frac{{\left( {d - 1} \right)\lambda '}}{r}} \right)^2
\end{equation}
\begin{multline}
{R^i} _\rho R^{j\rho }
= \frac{1}{{r^6 }}\left[ {k^2 - 2e^{ - 2\lambda } k\left( {r\left( {\nu ' - \lambda '} \right) + \left( {d - 2} \right)} \right)} \right. \\
\left. { + e^{ - 4\lambda } \left( {r^2 \left( {\nu ' - \lambda '} \right)^2 + 2\left( {d - 2} \right)r\left( {\nu ' - \lambda '} \right) + \left( {d - 2} \right)^2 } \right)} \right]\tilde g^{ij}.
\end{multline}
$R^{\mu \rho \nu \sigma} R_{\rho \sigma }$ non-vanishing terms:
\begin{multline}
R^{t\mu t\nu } R_{\mu \nu }
= \left( {d - 1} \right) e^{ - 2\nu - 4\lambda } \Biggl[ { - \frac{\left( {\nu '' - \lambda '\nu ' + \nu '^2 } \right)^2}{\left( {d - 1} \right)} + \frac{{\lambda '\left( {\nu '' - \lambda '\nu ' + \nu '^2 } \right)}}{r} } \\
{- \frac{{\nu '\left( {\nu ' - \lambda '} \right)}}{{r^2 }} - \frac{{\left( {d - 2} \right)\nu '}}{{r^3 }}} + e^{2\lambda } \frac{{k\nu '}}{{r^3 }} \Biggr]
\end{multline}
\begin{multline}
R^{r\mu r\nu } R_{\mu \nu }
= \left( {d - 1} \right) e^{ - 6\lambda } \Biggl[ {\frac{\left( {\nu '' - \lambda '\nu ' + \nu '^2 } \right)^2}{\left( {d - 1} \right)} + \frac{{\nu '\left( {\nu '' - \lambda '\nu ' + \nu '^2 } \right)}}{r} } \\
  {- \frac{{\lambda '\left( {\nu ' - \lambda '} \right)}}{{r^2 }} - \frac{{\left( {d - 2} \right)\lambda '}}{{r^3 }}} + e^{ 2\lambda } \frac{{k\lambda '}}{{r^3 }} \Biggr]
\end{multline}
\begin{multline}
R^{i\mu j\nu } R_{\mu \nu }
= \frac{{e^{ - 4\lambda } }}{{r^3 }}\left[ {\left( {\nu '' - \lambda '\nu ' + \nu '^2 } \right)\left( {\nu ' - \lambda '} \right) + \frac{{\left( {d - 1} \right)\left( {\nu '^2 + \lambda '^2 } \right)}}{r} }\right. \\ \left. {+ \frac{{\left( {d - 2} \right)\left( {\nu ' - \lambda '} \right)}}{{r^2 }} + \frac{{\left( {d - 2} \right)^2 }}{{r^3 }}} \right]\tilde g^{ij}
- \frac{{e^{ - 2\lambda } }}{{r^6 }}k\left( {r\left( {\nu ' - \lambda '} \right) + 2\left( {d - 2} \right)} \right)\tilde g^{ij} + \frac{{k^2 }}{{r^6 }}\tilde g^{ij}.
\end{multline}
${R^\mu} _{\kappa \lambda \rho } R^{\nu \kappa \lambda \rho }$ non-vanishing terms:
\begin{equation}
{R^t} _{\mu \nu \rho } R^{t\mu \nu \rho }
= - 2e^{ - 2\nu - 4\lambda } \left( {\left( {\nu '' - \lambda '\nu ' + \nu '^2 } \right)^2 + \frac{{\left( {d - 1} \right)\nu '^2 }}{{r^2 }}} \right) \\
\end{equation}
\begin{equation}
{R^r} _{\mu \nu \rho } R^{r\mu \nu \rho }
= 2e^{ - 6\lambda } \left( {\left( {\nu '' - \lambda '\nu ' + \nu '^2 } \right)^2 + \frac{{\left( {d - 1} \right)\lambda '^2 }}{{r^2 }}} \right) \\
\end{equation}
\begin{equation}
{R^i} _{\mu \nu \rho } R^{j\mu \nu \rho }
= \frac{{2e^{ - 4\lambda } \left( {\nu '^2 + \lambda '^2 } \right)}}{{r^4 }}\tilde g^{ij} + \frac{2}{{r^6 }}\left( {\frac{k}{{d - 2}} - e^{ - 2\lambda } } \right)^2 \left( {d - 2} \right)\tilde g^{ij}.
\end{equation}

\subsection*{Electromagnetic Intermediate Quantities}
\label{subsec:em}
We give some intermediate steps in the calculation of the matter energy tensor and the solution of the Maxwell equations:
\be
\begin{aligned}
F_{\mu \nu } F^{\mu \nu } &=& - 2f\left( r \right)^2 \\
F^{\mu \nu } F_{\nu \lambda } F^{\lambda \rho } F_{\rho \mu } &=& 2f\left( r \right)^4
\end{aligned}
\ee
\be
\begin{aligned}
{F_\mu} ^t F^{\mu t} &=& e^{ - 2\nu } f\left( r \right)^2 \\
{F_\mu} ^r F^{\mu r} &=& - e^{ - 2\lambda } f\left( r \right)^2
\end{aligned}
\ee
\be
\begin{aligned}
F^{t\rho } F_{\rho \sigma } F^{\sigma \tau } {F_\tau} ^t &= - e^{ - 2\nu } f\left( r \right)^4 \\
F^{t\rho } F_{\rho \sigma } F^{\sigma \tau } {F_\tau} ^t &= e^{ - 2\lambda } f\left( r \right)^4
\end{aligned}
\ee
\be
\begin{aligned}
F^{t\mu } F_{\mu \nu } F^{\nu r} = e^{ - \left( {\nu + \lambda } \right)} f\left( r \right)^3
\end{aligned}
\ee
All other relevant terms are vanishing.

\section{Thermodynamic Derivatives}
\label{sec:thder}
In this appendix we present the thermodynamic derivatives that allow us to compute the thermodynamic charge, entropy and mass used throughout the paper. Let us begin with the expressions in the $r_H$, $Q$ parametrization. For the grand canonical ensemble we used that:
\begin{eqnarray}
\mathcal{Q} &=& - \frac{1}{\beta }\left( {\frac{{\left( {\frac{{\partial
I}}{{\partial r_H }}} \right)_Q }}{{\left( {\frac{{\partial \phi }}{{\partial
r_H }}} \right)_Q - \left( {\frac{{\partial \phi }}{{\partial Q}}}
\right)_{r_H } \frac{{\left( {\frac{{\partial \beta }}{{\partial r_H }}}
\right)_Q }}{{\left( {\frac{{\partial \beta }}{{\partial Q}}} \right)_{r_H }
}}}} + \frac{{\left( {\frac{{\partial I}}{{\partial Q}}} \right)_{r_H }
}}{{\left( {\frac{{\partial \phi }}{{\partial Q}}} \right)_{r_H } - \left(
{\frac{{\partial \phi }}{{\partial r_H }}} \right)_Q \frac{{\left(
{\frac{{\partial \beta }}{{\partial Q}}} \right)_{r_H } }}{{\left(
{\frac{{\partial \beta }}{{\partial r_H }}} \right)_Q }}}}} \right),\\
S &=& \beta \left( {\frac{{\left( {\frac{{\partial I}}{{\partial r_H }}} \right)_Q
}}{{\left( {\frac{{\partial \beta }}{{\partial r_H }}} \right)_q - \left(
{\frac{{\partial \beta }}{{\partial Q}}} \right)_{r_H } \frac{{\left(
{\frac{{\partial \phi }}{{\partial r_H }}} \right)_Q }}{{\left(
{\frac{{\partial \phi }}{{\partial Q}}} \right)_{r_H } }}}} + \frac{{\left(
{\frac{{\partial I}}{{\partial Q}}} \right)_{r_H } }}{{\left( {\frac{{\partial
\beta }}{{\partial Q}}} \right)_{r_H } - \left( {\frac{{\partial \beta
}}{{\partial r_H }}} \right)_Q \frac{{\left( {\frac{{\partial \phi }}{{\partial
Q}}} \right)_{r_H } }}{{\left( {\frac{{\partial \phi }}{{\partial r_H }}}
\right)_Q }}}}} \right) - I,\quad\\
{\rm E} &=& \frac{{I + S}}{\beta } + \mathcal{Q}\phi.
\end{eqnarray}
The expressions for the specific heat and electrical permittivity are given by:
\begin{equation}
C_\phi = T_H \left( {\frac{{\partial S}}{{\partial T_H}}} \right)_\phi =
T_H\frac{{\frac{{dS}}{{dr_H }}}}{{\left( {\frac{{\partial T_H}}{{\partial r_H }}}
\right)_Q - \left( {\frac{{\partial T_H}}{{\partial Q}}} \right)_{r_H }
\frac{{\left( {\frac{{\partial \phi }}{{\partial r_H }}} \right)_Q }}{{\left(
{\frac{{\partial \phi }}{{\partial Q}}} \right)_{r_H } }}}}
\end{equation}
and
\be
\varepsilon _T = \left( {\frac{{\partial Q}}{{\partial \phi }}} \right)_{T_H }
= \frac{1}{{\left( {\frac{{\partial \phi }}{{\partial Q}}} \right)_{r_H } -
\left( {\frac{{\partial \phi }}{{\partial r_H }}} \right)_Q \frac{{\left(
{\frac{{\partial T_H }}{{\partial Q}}} \right)_{r_H } }}{{\left(
{\frac{{\partial T_H }}{{\partial r_H }}} \right)_Q }}}}.
\ee
The above expressions can also be obtained in the $r_H$, $T_H$ parametrization for which we have provided an thermodynamic analysis for the $\varepsilon=0$ case in appendix \ref{sec:rt}:
\begin{eqnarray}
\mathcal{Q} &=& -T_H \frac{{\left( {\frac{{\partial I}}{{\partial r_H }}} \right)_{T_H }
}}{{\left( {\frac{{\partial \phi }}{{\partial T_H }}} \right)_{r_{\rm H} } }},\\
S &=& - T_H \left( {\left( {\frac{{\partial I}}{{\partial T_H }}} \right)_{r_H } -
\left( {\frac{{\partial I}}{{\partial r_H }}} \right)_{T_H } \frac{{\left(
{\frac{{\partial \phi }}{{\partial T_H }}} \right)_{r_{\rm H} } }}{{\left(
{\frac{{\partial \phi }}{{\partial r_H }}} \right)_{T_H } }}} \right)-I,\\
{\rm E} &=& \frac{{I + S}}{\beta } + \mathcal{Q}\phi.
\end{eqnarray}
Furthermore, we provide the expressions for the specific heat and isothermal electric permittivity in the $r_H$, $T_H$ parametrization:
\begin{eqnarray}
C_\phi &=& T_H \left( {\frac{{\partial S}}{{\partial T_H }}} \right)_\phi = -
T_H \frac{{dS}}{{dr_H }}\frac{{\left( {\frac{{\partial \phi }}{{\partial T_H }}}
\right)_{r_{\rm H} } }}{{\left( {\frac{{\partial \phi }}{{\partial r_H }}}
\right)_{T_H } }},\\
\varepsilon _T &=& \left(\frac{\partial Q}{\partial \phi}\right)_{T_H} = \frac{\left(\frac{\partial Q}{\partial r_H}\right)_{T_H}}{\left(\frac{\partial \phi}{\partial r_H} \right)_{T_H}}.
\end{eqnarray}

\section{The $\varepsilon=0$ case in $r_H$, $T_H$ parametrization}
\label{sec:rt}

Even though we discussed throughout the text the practical usefulness of the $r_H$, $Q$ parametrization in analyzing the thermodynamics for non-zero $\varepsilon$ due to the highly non-trivial dependence on $Q$ in our expressions, this is not the case when $\varepsilon=0$. It turns out that the analytic expressions we can obtain in this case are much simpler in the $r_H$, $T_H$ parametrization for $\varepsilon=0$ and it is instructive to include the main results for the global and local thermodynamics. We note that the analysis in this parametrization for $\epsilon=0$ has been performed in \cite{Dey:2007vt,Cvetic:2001bk}. This provides us with a non-trivial check of our interpretations of the $\varepsilon=0$ analysis and further support for our interpretation of the $\varepsilon \neq 0$ results.

First we need to solve for $\mu$ and $Q$ in terms of $r_H$ and $T_H$. We get:
\begin{multline}
\mu = - \frac{1}{{d - 2}}\left\{ {2\frac{{\left( {d - 1} \right)k{r_H} ^{d -
2} }}{{\kappa ^2 }} + 2\frac{{\left( {d - 3} \right)Dk^2 {r_H} ^{d - 4}
}}{{\left( {d - 2} \right)^2 }} - 2\frac{\left( {d - 1} \right)\Lambda
}{d}{r_H} ^d }\right. \\ \left. {- 4\pi T_H \left[ {\frac{{\left( {d - 1}
\right){r_H} ^{d - 1} }}{{\kappa ^2 }} + \frac{{2kD{r_H} ^{d - 3} }}{{d - 2}}}
\right] } \right\},
\end{multline}
\begin{multline}
Q^2 = \frac{{2r_H ^{d - 2} }}{{g^2 }}\left\{ {\frac{{\left( {d - 1}
\right)k{r_H} ^{d - 2} }}{{\kappa ^2 }} + \frac{{\left( {d - 4} \right)Dk^2
{r_H} ^{d - 4} }}{{\left( {d - 2} \right)^2 }} - \Lambda{r_H} ^d }\right. \\
\left. {- 4\pi T_H \left[ {\frac{{\left( {d - 1} \right){r_H} ^{d - 1}
}}{{\kappa ^2 }} + \frac{{2kD{r_H} ^{d - 3} }}{{d - 2}}} \right] } \right\}.
\end{multline}
We find that the condition for a massless black hole can be written as:
\begin{equation}
T_H^{M=0} = \frac{{ - \left( {d - 2} \right)\kappa ^2 \Lambda r_H ^4 + d\left( {d -
2} \right)kr_H ^2 + d\left( {d - 3} \right)^2 ck^2\kappa ^2 }}{{2\pi d\left( {d -
2} \right)\left[ {r_H ^3 + 2\left( {d - 3} \right)ck\kappa ^2 r_H} \right]}}.
\label{eq:Tmassless}
\end{equation}
One important thing here is that we need to ensure that $Q^2$ is always
positive, so that the charge that enters our Lagrangian is real. In the $r_H$, $Q$ parametrization this was
trivial; however, now it is not. The payback in our parametrization is that the
extremality bound has become trivial. We find that $T_H$ needs to satisfy:
\begin{equation}
T_H \le \frac{{ - \left( {d - 2} \right)\kappa ^2 \Lambda r_H ^4 + \left( {d
- 1} \right)\left( {d - 2} \right)kr_H ^2 + \left( {d - 1} \right)\left( {d -
3} \right)\left( {d - 4} \right) c k^2 \kappa ^2 }}{{4\pi \left( {d - 1}
\right)\left( {d - 2} \right)\left[ {r_H ^3 + 2\left( {d - 3} \right)ck\kappa
^2 r_H} \right]}}.
\label{eq:Tchargeless}
\end{equation}
Finally, we substitute $\mu$ and $Q$ in the free energy to find:
\begin{multline}
F = \frac{2}{{(d - 2)}}\left[ {\left( {\frac{\Lambda }{d}r_H ^d +
\frac{{\left( {d - 1} \right)\left( {d - 3} \right)k^2 cr_H ^{d - 4} }}{{(d -
2)}}} \right) }\right. \\ \left. { + 4\pi T_H \left( {\frac{{r_H ^{d - 1}
}}{{2\kappa ^2 }} - kc\left( {d - 1} \right)r_H ^{d - 3} } \right)} \right],
\end{multline}
from which we can obtain the following critical temperature:
\begin{equation}
T_c = - \frac{1}{{4\pi }}\frac{{\frac{\Lambda }{d}r_H ^4 + \frac{{\left( {d -
1} \right)\left( {d - 3} \right)k^2 c}}{{d - 2}}}}{{\frac{{r_H ^3 }}{{2\kappa ^2
}} - kc\left( {d - 1} \right)r_H }}.
\label{eq:Tglobal}
\end{equation}
\begin{figure}[p]
\begin{center}
$\begin{array}{c@{\hspace{0.1in}}c}
\includegraphics[angle=0,width=0.45\textwidth]{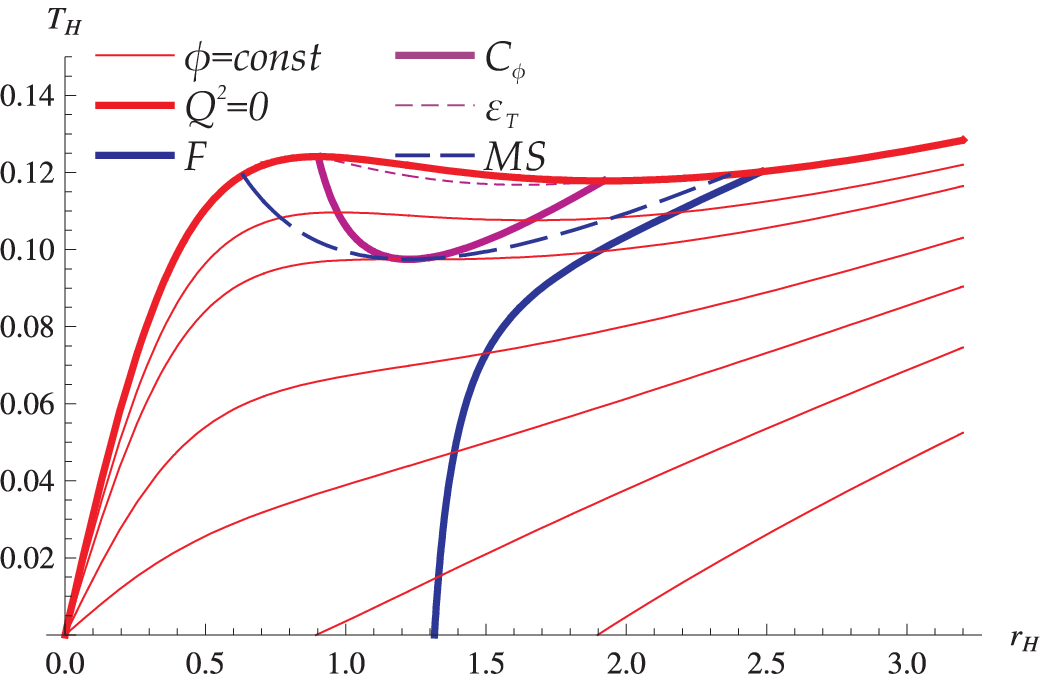} &
\includegraphics[angle=0,width=0.45\textwidth]{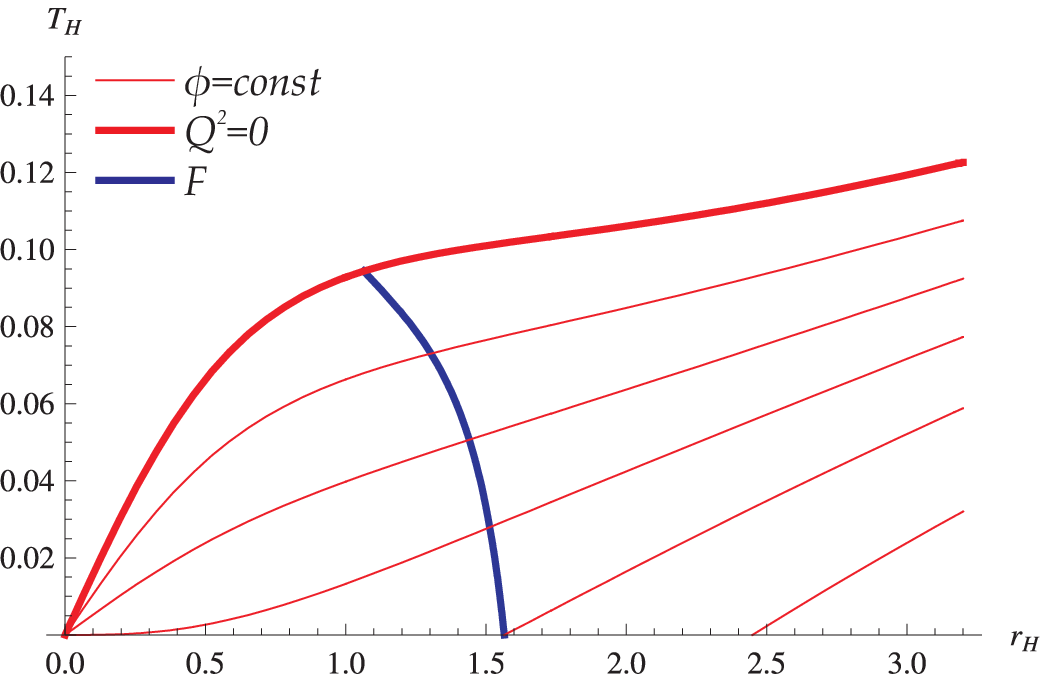} \\
\text{(a)}&\text{(b)}
\end{array}$
\end{center}
\caption{Thermal structure for $k=2$, $d=4$ and (a) $c<\frac{1}{6\kappa^4|\Lambda|}$, (b) $c>\frac{1}{6\kappa^4|\Lambda|}$. }
\label{fig:appk2c1-8}
\end{figure}
\begin{figure}[p]
\begin{center}
$\begin{array}{c@{\hspace{0.1in}}c}
\includegraphics[angle=0,width=0.45\textwidth]{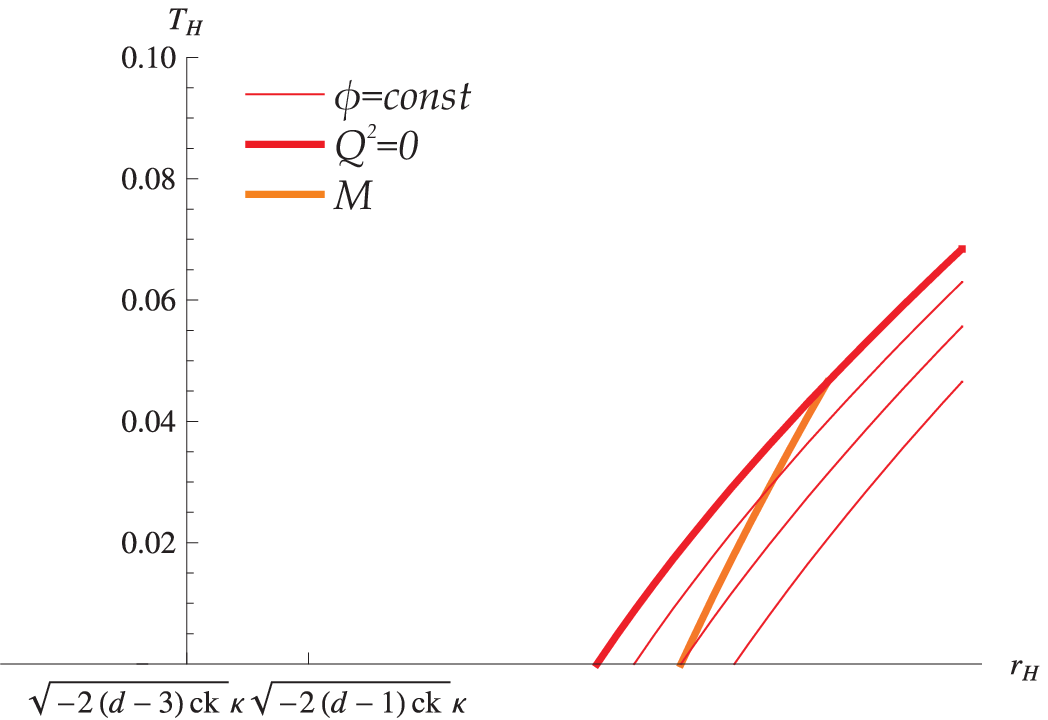} &
\includegraphics[angle=0,width=0.45\textwidth]{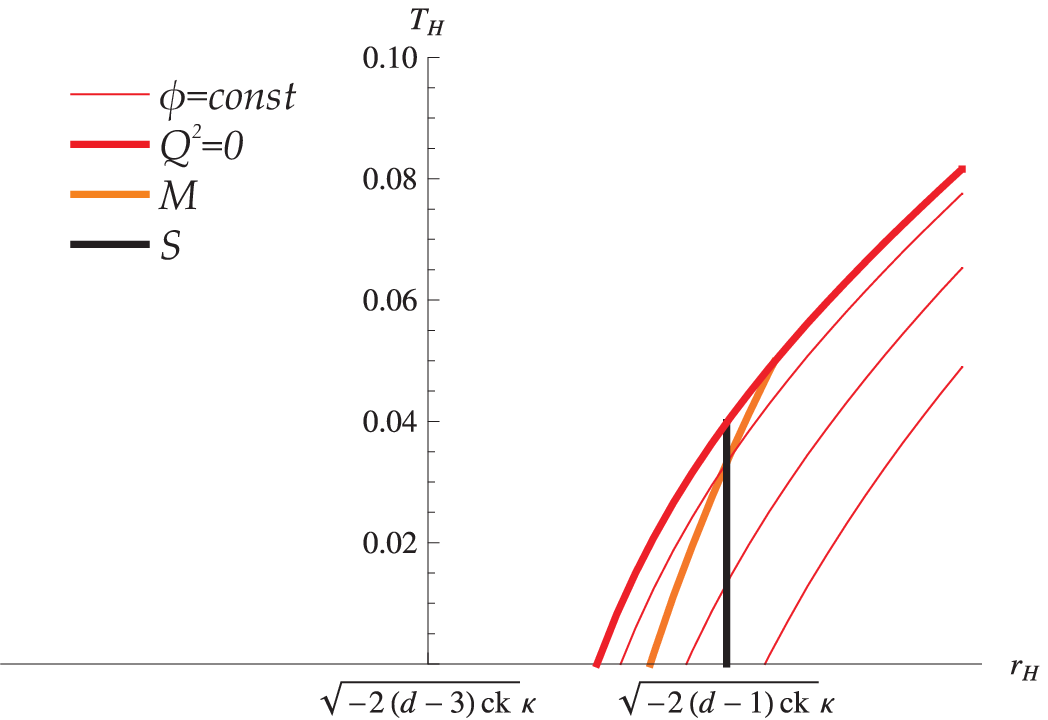} \\
\text{(a)}&\text{(b)}
\end{array}$
\end{center}
\caption{Thermal structure for $d\ge 4$, $k=-\left(d-2\right)$. The two regions of $c$ displayed are: (a) $c<\frac{d^2+d-8}{4\left(d-1\right)\left(d-2\right)\kappa^4|\Lambda|}$ and (b) $\frac{d^2+d-8}{4\left(d-1\right)\left(d-2\right)\kappa^4|\Lambda|}<c<\frac{d\left(d+1\right)}{4\left(d-1\right)\left(d-2\right)\kappa^4|\Lambda|}$. }
\label{fig:appk2c1}
\end{figure}
\begin{figure}[p]
\begin{center}
$\begin{array}{c@{\hspace{0.1in}}c}
\includegraphics[angle=0,width=0.45\textwidth]{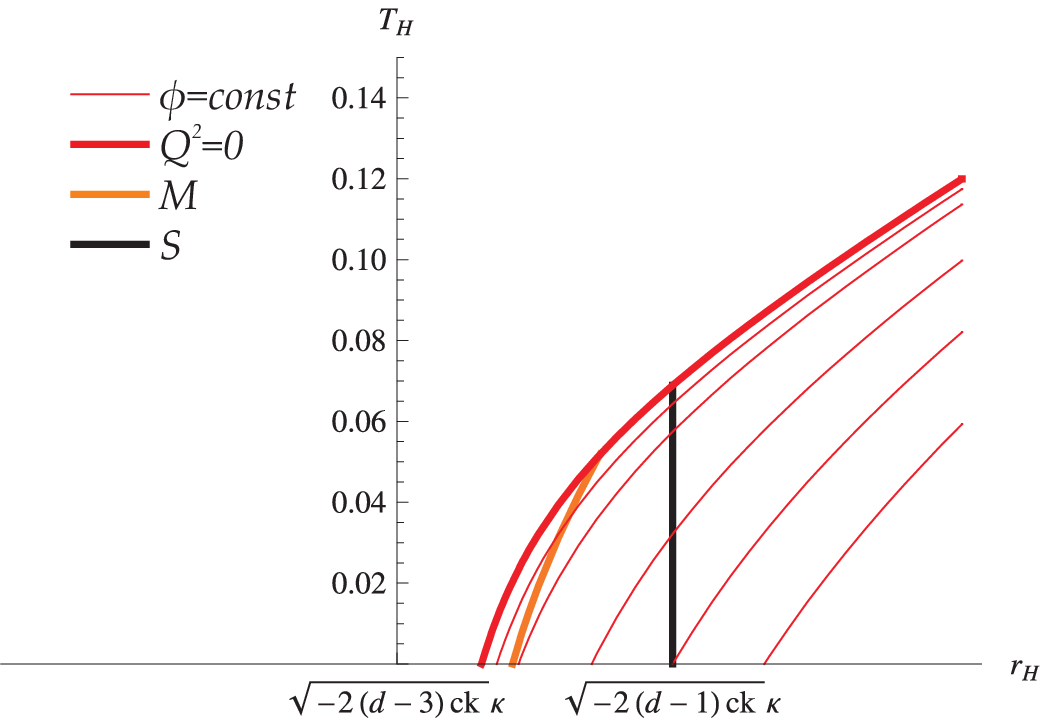} &
\includegraphics[angle=0,width=0.45\textwidth]{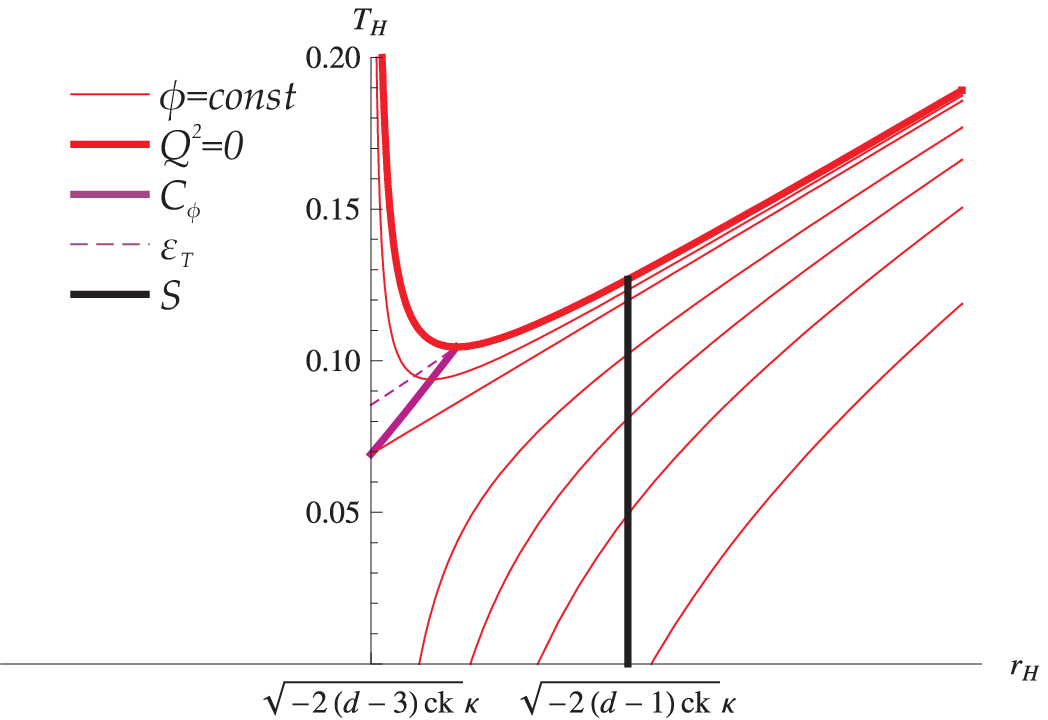} \\
\text{(a)}&\text{(b)}
\end{array}$
\end{center}
\caption{Thermal structure for $d\ge 4$, $k=-\left(d-2\right)$. The two regions of $c$ displayed are: (a) $\frac{d\left(d+1\right)}{4\left(d-1\right)\left(d-2\right)\kappa^4|\Lambda|}<c<\frac{d\left(d-1\right)}{4\left(d-2\right)\left(d-3\right)\kappa^4|\Lambda|}$ and (b) $c>\frac{d\left(d-1\right)}{4\left(d-2\right)\left(d-3\right)\kappa^4|\Lambda|}$. }
\label{fig:appkm2c3-4}
\end{figure}
Finally, we provide the expressions for the specific heat and isothermal electrical permittivity of our system. The thermodynamic derivative expressions for these quantities are given in appendix \ref{sec:thder}. The specific heat at constant electric potential is given by:
\be
\frac{C_\phi}{\Sigma_k} = - \frac{{8\pi ^2 \left( {d - 1} \right)^2 r_H ^{d - 2} T_H \left(
{r_H ^2 + 2c\left( {d - 3} \right)k\kappa ^2 } \right)^2 }}{{\kappa ^2 \left[
{2\left( {d - 1} \right)\pi r_H T_H \left( {r_H ^2 - 2c\left( {d - 3}
\right)k\kappa ^2 } \right) + \kappa ^2 \left( {c\frac{{\left( {d - 1}
\right)\left( {d - 3} \right)\left( {d - 4} \right)}}{{d - 2}}k^2 + \Lambda
r_H ^4 } \right)} \right]}}.
\ee
The electrical permittivity at constant temperature is given by:
\begin{multline}
\frac{\varepsilon _T}{\Sigma_k} = \frac{g}{{2\left( {d - 2} \right)r_H ^{d - 2} }} \\ \times \frac{{2\pi
r_H \left( {r^2 - \tilde k} \right) + T\left( {\frac{1}{2}k\frac{{d - 4}}{{d -
2}}\tilde k + \kappa ^2 \frac{\Lambda }{{d - 1}}r_H ^4 } \right)}}{{2\pi r_H
\left( {\left( {2d - 3} \right)r_H ^2 + \left( {2d - 5} \right)\tilde k}
\right) - T\left( {k\left( {\left( {d - 2} \right)r_H ^2 + \frac{{\left( {d -
3} \right)\left( {d - 4} \right)}}{{2\left( {d - 2} \right)}}\tilde k} \right)
- \kappa ^2 \Lambda r_H ^4 } \right)}}.
\end{multline}
The above expressions give rise to critical temperatures for the thermal and electrical local stability of our thermodynamic system. The critical temperatures indicate a change in sign of the specific heat and electrical permittivity and thus a change in stability. The critical temperature for thermal stability is given by:
\be
T_{C_\phi} = \kappa ^2 \frac{{\frac{{c\left( {d - 1} \right)\left( {d -
3} \right)\left( {d - 4} \right)}}{{d - 2}}k^2 + \Lambda r_H ^4 }}{{2\pi
\left( {d - 1} \right)\left( {r_H ^2 - 2\left( {d - 3} \right)ck\kappa ^2 }
\right)r_H }}.
\label{eq:Tlocal}
\ee
The critical temperature for electrical stability is given by:
\begin{eqnarray}
T^{(1)}_{\varepsilon_\phi} &=& \kappa ^2 \frac{{\frac{{c\left( {d - 1} \right)\left( {d -
3} \right)\left( {d - 4} \right)}}{{d - 2}}k^2 + \Lambda r_H ^4 }}{{2\pi
\left( {d - 1} \right)\left( {r_H ^2 - 2\left( {d - 3} \right)ck\kappa ^2 }
\right)r_H }},\\
T^{(2)}_{\varepsilon_\phi} &=& \frac{{\frac{\left(
{d - 1} \right)k}{{\kappa ^2 }}r_H ^2 + \frac{{c\left( {d - 3} \right)^2 \left( {d - 4} \right)}}{{d - 2}}k^2 -
\Lambda r_H ^4 }}{{2\pi r_H \left( {\frac{{2d - 3}}{{\kappa ^2 }}r_H ^2 +
2\left( {d - 3} \right)\left( {2d - 5} \right)c k} \right)}}.
\end{eqnarray}
Finally, we give the diagrams corresponding to those discussed in section \ref{sec:localgb}.

\section{Explicit regions}
\label{sec:figs}
\subsection{$c$-$\phi^2$ Plane}
Before proceeding to calculate the form of the curves separating regions with qualitatively different behavior in the $c$-$\phi^2$ plane, it is useful to write down the temperature as function of the horizon radius and the electrostatic potential, thus making it simple to have the form of a constant $\phi$ slice in the $r_H$, $T_H$ plane. Using equations (\ref{eq:phi}) and (\ref{eq:tgb}) we easily find that:
\be
T\left( {r_H ,\phi } \right) = \frac{{\kappa ^2 \left( {\frac{{d - 1}}{{\kappa ^2 }}kr_H ^2  + c\frac{{\left( {d - 1} \right)\left( {d - 3} \right)\left( {d - 4} \right)}}{{d - 2}}k^2  - \Lambda r_H ^4  - 2\left( {d - 2} \right)^2 r_H ^2 \phi^2 } \right)}}{{4\pi \left( {d - 1} \right)r_H \left( {r_H ^2  + 2\left( {d - 3} \right)ck\kappa ^2 } \right)}}.
\ee

\subsection*{$k>0$}

First of all, let's study the region $\bold{c>0}$. We can see in figures \ref{fig:k2c1-8}(a), (b) and \ref{fig:k3c1-8}, that a constant $\phi$ curve does not cross the $F$-curve if $\phi > \phi_0$, where $\phi_0$ is the potential for the curve that incudes the extremal black hole with vanishing free energy. So requiring that $T_c \left( {r_{E,F = 0} } \right) = 0$ we can find the radius of this black hole:
\be
r_{E,F = 0} ^4 = - \frac{{d\left( {d - 1} \right)\left( {d - 3} \right)ck^2 }}{{\left( {d - 2} \right)\Lambda }}.
\ee
So solving $T\left( {r_{E,F = 0} ,\phi _0 } \right) = 0$ we get:
\be
\phi _0 ^2 = \frac{{\left( {d - 1} \right)k}}{{2\left( {d - 2} \right)^2 \kappa ^2 }} + \sqrt {\frac{{\left( {d - 1} \right)\left( {d - 3} \right)}}{{d\left( {d - 2} \right)^3 }}} k\sqrt { - c\Lambda }.
\ee
The above curve obviously crosses the $\phi$ axis at $\phi _0 ^2 = \frac{{\left( {d - 1} \right)k}}{{2\left( {d - 2} \right)^2 \kappa ^2 }}$. Notice that for $d>4$ a constant $\phi$ curve with $\phi > \phi_0$, has two disconnected branches, one where the free energy is always positive, and one where the free energy is always negative, while for $d=4$ there is only one branch where the free energy is always negative.
\newline
\newline
Let's now study the temperature behavior as function of the potential. We can see in figures \ref{fig:k2c1-8}(a), (b) and \ref{fig:k3c1-8}, that all curves with $\phi>\phi_0$, where $\phi_0$ is some critical potential, contain extremal black holes of finite size. Actually, for $d=4$ these curves contain exactly one extremal black hole, while for $d>4$ they contain two. These two can be either both globally unstable, or one stable (the larger one) and one unstable. In order to specify this critical $\phi_0$, we need to find the radii of the extremal black holes solving $T\left( r_H ,\phi \right) = 0$. This can be written as:
\be
- \Lambda r_H^4 + \left( {\frac{{\left( {d - 1} \right)k}}{{\kappa ^2 }} - 2\left( {d - 2} \right)^2 \phi ^2 } \right)r_H^2 + \frac{{\left( {d - 1} \right)\left( {d - 3} \right)\left( {d - 4} \right)}}{{d - 2}}ck^2=0.
\ee
This is just a quadratic equation for $r_H^2$. We can always check if there are real roots using the discriminant. Obviously $\phi_0$ is the potential that sets the discriminant equal to zero. That leads us to:
\be
\phi _0 ^2 = \frac{{\left( {d - 1} \right)k}}{{2\left( {d - 2} \right)^2 \kappa ^2 }} + \sqrt {\frac{{\left( {d - 1} \right)\left( {d - 3} \right)\left( {d - 4} \right)}}{{d\left( {d - 2} \right)^5 }}} k\sqrt { - c\Lambda }.
\ee
Notice that if there are real roots, then they are both positive, as we noticed in figure \ref{fig:k3c1-8}. Notice also that for $d=4$, the above curve reduces to a constant $\phi$ curve. Finally, it crosses the $\phi$ axis at the same point as the previous curve.
\newline
\newline
Let's now consider the region $\bold{c<0}$. Observing figures \ref{fig:k2cm1-8}, \ref{fig:k3cm1-50}(a) and (b), \ref{fig:k3cm1-4}(a) and (b) and \ref{fig:k3cm1}, we see that a constant $\phi$ curve enters the negative entropy region, if $\phi < \phi_0$ where $\phi_0$ is the potential for the curve that contains the extremal black hole with zero entropy. From equation (\ref{eq:entropybound}), we know that for $ck<0$ all black holes with vanishing entropy have horizon radius:
\be
r_{S = 0} = \sqrt { - 2\left( {d - 1} \right)ck\kappa ^2 }.
\ee
So requiring $T_H\left( {r_{S=0} ,\phi _0 } \right) = 0 $ we find that $\phi_0$ equals:
\be
\phi _0 ^2 = \frac{k}{{4\left( {d - 2} \right)^3 \kappa ^2 }}\left( {d^2 + d - 8 + 4\left( {d - 1} \right)\left( {d - 2} \right)c\kappa ^4 \Lambda } \right).
\ee
\newline
\newline
Now in order to check whether a constant $\phi$ curve crosses the local stability curve, first we check the behavior of the temperature as $r \to r_{\min}$, where
\be
r_{\min } = \sqrt { - 2\left( {d - 3} \right)ck\kappa ^2 }.
\ee
We solve $\mathop {\lim }\limits_{r \to r_{\min } } T\left( {r,\phi _0 } \right)\left( {r - r_{\min } } \right) = 0$, to find:
\be
\phi _0 ^2 = \frac{k}{{4\left( {d - 2} \right)^3 \kappa ^2 }}\left( {d\left( {d - 1} \right) + 4\left( {d - 2} \right)\left( {d - 3} \right)c\kappa ^4 \Lambda } \right).
\ee
The above curve crosses the $\phi$ axis at $\phi _0 ^2 = \frac{{d\left( {d - 1} \right)k}}{{4\left( {d - 2} \right)^3 \kappa ^2 }}$, which is lower than the point where the curves lying in the $c>0$ region cross the $\phi$ axis, except for $d=4$ when they coincide. For potentials smaller than the above there are two black holes for every temperature, and for larger one or three black holes, except for $d=4$ when there is always one.
\newline
\newline
Notice that the behavior of a constant $\phi$ curve as $r \to r_{\min}$ determines whether this curve crosses the local stability curve an even or odd number of times. In \ref{fig:k2c1-8}(a) and \ref{fig:k3cm1-50}(a) and (b), we see that there are some cases when the local stability behavior changes without change in the behavior of the temperature for $r \to r_{\min } $. This happens for positive $c$ at $d=4$ and for negative $c$ at $d>4$. In both cases the local stability curve has a minimum temperature, and there are three black holes for the same temperature (for some temperatures) if $\phi < \phi_0$, where $\phi_0$ is the potential for the curve that contains the aforementioned minimum. So in order to specify the curve that separates the area with one black hole from the area of three black holes we need to solve
\be
\left. {\frac{{dT_{C_\phi} \left( r \right)}}{{dr}}} \right|_{r = r_{\min T,C_\phi = 0} } = 0,
\ee
\be
T\left( {r_{\min T,C_\phi = 0} ,\phi _0 } \right) = T_{C_\phi} \left( {r_{\min T,C_\phi = 0} } \right).
\ee
For $d>4$ this gives a quite complicated expression. Some important characteristics of this curve is that it crosses the curve separating the odd-even number of black holes areas at the point $\left(c = - \frac{{d\left( {d - 3} \right)}}{{4\left( {d - 1} \right)\left( {d - 2} \right)\kappa ^4 \Lambda }},\phi _0 ^2 = \frac{{d \left( {d^2 - 4d+5} \right)k}}{{\left( {d - 1} \right)\left( {d - 2} \right)^3 \kappa ^2 }}\right)$. For $c$ smaller than the above the local stability curve is one-to-one in $r_H,T_H$ and there is no constant $\phi$ curve containing three black holes with the same temperature. Finally, it crosses the $\phi$ axis at the same point as the curves at the $c>0$ region.

Finally, at $d=4$, $c>0$, it is easy to solve the relevant equations to get
\be
r_{\min T,C_\phi = 0} ^2 = 6ck\kappa ^2,
\ee
\be
\phi _0 ^2 = \frac{{3k}}{{8\kappa ^2 }}\left( {1 + 6ck\kappa ^2 } \right).
\ee
This curve crosses the $c$ axis at $c=-\frac{1}{6\kappa^4\Lambda}$. There are no constant $\phi$ curves containing three black holes at the same temperature for $c$ greater than the above.
\newline
\newline
Let's check a detail about the $d>4$, $c<0$ case. It is obvious that for some potentials the metastable black hole (the smallest one) has manifestly negative entropy. This happens for $\phi < \phi_0$, where $\phi_0$ is the potential for the curve crossing through the black hole with infinite specific heat and zero entropy. Solving $T_{C_\phi} \left( {r_{S = 0} } \right) = T\left( {r_{S = 0} ,\phi _0 } \right)$ we get
\be
\phi _0 ^2 = \frac{k}{{4\left( {d - 2} \right)^4 \kappa ^2 }}\left( {d^3 - 3d^2 + 4d - 8 + 4\left( {d - 1} \right)\left( {d - 2} \right)\left( {d - 4} \right)c\kappa ^4 \Lambda } \right).
\ee
The above curve is tangent at the curve separating the one black hole region from the three black holes region at $c = - \frac{{\left( {d - 3} \right)\left( {d - 4} \right)\left( {2d - 3} \right)}}{{4\left( {d - 1} \right)\left( {d - 2} \right)\left( {2d - 5} \right)\kappa ^4 \Lambda }}$.
\newline
\newline
Finally, let's be a little careful about the Hawking-Page phase transition for $c<0$. It is possible that a constant $\phi$ curve crosses the global stability curve; however, this happens at an area of locally unstable black holes or negative entropy black holes. The first case occurs if $c > - \frac{{d\left( {d - 3} \right)}}{{4\left( {d - 1} \right)\left( {d - 2} \right)\kappa ^4 \Lambda }}$ for $\phi > \phi_0$, where $\phi_0$ is the potential for the curve crossing the mutual point of the global stability and local stability curve. We can specify this point by solving$
\left. {\frac{{dT_{c} \left( r \right)}}{{dr}}} \right|_{r = r_{\min T,F = 0} } = 0$, and then $T\left( {r_{\min T,F = 0} ,\phi _0 } \right) = T_{c} \left( {r_{\min T,F = 0} } \right)$, to get a very complicated expression.

The second case occurs when $c < - \frac{{d\left( {d - 3} \right)}}{{4\left( {d - 1} \right)\left( {d - 2} \right)\kappa ^4 \Lambda }}$ and $\phi > \phi_0$ where $\phi_0$ is the potential for the curve crossing the mutual point of the global stability and zero entropy curve. We need to solve $T_{F = 0} \left( {r_{S = 0} } \right) = T\left( {r_{S = 0} ,\phi _0 } \right)$ to get
\be
\phi _0 ^2 = \frac{k}{{4d\left( {d - 2} \right)^2 \kappa ^2 }}\left( {d\left( {d + 1} \right) + 4\left( {d - 1} \right)\left( {d - 2} \right)c\kappa ^4 \Lambda } \right).
\ee

The two curves above meet at $\left(c = - \frac{{d\left( {d - 3} \right)}}{{4\left( {d - 1} \right)\left( {d - 2} \right)\kappa ^4 \Lambda }}, \phi _0 ^2 = \frac{{\left( {d - 1} \right)k}}{{2\left( {d - 2} \right)^2 \kappa ^2 }}\right)$. At this $c$ they are also parallel, thus forming a continuous and smooth curve. For $c$ equal to the above the entropy bound occurs at exactly the minimum of the global stability curve.

One final detail. If one extends the curves above, in the complementary regions of $c$ we used, there is an area created, containing curves with three Hawking-Page phase transitions. However, always only one of them happens in a region with both positive entropy and positive specific heat. Specifically in the relevant region for $c > - \frac{{d\left( {d - 3} \right)}}{{4\left( {d - 1} \right)\left( {d - 2} \right)\kappa ^4 \Lambda }}$, the two black holes smaller in radius, have negative specific heat, and the smallest has negative entropy, while in the case $c < - \frac{{d\left( {d - 3} \right)}}{{4\left( {d - 1} \right)\left( {d - 2} \right)\kappa ^4 \Lambda }}$, the two smaller ones have negative entropy and the smallest has negative specific heat. In this region the curve separating odd-even number of black holes appears, and once one enters in the region of two black holes, there are actually only two positions of Hawking-Page phase transition, again only one being valid, as the other is in a region with negative entropy. This happens for $c$ as small as $c = \frac{{d\left( {d - 1} \right)\left( {2d - 5} \right)}}{{4\left( {d - 2} \right)\left( {d - 3} \right)\left( {2d - 3} \right)\kappa ^4 \Lambda }}$, where our curve ends on the odd-even separating curve.

\subsection*{$k<0$}

Observing figures \ref{fig:km2c1-4}(a), (b) and \ref{fig:km2c17}(a), we see that a constant $\phi$ curve enters the negative mass region if $\phi < \phi_0$, where $\phi_0$ is the potential for the curve that contains the extremal massless black hole. We can find the horizon radius of the extremal massless black hole using equation (\ref{eq:Tmassless}):
\be
r_{E,M=0} = \frac{{k d}}{{2\kappa ^2 \Lambda }}\left( {1 + \sqrt {\frac{{4\left( {d - 3} \right)^2 }}{{d\left( {d - 2} \right)}}c\kappa ^4 \Lambda } } \right).
\ee
So requiring that $T_H\left( {r_{E,M=0} ,\phi _0 } \right) = 0 $, we find that $\phi_0$ equals:
\be
\phi _0 ^2 = - \frac{k}{{2\left( {d - 2} \right)^2 \kappa ^2 }}\left( {1 + \frac{{4\frac{d - 3}{d} c\kappa ^4 \Lambda }}{{1 + \sqrt {1 + 4\frac{\left( {d - 3} \right)^2}{d\left( {d - 2} \right)} c\kappa ^4 \Lambda } }}} \right).
\ee
This curve crosses the $c$ axis at $c = - \frac{{d\left( {d - 1} \right)}}{{4\left( {d - 2} \right)\left( {d - 3} \right)\kappa ^4 \Lambda }}$. For $c$ larger than the above there are no negative mass black holes.
\newline
\newline
Observing figures \ref{fig:km2c1-4}(b), \ref{fig:km2c17}(a) and (b) we see that a constant $\phi$ curve enters the negative entropy region, if $\phi < \phi_0$ where $\phi_0$ is the potential for the curve that contains the extremal black hole with zero entropy. From equation (\ref{eq:entropybound}), we know that for $ck<0$ all black holes with vanishing entropy have horizon radius:
\be
r_{S = 0} = \sqrt { - 2\left( {d - 1} \right)ck\kappa ^2 }.
\ee
So requiring $T_H\left( {r_{S=0} ,\phi _0 } \right) = 0 $ we find that $\phi_0$ equals:
\be
\phi _0 ^2 = \frac{k}{{4\left( {d - 2} \right)^3 \kappa ^2 }}\left( {d^2 + d - 8 + 4\left( {d - 1} \right)\left( {d - 2} \right)c\kappa ^4 \Lambda } \right).
\ee
The above curve crosses the c axis at $c = - \frac{{d^2 + d - 8}}{{4\left( {d - 1} \right)\left( {d - 2} \right)\kappa ^4 \Lambda }}$. This is the minimum $c$ for which there are negative entropy black holes.
\newline
\newline
Now trying to combine the above, we check whether a constant $\phi$ curve enters a region of negative mass black holes with positive entropy. If we observe figure \ref{fig:km2c1-4}(b) this happens if $\phi < \phi_0$, where $\phi_0$ is the potential for the curve that contains the massless black hole with zero entropy. Thus solving $T_H\left( {r_{S = 0} ,\phi _0 } \right) = T_H^{M = 0} \left( {r_{S = 0} } \right)$, we get:
\be
\phi _0 ^2 = - \frac{k}{{4d\left( {d - 2} \right)^2 \kappa ^2 }}\left( {d\left( {d + 1} \right) + 4\left( {d - 1} \right)\left( {d - 2} \right)c\kappa ^4 \Lambda } \right).
\ee
This curve crosses the massless curve and the entropy curve, at the same point $\left(
c = - \frac{{d\left( {d^2 - 5} \right)}}{{4\left( {d - 1} \right)^2 \left( {d - 2} \right)\kappa ^4 \Lambda }}, \phi _0 ^2 = - \frac{k}{{\left( {d - 1} \right)\left( {d - 2} \right)^2 \kappa ^2 }}\right)$. This value of $c$ is the minimum for which there are positive mass black holes with negative entropy. This curve also crosses the $c$ axis at $c = - \frac{{d\left( {d + 1} \right)}}{{4\left( {d - 1} \right)\left( {d - 2} \right)\kappa ^4 \Lambda }}$. This is the maximum $c$ for which negative mass black holes with positive entropy exist.
\newline
\newline
Finally, in order to know if a constant $\phi$ curve crosses the local stability curve we need just to specify the behavior of the temperature as horizon radius approaches the minimum radius
\be
r_{\min } = \sqrt { - 2\left( {d - 3} \right)ck\kappa ^2 },
\ee
as we can see in figure \ref{fig:km2c17}(b). If it the curve goes to infinity, that means that it had a minimum at some horizon radius. There it has crossed the local stability curve. If is goes to minus infinity then from the figure we can see that it is one-to-one, and obviously it has not crossed the local stability curve. So the critical $\phi_0$ is the one for which $\mathop {\lim }\limits_{r \to r_{\min } } T\left( {r,\phi _0 } \right)\left( {r - r_{\min } } \right) = 0$. Solving this we get
\be
\phi _0 ^2 = \frac{k}{{4\left( {d - 2} \right)^3 \kappa ^2 }}\left( {d\left( {d - 1} \right) + 4\left( {d - 2} \right)\left( {d - 3} \right)c\kappa ^4 \Lambda } \right).
\ee
This curve crosses the $c$ axis at $c = - \frac{{d\left( {d - 1} \right)}}{{4\left( {d - 2} \right)\left( {d - 3} \right)\kappa ^4 \Lambda }}$. This is the minimum $c$ for which there are locally unstable black holes. It coincides with the minimum $c$ for which there are no negative mass black holes.

\subsection{$\varepsilon$-$\phi^{-2}$ Plane}
\subsection*{$k>0$}
For $\bold{\varepsilon<0}$ the $\varepsilon$-bound excludes the region of the black holes where the small $\chi$ expansion is good. That allows us to specify the regions of qualitatively different thermal behavior approximately, using the small $\chi$ expansion.

First let's specify which constant $\phi$ curves contain an extremal black hole. We can see from figure \ref{fig:em01k2}(a), that extremal black holes are contained in constant $\phi$ curves only if $\phi>\phi_E$, where $\phi_E$ is the potential of the curve containing the extremal black hole that saturates the $\varepsilon$-bound. Using the zeroth order forms for the potential (\ref{eq:phi}) and the temperature (\ref{eq:tgb}), we acquire the result:
\be
\phi _E = \frac{1}{{\left( {d - 2} \right)\kappa }}\sqrt {\frac{{\left( {d - 1}
\right)k}}{{2\left( {- 108g^4 \Lambda \varepsilon + 1} \right)}}},
\label{eq:appphie0}
\ee
while if one uses the first order corrected ones (\ref{eq:Teps}) and ({\ref{eq:phieps}), one finds:
\be
\phi _E = \frac{{85d - 116}}{{3\left( {d - 2} \right)\left( {3d - 4}
\right)\kappa }}\sqrt {\frac{{\left( {d - 1} \right)k}}{{6\left( {- 2916g^4
\Lambda \varepsilon + 29} \right)}}}.
\label{eq:appphie1}
\ee
Notice that both results are of the form:
\be
\phi_E = \frac{A}{{\sqrt {\varepsilon - \varepsilon _1^+ } }}
\ee
and actually are very close to the actual numbers we calculated numerically, taking into account the full dependence on $\varepsilon$.

Similarly in order to calculate which potentials cross the $F$-curve, thus allowing for Hawking-Page phase transitions, we need to specify the potential containing the black hole with zero free energy that saturates the $\varepsilon$-bound, $\phi_{HP}$. Then a constant $\phi$ curve crosses the $F$-curve provided that $\phi<\phi_{HP}$. Once again using the zeroth order result for the potential and the free energy we get:
\be
\phi _{HP} = \frac{1}{{\left( {d - 2} \right)\kappa }}\sqrt {\frac{{d\left( {d
- 1} \right)k}}{{2\left( {108\left( {d - 2} \right)g^4 \Lambda \varepsilon +
d} \right)}}},
\ee
while at first order:
\be
\phi _{HP} = \frac{{85d - 116}}{{3\left( {d - 2} \right)\left( {3d - 4}
\right)\kappa }}\sqrt {\frac{{d\left( {d - 1} \right)k}}{{6\left( {2916\left(
{d - 2} \right)g^4 \Lambda \varepsilon + \frac{{d\left( {83d - 112}
\right)}}{{3d - 4}}} \right)}}}.
\ee
We again observe that both results are of the form:
\be
\phi = \frac{A}{{\sqrt {-\left(\varepsilon - \varepsilon _0 \right)} }},
\ee
but now the value of $\varepsilon_0$ depends on the number of dimensions. It looks like $\varepsilon_0$ will equal $-\varepsilon_1^+$ at the limit of infinite number of dimensions.

We can specify similarly the curve separating regions with one and two black holes to find at zeroth order:
\be
\phi _{LS} = \frac{1}{{\left( {d - 2} \right)\kappa }}\sqrt {\frac{{\left( {d - 1}
\right)k}}{{2\left( {108g^4 \Lambda \varepsilon + 1} \right)}}},
\label{eq:appphie0}
\ee
Notice that this curve is of the same form as the curve specifying the existence of extremal black holes, where we have substituted $\varepsilon$ with $-\varepsilon$. Thus the positions of the poles of the two curves are opposite, and so are the inclinations of the two curves in the $\varepsilon$-$\phi^{-2}$ plane. This agrees very well with our numerical results, where we have taken into account the full $\varepsilon$ dependence.

We further notice that the inclination of the curve separating Hawking-Page and non-Hawking-Page constant $\phi$ curves in the $\varepsilon-\frac{1}{\phi^2}$ plane is a fraction $\frac{d-2}{2}$ of that of the curve separating one and two black hole solutions, the later being equal to the opposite of the curve separating constant $\phi$ curves containing extremal black holes. This is consistent with our numerical results.

We can specify specify the values of $\varepsilon _1 ^ +$ and $\varepsilon _2 ^ +$ in figure \ref{fig:ek2}. $\varepsilon _1 ^ +$ is the value of $\varepsilon$ that the extremality curve crosses the $\frac{1}{\phi^2}$ axis, thus it is the minimum $\varepsilon$, that extremal black holes exist. Using the first order result we acquired above we find:
\be
\varepsilon _1 ^ +   = \frac{{29}}{{2916g^4 \Lambda }}.
\ee
Finally, $\varepsilon _2 ^ +$ is the value of $\varepsilon$ where the Hawking-Page curve crosses the extremality curve and ends. Again using the first order results we got above we find:
\be
\varepsilon _2 ^ +   = \frac{{2d}}{{2916\left( {3d - 4} \right)g^4 \Lambda }}.
\ee

Unfortunately for $\bold{\varepsilon>0}$ the critical constant $\phi$ curves cross the $F$-curve and $C_\phi$-curve in regions where the small $\chi$ expansion is not good, thus making the numerical calculation of the different regions necessary.

\subsection*{$k<0$}

As we can see in figure \ref{fig:em003km2}(a) and (b) for $\bold{\varepsilon<0}$, the constant $\phi$ curves are again separated in two classes by the curve containing the extremal black hole that saturates the $\varepsilon$-bound. Repeating the same exact procedure as in the $k>0$ we find the same exact result given by equations (\ref{eq:appphie0}) and (\ref{eq:appphie1}). Notice though that now the curves containing extremal black holes are those with $\phi<\phi_E$.

Finally, we can calculate which constant $\phi$ curves contain massless black holes. They are those with $\phi<\phi_M$, where $\phi_M$ is the potential of the curve containing the massless black hole that saturates the $\varepsilon$-bound. Using the expressions for the thermodynamic energy (\ref{eq:mueps}) we get at zero order:
\be
\phi _M = \frac{1}{\kappa }\sqrt { - \frac{{d\left( {d - 1} \right)k}}{{2\left( {d - 2} \right)\left( {108\left( {d - 2} \right)g^4 \Lambda \varepsilon + d} \right)}}}
\ee
and at first order in $\varepsilon$:
\be
\phi _M = \frac{{85d - 116}}{{3\left( {d - 2} \right)\left( {3d - 4} \right)\kappa }}\sqrt { - \frac{{d\left( {d - 1} \right)k}}{{6\left( {2916\left( {d - 2} \right)g^4 \Lambda \varepsilon + \frac{{d\left( {83d - 112} \right)}}{{3d - 4}}} \right)}}}.
\ee
Notice that strangely these expression are identical with those for $\phi_{HP}$ for the $k>0$ case, having substituted $k$ with $-k$. The inclination of this curve in the $\varepsilon-\frac{1}{\phi^2}$ plane is a fraction $\frac{d-2}{2}$ of that of the curve separating regions with extremal and without extremal black holes.

Let's specify the values $\varepsilon _1 ^ -$ and $\varepsilon _2 ^ -$ in figure \ref{fig:ekm2}. $\varepsilon _2 ^ -$ is the value of $\varepsilon$ that the extremality curve crosses the $\frac{1}{\phi^2}$ axis. As in the $k>0$ case we find:
\be
\varepsilon _2 ^ -   = \frac{{29}}{{2916g^4 \Lambda }} = \varepsilon _1 ^ +.
\ee
$\varepsilon _1 ^ -$, is the value where the curve separating constant $\phi$ curves containing and non-containing massless black holes meets the extremality curve and ends. Using the first order results we acquired above we find:
\be
\varepsilon _1 ^ -   = \frac{{d\left( {85d - 114} \right)}}{{2916\left( {3d - 4} \right)g^4 \Lambda }}.
\ee

Once again we cannot acquire the critical curves using the small $\chi$ expansion in the $\bold{\varepsilon>0}$ case.

\end{document}